\documentclass[twocolumn, twocolappendix, trackchanges]{aastex631}
\usepackage{cancel, gensymb, graphicx, graphbox, float, parskip, textcomp}
\usepackage{natbib}
\usepackage{color}
\usepackage[super]{nth}
\usepackage{units}
\usepackage{textgreek}
\usepackage{rotating}
\usepackage[T1]{fontenc}    
\usepackage{hyperref}

\newcommand{\um}{\text{ \textmu m}}

\newcommand{\teff}{T_{\text{eff}}}
\newcommand{\brgamma}{Br-\textgamma~}

\shorttitle{Estimate of a High Galactic Center Young Star Binary Fraction}
\shortauthors{Gautam et al.}

\defcitealias{Gautam:2019}{G19}

\begin{document}
\title{An Estimate of the Binary Star Fraction Among Young Stars\\at the Galactic Center: Possible Evidence of a Radial Dependence}

\author[0000-0002-2836-117X]{Abhimat K. Gautam}
\affiliation{Department of Physics and Astronomy, University of California, Los Angeles, USA}

\author[0000-0001-9554-6062]{Tuan Do}
\affiliation{Department of Physics and Astronomy, University of California, Los Angeles, USA}

\author[0000-0003-3230-5055]{Andrea M. Ghez}
\affiliation{Department of Physics and Astronomy, University of California, Los Angeles, USA}

\author[0000-0003-3765-8001]{Devin S. Chu}
\affiliation{Department of Physics and Astronomy, University of California, Los Angeles, USA}

\author[0000-0003-2874-1196]{Matthew W. Hosek Jr.}
\altaffiliation{Brinson Prize Fellow}
\affiliation{Department of Physics and Astronomy, University of California, Los Angeles, USA}

\author[0000-0001-5972-663X]{Shoko Sakai}
\affiliation{Department of Physics and Astronomy, University of California, Los Angeles, USA}

\author[0000-0002-9802-9279]{Smadar Naoz}
\affiliation{Department of Physics and Astronomy, University of California, Los Angeles, USA}
\affiliation{Mani L. Bhaumik Institute for Theoretical Physics, University of California, Los Angeles, USA}

\author[0000-0002-6753-2066]{Mark R. Morris}
\affiliation{Department of Physics and Astronomy, University of California, Los Angeles, USA}

\author[0000-0001-5800-3093]{Anna Ciurlo}
\affiliation{Department of Physics and Astronomy, University of California, Los Angeles, USA}

\author[0009-0004-0026-7757]{Zo{\"e} Haggard}
\affiliation{Department of Physics and Astronomy, University of California, Los Angeles, USA}

\author[0000-0001-9611-0009]{Jessica R. Lu}
\affiliation{Department of Astronomy, University of California, Berkeley, USA}

\correspondingauthor{Abhimat K. Gautam}
\email{abhimat@ucla.edu}

\begin{abstract}
    \raggedright
    We present the first estimate of the intrinsic binary fraction of young stars across the central $\approx 0.4$~pc surrounding the supermassive black hole (SMBH) at the Milky Way Galactic center (GC). This experiment searched for photometric variability in 102 spectroscopically confirmed young stars, using 119 nights of $10''$-wide adaptive optics imaging observations taken at W. M. Keck Observatory over 16 years in the $K'$- $[2.1 \um]$ and $H$- $[1.6 \um]$ bands.
    We photometrically detected three binary stars, all of which are situated more than $1''$ (0.04 pc) from the SMBH and one of which, S2-36, is newly reported here with spectroscopic confirmation. All are contact binaries or have photometric variability originating from stellar irradiation.
    To convert the \textit{observed} binary fraction into an estimate of the \textit{underlying} binary fraction, we determined the experimental sensitivity through detailed light curve simulations, incorporating photometric effects of eclipses, irradiation, and tidal distortion in binaries. The simulations assumed a population of young binaries, with stellar ages (4 Myr) and masses matched to the most probable values measured for the GC young star population, and underlying binary system parameters (periods, mass ratios, and eccentricities) similar to those of local massive stars. As might be expected, our experimental sensitivity decreases for eclipses narrower in phase.
    The detections and simulations imply that the young, massive stars in the GC have a stellar binary fraction $\geq 71\%$ (68\% confidence), or $\geq 42\%$ (95\% confidence). This inferred GC young star binary fraction is consistent with that typically seen in young stellar populations in the solar neighborhood.
    Furthermore, our measured binary fraction is significantly higher than that recently reported by \citet{Chu:2023} based on radial velocity measurements for stars $\lesssim 1''$ of the SMBH. Constrained with these two studies, the probability that the same underlying young star binary fraction extends across the entire region is $< 1.4\%$. This tension provides support for a radial dependence of the binary star fraction and, therefore, for the dynamical predictions of binary merger and evaporation events close to the SMBH.
\end{abstract}

\section{Introduction} 
\label{sec:introduction}
At a distance of $\approx 8$ kpc, the Milky Way Galactic center (GC) hosts the closest nuclear star cluster, with a supermassive black hole (SMBH) located at the location of the radio source Sgr A* and having a mass of $\approx 4 \times 10^6$ $M_{\odot}$ \citep{Ghez:2008, Schodel:2009, Gillessen:2009a, Boehle:2016, Gillessen:2017, GRAVITY-Collaboration:2018, Do:2019a, GRAVITY-Collaboration:2019}. A dense stellar population surrounds the SMBH, with $\approx 3 \times 10^6$ $M_{\odot}$ of stellar mass enclosed in just the central parsec \citep[e.g.,][]{Feldmeier:2014}. Adaptive optics equipped 8--10 m class near-infrared (NIR) telescopes have allowed spectroscopic studies that have revealed a population of more than 100 young, massive stars within the central parsec \citep{Bartko:2010, Pfuhl:2011, Do:2013a, Stostad:2015}. The population of young stars surrounding the SMBH within a radius of $\approx 0.5$~pc constitutes the \emph{Young Nuclear Cluster} (YNC).

The YNC is composed of stars of age 3 -- 8 MYr with a top-heavy initial mass function (IMF) \citep{Bartko:2010, Lu:2013}. At least 20\% of the young stars have orbits around the SMBH making up a disk structure, known as the clockwise disk, with the inner edge of the disk extending down to $\approx 0.03$ pc from the SMBH  \citep{Levin:2003, Genzel:2003, Paumard:2006, Lu:2009, Bartko:2009, Yelda:2014, Jia:2023}. \citet{Naoz:2018} predict that disk membership may be even higher than what has been inferred from previous studies which did not account for velocity shifts from stellar binaries. Closer towards the SMBH is the S-star population, which are largely in highly eccentric orbits and have a wide range of orbit orientations around the SMBH \citep{Ghez:2003, Eisenhauer:2005}. The young stars in the S-star population are observed to be B-type main-sequence stars \citep[e.g.][]{Habibi:2017}.

Open questions still remain about the young stars in the YNC, namely: how did these young stars form at the GC and how is the young star population shaped by interactions within the high stellar density environment at the GC? A measurement of the young stars' binary fraction is a powerful method to address both questions since the binary fraction constrains the \emph{star formation history} of the young GC stars and the \emph{dynamical evolution} of the stars in the GC environment in the following ways:
\begin{itemize}
    \item \textbf{Constraints on the formation of young GC stars:}
    The young cluster age, the proximity of the YNC to the central SMBH, and the lack of young stars outside the central $\approx 0.5$ pc YNC radius suggests that a recent star formation event has occurred \emph{in situ} \citep[e.g.,][]{Nayakshin:2005b, Morris:2023}.
    While the formation of young stars in the GC is challenged by strong tidal forces from the SMBH \citep{Morris:1993a}, the young star clockwise disk hints at a possible avenue for star formation in the GC, where young stars formed through fragmentation of a previous accretion disk surrounding the SMBH \citep[e.g.][]{Levin:2003, Milosavljevic:2004, Nayakshin:2005a}. The degree of multiplicity in a young star population constrains fragmentation during star formation \citep[e.g.,][]{Duchene:2013}, so a measurement of the binary fraction of GC young stars is crucial to test whether disk formation or other \textit{in situ} star formation models are indeed viable explanations for the presence of young stars at the GC.
    
    \item \textbf{Constraints on dynamical evolution of young GC stars:}
    The high stellar densities and the large mass of the SMBH at the GC are expected to lead to dynamical phenomena that leave an observable imprint on the GC stellar binary fraction.
    Frequent close interactions with the high stellar densities are expected to result in loosely bound (i.e., \emph{dynamically soft}) binaries to get more loosely bound over time and eventually \emph{evaporate}, where the stellar binary's component members become unbound from each other \citep{Alexander:2014, Rose:2020}.
    Stellar binaries also form a hierarchical triple with the central SMBH. In such a setup, the eccentric Kozai-Lidov (EKL) effect predicts the orbits of stellar binaries to enter occasional periods of high eccentricity, leading to possible mergers of the component binary stars \citep{Naoz:2016, Stephan:2016, Stephan:2019}. Both processes, binary evaporation and mergers due to the EKL mechanism, are expected to \emph{reduce} the GC stellar binary fraction, leaving the most prominent imprint at close distances to the central SMBH. In this context, the GC stellar binary fraction is important to measure in order to constrain the frequency and degree of these dynamical predictions.
  
\end{itemize}

In this experiment, we measure the binary fraction of young, massive GC stars using near-infrared adaptive optics photometry. Previous studies measuring the eclipsing or radial velocity (RV) binary fraction in the GC \citep{Pfuhl:2014, Gautam:2019} and in the nearby Arches massive star cluster \citep{Clark:2023} have found consistency with the observed binary fractions of local OB star populations, such as with the eclipsing binary fractions of \citet{Lefevre:2009}.
These results suggest that the young GC stars likely have a high binary fraction similar to that of local B stars \citep[$\approx 60\%$--70\%; e.g.,][]{Duchene:2013, Moe:2013, Moe:2015b} and of local O stars \citep[$69\% \pm 9\%$;][]{Sana:2012}. 
However, since the inferred binary fraction can be affected by observational cadence and experimental precision, previous GC studies have been limited in their constraints on the underlying \emph{intrinsic} binary fraction of GC young stars.
\citet{Chu:2023} used spectroscopic observations to constrain the intrinsic binary fraction of young B stars in the S-star population, located $\lesssim 1''$ from Sgr A* (corresponding to $\lesssim 0.04$ pc from the SMBH), finding that it can be at most 47\% with their null detection.
This study was sensitive to a median of $\approx 50$ \nicefrac{km}{s} in radial velocity (RV) variation, corresponding to a median upper limit on binary companion mass of $3.7 M_\odot$ for the 16 young stars in their sample.
At larger distances from the SMBH, dynamical models predict higher binary fractions \citep[e.g.,][]{Stephan:2016}.
However, the binary fraction at these larger distances from the SMBH had not yet been measured prior to the results presented in this work, in which we find evidence for a high young star binary fraction at the GC outside the young S-star cluster.

Section~\ref{sec:observations_dataset} provides an overview of the observations used in our experiment and a description of the stellar sample used for our analysis.
Section~\ref{sec:methodology_and_results} gives the details of our methods and provides our results.
To search for binary systems in our sample we performed a periodicity search on the stellar light curves. The implementation details of our periodicity search are provided in Section~\ref{sub:photometric_periodicity_search}, while Section~\ref{ssub:results_detection_of_periodic_signals} presents the detection resulting from our periodicity search experiment.
We next describe the sensitivity of our experiment with the methods detailed in Section~\ref{sub:binary_star_fraction_determination}. An estimate of our experiment's sensitivity allow a measure of the underlying young star binary fraction from our detections, detailed in Section~\ref{ssub:results_bin_frac}.
We put the inference of a high young GC star binary fraction in context of the current leading star formation and dynamical evolution models for the GC environment in Section~\ref{sec:discussion}.
Finally, we summarize our conclusions in Section~\ref{sec:conclusions}.


\begin{figure*}[htb!]
  \centering
    \includegraphics[width=\textwidth]{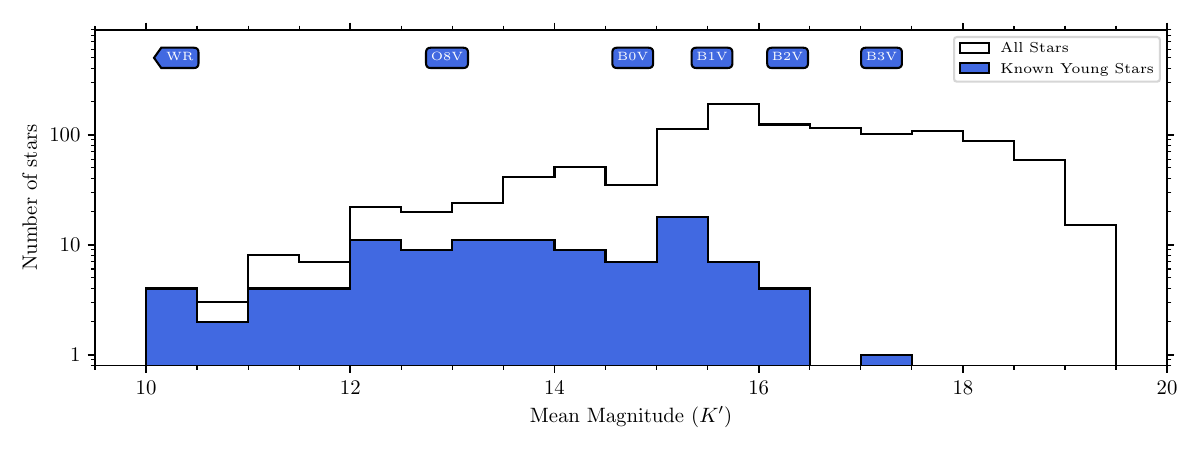}
  \caption{The distribution of observed fluxes of this experiment's stellar sample in $K'$-band. The population of known young stars is plotted as the solid blue histogram. For reference, expected fluxes of Wolf-Rayet and main sequence spectral types for young stars at the Galactic center are indicated above the histogram, assuming mean extinction towards the GC as measured by \citet{Schodel:2010}. Expected young star fluxes are derived for a 4 Myr population using \textsc{SPISEA} \citep{Hosek:2020} following the procedure detailed in \S~\ref{par:stellar_params}, with $T_{\text{eff}}$ values matched to spectral type using \citet{Pecaut:2013}. The histogram of all stars' observed fluxes are indicated in white.}
  \label{fig:sample_mag_dist}
\end{figure*}

\begin{figure*}[htb!]
  \centering
    \includegraphics[width=\textwidth]{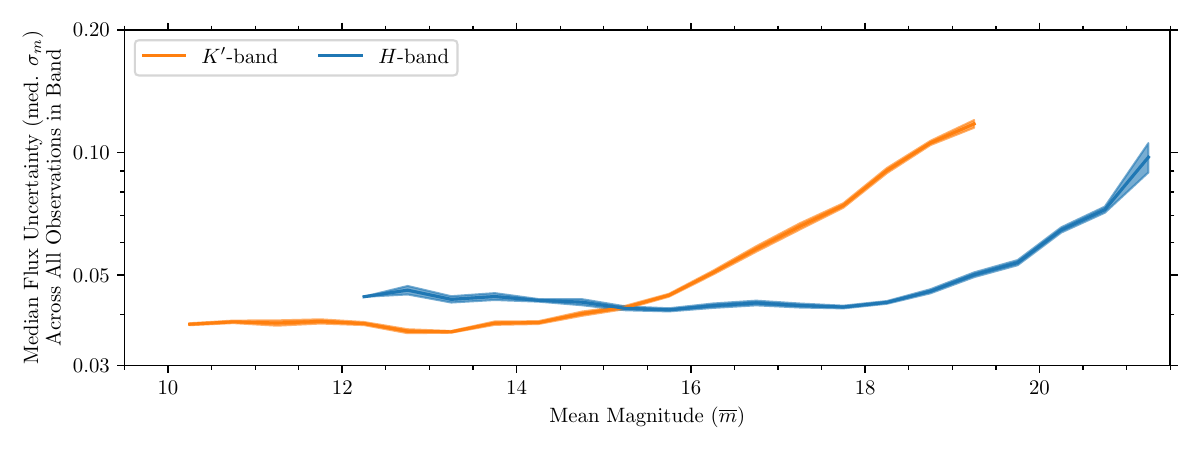}
  \caption{The median photometric precision of this experiment's stellar sample in $K'$-band (orange solid line) and in $H$-band (blue solid line) plotted as a function of stellar brightness in each band, computed in bins of mean stellar magnitude. The shaded colored regions indicate the median absolute deviation[$\sigma_m$] $/ \sqrt{N_{\text{stars}}}$ in each bin. In $K'$-band, the median photometric uncertainty across all observations is $\approx 3.5\%$ out to $m_{K'} \approx 15$. While in the $H$-band, the median photometric uncertainity is $\approx 4 \%$ mag out to $m_{H} \approx 18$. Fainter stars typically have higher photometric uncertainties.}
  \label{fig:mag_magerr}
\end{figure*}

\section{Observations and Dataset} 
\label{sec:observations_dataset}

Our binary fraction experiment employed \emph{Galactic Center Orbits Initiative} (GCOI; PI: A.~M.~Ghez) adaptive optics (AO) imaging observations of the Galactic center. The experiment's images were obtained at the 10-m W. M. Keck II telescope with the NIRC2 near-infrared (NIR) facility imager (PI: K. Matthews) and were used to measure the stellar fluxes used in this work. The photometric analysis used data taken between 2006--2022 in the $K'$- ($\lambda_0 = 2.124 \um$, $\Delta \lambda = 0.351 \um$) and $H$-bandpass ($\lambda_0 = 1.633 \um$, $\Delta \lambda = 0.296 \um$). This work built off of the 2006--2017 $K'$-band dataset presented by \citet{Gautam:2019} (hereafter denoted as \citetalias{Gautam:2019}), with the following additions:
\begin{itemize}
  \item Newer $K'$-band observations performed since the analysis presented in \citetalias{Gautam:2019}. These observations were conducted in 2018--2022.
  \item Additional observations taken in the $H$-bandpass ($\lambda_0 = 1.633 \um$, $\Delta \lambda = 0.296 \um$), conducted in 2009 and in 2017--2022.
  \item Several nights of $K'$-band observations conducted in 2006, 2009, 2014--2017 that were not previously included in \citetalias{Gautam:2019}. These nightly combined images consisted of fewer individual frames, and therefore are not as sensitive to fainter stars and typically have larger photometric flux uncertainties. Despite these challenges, including these nights in our experiment's dataset allowed greater sensitivity to periodic variability for the bright stars detected in the shallow data.
\end{itemize}
Table~\ref{tab:observations} lists all nights of observations used in this experiment, along with metrics to describe the quality of each observation.

\subsection{Image reduction and photometric calibration} 
\label{sub:image_red_phot_calib}

This experiment used the same image reduction techniques for both $K'$- and $H$-band data as reported in \citetalias{Gautam:2019}, with use of the Keck AO Imaging (KAI) pipeline \citep{KAI:1_0_0}. In order to assign an observation time to our photometric observations, we followed the \citetalias{Gautam:2019} method to derive a weighted MJD time based on the weights of the individual frames used to construct the final combined images. The time precision allowed for more robust detections of periodic signals and of possible aliases introduced by our observing cadence.

In order to detect stellar sources in our imaging data, we deployed the point spread function (PSF) fitting routine StarFinder \citep{Diolaiti:2000a}, with additional improvements and optimizations made with ``Single-PSF'' mode StarFinder in AIROPA \citep{Witzel:2016}. The improvements largely resulted in the detection of fainter stars located in the PSF halos of bright stars in our experiment's field of view, and fewer detections of artifact sources near the edge of our field of view \citep{Terry:2023}. Appendix~\ref{sec:use_of_airopa_single_psf_mode} offers a detailed explanation to describe the advantages of switching to AIROPA Single-PSF mode for our photometric experiment over the $10''$ NIRC2 imager field of view compared to the StarFinder ``Legacy'' mode used in \citetalias{Gautam:2019}.

Photometric calibrations were carried out using the methods presented by \citetalias{Gautam:2019}, with a few modifications. The modifications included updating the photometric calibrator stars used, deriving initial calibration reference fluxes for the photometric calibrator stars, and extending the calibration methodology to the new $H$-band images in this work.
The updated procedure is summarized below.

In this work, we selected the following photometric calibrator stars: IRS~16NW, S3-22, S2-22, S4-3, S1-1, S1-21, S1-12, S2-2, S3-88, S2-75, S3-36, and S1-33. These calibrator stars are the same as those used by \citetalias{Gautam:2019}, but with the removal of the calibrator stars S1-17, S1-34, S3-370, S0-14, and S2-63, and the addition of the calibrator stars S2-22, S1-12, S2-2, S3-88, S2-75, and S1-33.
The removed calibrator stars are those that are no longer photometrically stable in new $K'$-band observations taken after 2017, and are replaced by other stars that are photometrically stable throughout the entire 2006--2022 time baseline of this experiment. The new set of calibrator stars was selected using the criteria developed in \citetalias{Gautam:2019}, which selected stars for photometric calibration that are isolated and distributed throughout the experiment field of view.

For the newly selected calibrator stars, we derived $K'$- and $H$-band reference flux measurements using absolute photometry from the \citet{Schodel:2010} photometric catalog transformed to the Keck NIRC2 bandpasses used in this work. The bandpass-corrected reference flux measurements of the photometric calibrator stars and their photometric flux properties in our experiment are listed in Table~\ref{tab:bandpass_corr} in Appendix~\ref{sec:phot_calibration_details}.
After photometric calibration, flux measurements of all our experiment's detected stars are determined in the same method as \citetalias{Gautam:2019}. $H$-band flux measurements are determined following the same photometric calibration and local photometric correction procedure as used for our $K'$-band dataset.


\begin{figure}[h]
  \centering
    \includegraphics[width=.49\textwidth]{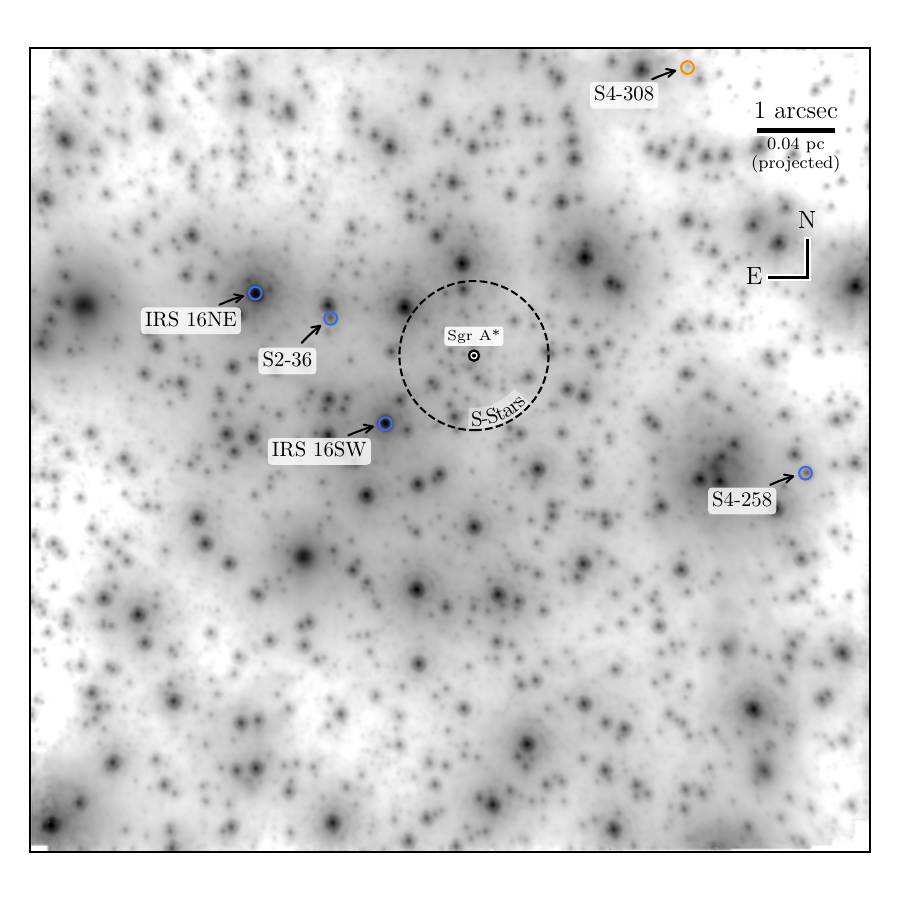}
  \caption{The field of view of this experiment. The background image is from the 2022-07-22 $K'$-band observation, and is an example of the $\approx 10''$-wide experiment field of view. Sgr A*, the location of the GC SMBH, is labeled, along with the approximate location of the S-star cluster in the $1''$ surrounding Sgr A*.
    Blue circles indicate the four known stellar binaries in the central half parsec surrounding the GC SMBH. All known binaries are spectroscopically confirmed young stars. IRS~16SW, S2-36, and S4-258 are photometric and spectroscopic binaries \citep[Gautam et al. in prep.]{Ott:1999, Peeples:2007, Rafelski:2007, Pfuhl:2014, Gautam:2019}, and are each detected in this work's photometric sample. IRS~16NE is a spectroscopic binary \citep{Pfuhl:2014} and is not a part of this experiment's sample since it is a resolved (i.e., non-point) source in our $K'$-band images \citepalias[as described by][]{Gautam:2019}. As a resolved source, this experiment's PSF fitting routine is unable to estimate accurate flux for IRS~16NE.
    S4-308 is an unknown age star for which we detect possible periodic flux variability. Future spectroscopic observations of S4-308 can verify if it is a young, early-type star and if it is a stellar binary.}
  \label{fig:gc_binaries_fov}
\end{figure}

\subsection{Stellar sample} 
\label{sub:stellar_sample}

The alignment of the stellar detections into a common reference frame \citep{Sakai:2019} and the matching of stellar detections was carried out using the same methods as detailed in previous GCOI publications: \citet{Ghez:2008} and \citet{Jia:2019}. We carried out this procedure separately in $K'$- and $H$-band.
Assigning $K'$-band stars to names in the GCOI catalog was conducted by matching with both proper motion and $K'$ band flux. For $H$-band data, we first applied a bulk $H-K'$ offset of 2.05 to $K'$-band stellar fluxes in the GCOI catalog in order to match by flux.
For stars with $H-K'$ colors very discrepant from the bulk population (e.g., stars with bluer colors due to being situated foreground of the GC), we performed naming of stars in $H$-band data based only on matching proper motion to their $K'$-band counterparts.
This proper-motion-only matching for $H$-band stars was performed for 40 stars in our sample.

Our experiment's dataset consisted of 100 observations in $K'$-band and 19 observations in $H$-band. The $K'$ observations were taken between 2006.336 and 2022.634 (total time baseline of 16.30 years) and the $H$ observations were taken between 2009.556 and 2022.634 (total time baseline of 13.08 years).
Individual observations each consist of combined images from individual frames taken over a single night. Therefore the shortest gaps between individual observations in our dataset are $\approx 1$ day. All observations used in this experiment along with their quality are listed in Table~\ref{tab:observations}.

The stars included in our experiment's dataset are those that are detected and matched across at least 30 $K'$-band nights. For sources matched across fewer than 30 nights, it was difficult to confidently determine by visual inspection if the source was indeed a real stellar source, or instead an {\em artifact source} from a nearby, brighter star \citep[detailed in][]{Jia:2019, Gautam:2019}.

The stellar sample is 1129 stars in our experiment's field of view. The stars' mean magnitudes in $K'$-band, $\overline{K'}$, span from 10.03 (star IRS~16C) to 19.36 (star S5-345), and the distribution of observed flux in $K'$-band is plotted in Figure~\ref{fig:sample_mag_dist}. 168 stars in our sample do not have detections in $H$-band. Stars not detected in $H$-band are largely faint stars that fall below the sensitivity limit of our $H$-band observations. 25 of the stars not included in this experiment's $H$-band dataset are those whose detections were dropped due to confusion with another nearby star, where the proximity of two coincident stellar PSFs result in the detection of only one merged PSF during the point source detection step \citep[the process used to identify stellar confusion is detailed by][]{Jia:2019}.
The dataset's photometric precision in $K'$- and $H$-bands are shown in Figure~\ref{fig:mag_magerr}, and an example image from this experiment to demonstrate the field of view is shown in Figure~\ref{fig:gc_binaries_fov}. The full 1129 star sample is collected in Table~\ref{tab:stellar_sample}.
There are 102 known young stars in our sample: 101 confirmed to be early-type (Wolf-Rayet (WR), O-type, or B-type) with spectroscopic observations reported in previous publications \citep{Blum:2003, Paumard:2006, Bartko:2009, Gillessen:2009a, Pfuhl:2011, Do:2013a, Feldmeier-Krause:2015, Chu:2020}, and the star S2-36 with spectroscopic observations reported in this work (\S~\ref{sec:S2-36_spec_binary_confirmation}).
The known young stars used in our analysis are listed in Table~\ref{tab:young_stellar_sample}.


\startlongtable
\begin{deluxetable*}{lD|L|rrD|r}
    \tablehead{
        \colhead{Date (UT)} &
        \multicolumn2c{MJD} &
        \colhead{Bandpass}  &
        \colhead{Sample stars} &
        \colhead{Bright stars}   &
        \multicolumn2c{Median} &
        \colhead{Known} \\
        \colhead{}  &
        \multicolumn2c{}    &
        \colhead{}  &
        \colhead{detected}  &
        \colhead{($\overline{m}_{K'} \leq 16$)}  &
        \multicolumn2c{phot. unc.} &
        \colhead{young stars}    \\
        \colhead{}  &
        \multicolumn2c{}    &
        \colhead{}  &
        \colhead{}  &
        \colhead{detected}  &
        \multicolumn2c{bright stars} &
        \colhead{detected}
    }
    \tablecaption{Observations used in this experiment\label{tab:observations}}
    \decimals
    \startdata
    2006-05-03                      & 53858.512 & K' & 892 & 440 & 0.039 & 86 \\
    2006-05-21 \tablenotemark{n}    & 53876.453 & K' & 403 & 295 & 0.072 & 71 \\
    2006-06-20 \tablenotemark{n}    & 53906.393 & K' & 873 & 435 & 0.055 & 88 \\
    2006-06-21                      & 53907.411 & K' & 909 & 447 & 0.050 & 89 \\
    2006-07-17                      & 53933.344 & K' & 882 & 430 & 0.044 & 87 \\
    2007-05-17                      & 54237.551 & K' & 947 & 444 & 0.067 & 91 \\
    2007-08-10 \tablenotemark{n}    & 54322.315 & K' & 792 & 419 & 0.041 & 86 \\
    2007-08-12                      & 54324.304 & K' & 596 & 391 & 0.053 & 74 \\
    2008-05-15                      & 54601.492 & K' & 956 & 467 & 0.040 & 95 \\
    2008-07-24                      & 54671.323 & K' & 750 & 436 & 0.045 & 91 \\
    2009-05-01                      & 54952.543 & K' & 937 & 444 & 0.025 & 92 \\
    2009-05-02                      & 54953.517 & K' & 917 & 446 & 0.034 & 97 \\
    2009-05-04                      & 54955.552 & K' & 923 & 452 & 0.028 & 95 \\
    2009-07-22 \tablenotemark{n}    & 55034.306 & K' & 570 & 377 & 0.042 & 81 \\
    2009-07-22 \tablenotemark{n}    & 55034.318 & H & 200 & 199 & 0.089 & 62 \\
    2009-07-24                      & 55036.330 & K' & 886 & 426 & 0.032 & 88 \\
    2009-09-09                      & 55083.249 & K' & 908 & 453 & 0.043 & 92 \\
    2010-05-04                      & 55320.545 & K' & 897 & 439 & 0.039 & 91 \\
    2010-05-05                      & 55321.583 & K' & 904 & 449 & 0.036 & 91 \\
    2010-07-06                      & 55383.351 & K' & 862 & 438 & 0.039 & 89 \\
    2010-08-15                      & 55423.283 & K' & 810 & 440 & 0.040 & 86 \\
    2011-05-27                      & 55708.505 & K' & 694 & 413 & 0.044 & 84 \\
    2011-07-18                      & 55760.344 & K' & 910 & 455 & 0.038 & 91 \\
    2011-08-23                      & 55796.280 & K' & 997 & 462 & 0.033 & 90 \\
    2011-08-24                      & 55797.275 & K' & 883 & 443 & 0.041 & 91 \\
    2012-05-15                      & 56062.519 & K' & 916 & 449 & 0.037 & 87 \\
    2012-05-18                      & 56065.494 & K' & 760 & 423 & 0.041 & 87 \\
    2012-07-24                      & 56132.311 & K' & 893 & 447 & 0.041 & 85 \\
    2013-04-26                      & 56408.563 & K' & 811 & 427 & 0.069 & 89 \\
    2013-04-27                      & 56409.566 & K' & 828 & 416 & 0.035 & 88 \\
    2013-07-20                      & 56493.325 & K' & 996 & 458 & 0.030 & 90 \\
    2014-03-19 \tablenotemark{n}    & 56735.639 & K' & 296 & 261 & 0.116 & 67 \\
    2014-03-20 \tablenotemark{n}    & 56736.631 & K' & 757 & 394 & 0.045 & 84 \\
    2014-04-18 \tablenotemark{n}    & 56765.608 & K' & 488 & 342 & 0.033 & 79 \\
    2014-04-19 \tablenotemark{n}    & 56766.634 & K' & 186 & 185 & 0.069 & 55 \\
    2014-05-11 \tablenotemark{n}    & 56788.587 & K' & 669 & 380 & 0.043 & 80 \\
    2014-05-19                      & 56796.524 & K' & 861 & 443 & 0.039 & 86 \\
    2014-07-03 \tablenotemark{n}    & 56841.343 & K' & 725 & 387 & 0.028 & 79 \\
    2014-07-04 \tablenotemark{n}    & 56842.392 & K' & 698 & 415 & 0.034 & 85 \\
    2014-08-03 \tablenotemark{n}    & 56872.271 & K' & 403 & 316 & 0.061 & 75 \\
    2014-08-04 \tablenotemark{n}    & 56873.286 & K' & 930 & 461 & 0.047 & 90 \\
    2014-08-06                      & 56875.290 & K' & 933 & 464 & 0.045 & 89 \\
    2015-03-31 \tablenotemark{n}    & 57112.613 & K' & 726 & 393 & 0.079 & 83 \\
    2015-04-01 \tablenotemark{n}    & 57113.585 & K' & 672 & 401 & 0.089 & 84 \\
    2015-04-02 \tablenotemark{n}    & 57114.586 & K' & 814 & 427 & 0.055 & 89 \\
    2015-05-14 \tablenotemark{n}    & 57156.532 & K' & 600 & 373 & 0.047 & 80 \\
    2015-08-09                      & 57243.298 & K' & 790 & 443 & 0.050 & 92 \\
    2015-08-10                      & 57244.291 & K' & 1012 & 464 & 0.052 & 95 \\
    2015-08-11                      & 57245.302 & K' & 1030 & 477 & 0.034 & 98 \\
    2016-05-03                      & 57511.515 & K' & 889 & 440 & 0.038 & 91 \\
    2016-07-12 \tablenotemark{n}    & 57581.300 & K' & 549 & 368 & 0.077 & 80 \\
    2016-07-13                      & 57582.363 & K' & 776 & 435 & 0.038 & 87 \\
    2017-05-04                      & 57877.534 & K' & 809 & 423 & 0.042 & 89 \\
    2017-05-05                      & 57878.523 & K' & 849 & 450 & 0.034 & 91 \\
    2017-05-07 \tablenotemark{n}    & 57880.559 & H & 940 & 482 & 0.027 & 97 \\
    2017-07-18                      & 57952.402 & K' & 761 & 410 & 0.039 & 89 \\
    2017-07-27                      & 57961.301 & K' & 655 & 388 & 0.038 & 86 \\
    2017-08-08                      & 57973.256 & K' & 382 & 318 & 0.047 & 75 \\
    2017-08-09                      & 57974.321 & K' & 891 & 445 & 0.024 & 93 \\
    2017-08-10                      & 57975.286 & K' & 838 & 440 & 0.040 & 94 \\
    2017-08-11                      & 57976.283 & K' & 961 & 472 & 0.045 & 93 \\
    2017-08-13 \tablenotemark{n}    & 57978.275 & H & 903 & 471 & 0.051 & 99 \\
    2017-08-23 \tablenotemark{n}    & 57988.244 & H & 427 & 370 & 0.063 & 87 \\
    2017-08-23                      & 57988.268 & K' & 885 & 441 & 0.034 & 92 \\
    2017-08-24 \tablenotemark{n}    & 57989.266 & H & 785 & 458 & 0.044 & 95 \\
    2017-08-24                      & 57989.268 & K' & 732 & 443 & 0.033 & 90 \\
    2017-08-26 \tablenotemark{n}    & 57991.252 & H & 832 & 455 & 0.039 & 95 \\
    2017-08-26                      & 57991.255 & K' & 878 & 449 & 0.037 & 91 \\
    2018-03-17 \tablenotemark{n}    & 58194.634 & K' & 897 & 450 & 0.035 & 92 \\
    2018-03-17 \tablenotemark{n}    & 58194.636 & H & 902 & 469 & 0.024 & 99 \\
    2018-03-22 \tablenotemark{n}    & 58199.621 & H & 679 & 426 & 0.051 & 90 \\
    2018-03-22 \tablenotemark{n}    & 58199.621 & K' & 848 & 433 & 0.032 & 90 \\
    2018-03-30 \tablenotemark{n}    & 58207.623 & K' & 805 & 443 & 0.030 & 90 \\
    2018-03-30 \tablenotemark{n}    & 58207.629 & H & 817 & 457 & 0.021 & 95 \\
    2018-05-19 \tablenotemark{n}    & 58257.545 & H & 320 & 299 & 0.053 & 78 \\
    2018-05-19 \tablenotemark{n}    & 58257.546 & K' & 602 & 377 & 0.049 & 84 \\
    2018-05-24 \tablenotemark{n}    & 58262.517 & K' & 689 & 398 & 0.041 & 89 \\
    2018-05-24 \tablenotemark{n}    & 58262.530 & H & 642 & 412 & 0.050 & 91 \\
    2018-09-03 \tablenotemark{n}    & 58364.258 & H & 769 & 440 & 0.056 & 91 \\
    2019-04-19 \tablenotemark{n}    & 58592.571 & K' & 731 & 402 & 0.033 & 85 \\
    2019-04-20 \tablenotemark{n}    & 58593.582 & K' & 934 & 450 & 0.027 & 93 \\
    2019-05-13 \tablenotemark{n}    & 58616.507 & K' & 816 & 447 & 0.034 & 89 \\
    2019-05-13 \tablenotemark{n}    & 58616.509 & H & 948 & 481 & 0.029 & 98 \\
    2019-05-23 \tablenotemark{n}    & 58626.498 & K' & 846 & 443 & 0.027 & 89 \\
    2019-06-25 \tablenotemark{n}    & 58659.459 & K' & 620 & 362 & 0.050 & 83 \\
    2019-06-30 \tablenotemark{n}    & 58664.457 & K' & 897 & 434 & 0.026 & 90 \\
    2019-08-14 \tablenotemark{n}    & 58709.292 & K' & 860 & 427 & 0.031 & 89 \\
    2019-08-18 \tablenotemark{n}    & 58713.244 & K' & 748 & 397 & 0.037 & 86 \\
    2019-08-19 \tablenotemark{n}    & 58714.316 & K' & 803 & 420 & 0.035 & 85 \\
    2020-07-07 \tablenotemark{n}    & 59037.359 & K' & 879 & 434 & 0.026 & 87 \\
    2020-08-09 \tablenotemark{n}    & 59070.310 & K' & 727 & 423 & 0.030 & 85 \\
    2021-04-29 \tablenotemark{n}    & 59333.568 & K' & 1003 & 478 & 0.044 & 92 \\
    2021-05-13 \tablenotemark{n}    & 59347.537 & K' & 728 & 403 & 0.046 & 83 \\
    2021-05-14 \tablenotemark{n}    & 59348.524 & K' & 849 & 444 & 0.027 & 86 \\
    2021-07-13 \tablenotemark{n}    & 59408.337 & K' & 763 & 413 & 0.030 & 87 \\
    2021-07-14 \tablenotemark{n}    & 59409.339 & K' & 609 & 383 & 0.025 & 82 \\
    2021-07-14 \tablenotemark{n}    & 59409.347 & H & 401 & 337 & 0.038 & 81 \\
    2021-08-12 \tablenotemark{n}    & 59438.253 & K' & 919 & 460 & 0.039 & 90 \\
    2021-08-13 \tablenotemark{n}    & 59439.283 & K' & 871 & 453 & 0.035 & 91 \\
    2021-08-14 \tablenotemark{n}    & 59440.294 & K' & 733 & 408 & 0.031 & 85 \\
    2021-08-15 \tablenotemark{n}    & 59441.275 & K' & 999 & 477 & 0.057 & 91 \\
    2021-08-21 \tablenotemark{n}    & 59447.273 & K' & 963 & 471 & 0.081 & 90 \\
    2022-05-14 \tablenotemark{n}    & 59713.527 & K' & 934 & 468 & 0.048 & 92 \\
    2022-05-15 \tablenotemark{n}    & 59714.530 & K' & 702 & 404 & 0.041 & 84 \\
    2022-05-21 \tablenotemark{n}    & 59720.498 & H & 939 & 482 & 0.040 & 97 \\
    2022-05-21 \tablenotemark{n}    & 59720.499 & K' & 1020 & 479 & 0.030 & 91 \\
    2022-05-25 \tablenotemark{n}    & 59724.507 & K' & 869 & 449 & 0.032 & 87 \\
    2022-05-25 \tablenotemark{n}    & 59724.511 & H & 906 & 481 & 0.039 & 97 \\
    2022-07-16 \tablenotemark{n}    & 59776.317 & K' & 698 & 406 & 0.038 & 83 \\
    2022-07-19 \tablenotemark{n}    & 59779.354 & K' & 953 & 461 & 0.027 & 89 \\
    2022-07-22 \tablenotemark{n}    & 59782.329 & K' & 981 & 472 & 0.039 & 92 \\
    2022-08-14 \tablenotemark{n}    & 59805.302 & K' & 878 & 437 & 0.034 & 86 \\
    2022-08-15 \tablenotemark{n}    & 59806.286 & K' & 1038 & 490 & 0.043 & 92 \\
    2022-08-16 \tablenotemark{n}    & 59807.272 & H & 773 & 469 & 0.031 & 97 \\
    2022-08-16 \tablenotemark{n}    & 59807.275 & K' & 879 & 452 & 0.030 & 90 \\
    2022-08-19 \tablenotemark{n}    & 59810.271 & K' & 918 & 467 & 0.043 & 91 \\
    2022-08-19 \tablenotemark{n}    & 59810.274 & H & 827 & 466 & 0.049 & 95 \\
    2022-08-20 \tablenotemark{n}    & 59811.280 & H & 817 & 462 & 0.037 & 95 \\
    2022-08-20 \tablenotemark{n}    & 59811.282 & K' & 879 & 443 & 0.031 & 87 \\
    \enddata
    \tablenotetext{n}{ Denotes a photometric observation not previously reported in \citetalias{Gautam:2019}.}
\end{deluxetable*}

\startlongtable
\begin{deluxetable*}{l|DD|rr|DDD}
    \tablehead{
        \colhead{Star} &
        \multicolumn2c{$\overline{m}_{K'}$} & \multicolumn2c{$\overline{m}_{H}$} &
        \colhead{$K'$} & \colhead{$H$} &
        \multicolumn2c{$x_0$} & \multicolumn2c{$y_0$} & \multicolumn2c{$t_0$} \\
        \colhead{} &
        \multicolumn2c{} & \multicolumn2c{} &
        \colhead{Nights} & \colhead{Nights} &
        \multicolumn2c{($''$ E of Sgr A*)} & \multicolumn2c{($''$ N of Sgr A*)} & \multicolumn2c{}
    }
    \tablecaption{Known young star sample\label{tab:young_stellar_sample}}
\decimals
\startdata
IRS 16C & 10.03 & 12.07 & 100 & 19 & 1.05 & 0.55 & 2009.989 \\
IRS 16SW & 10.08 & 12.16 & 100 & 19 & 1.11 & -0.95 & 2009.820 \\
IRS 16NW & 10.28 & 12.34 & 100 & 19 & 0.08 & 1.22 & 2010.047 \\
IRS 33E & 10.29 & 12.51 & 100 & 19 & 0.71 & -3.14 & 2010.182 \\
S2-17 & 10.69 & 12.74 & 100 & 19 & 1.34 & -1.88 & 2010.154 \\
IRS 16CC & 10.99 & 13.46 & 100 & 19 & 1.98 & 0.60 & 2010.135 \\
IRS 16SW-E & 11.22 & 14.36 & 92 & 19 & 1.90 & -1.12 & 2010.045 \\
IRS 33N & 11.29 & 13.48 & 100 & 19 & -0.03 & -2.24 & 2010.161 \\
S6-63 & 11.31 & 13.44 & 87 & 17 & 1.87 & -6.31 & 2011.057 \\
S1-24 & 11.42 & 13.62 & 73 & 19 & 0.74 & -1.65 & 2010.125 \\
S5-183 & 11.59 & 13.49 & 100 & 19 & 4.59 & -3.44 & 2010.024 \\
IRS 34W & 11.67 & 14.85 & 70 & 8 & -4.07 & 1.55 & 2010.189 \\
S2-6 & 11.90 & 14.11 & 100 & 19 & 1.66 & -1.33 & 2010.067 \\
S3-5 & 11.93 & 14.25 & 100 & 19 & 2.96 & -1.15 & 2009.910 \\
S2-4 & 12.04 & 14.39 & 99 & 19 & 1.52 & -1.46 & 2010.007 \\
S3-10 & 12.13 & 14.09 & 100 & 19 & 3.34 & -1.11 & 2009.944 \\
S3-2 & 12.16 & 14.36 & 100 & 19 & 3.09 & 0.55 & 2010.014 \\
IRS 9W & 12.19 & 14.59 & 99 & 19 & 2.90 & -5.59 & 2010.247 \\
S1-3 & 12.20 & 14.31 & 98 & 19 & 0.32 & 0.88 & 2010.207 \\
S6-89 & 12.30 & 14.61 & 44 & 9 & 5.45 & 3.00 & 2009.877 \\
S7-5 & 12.33 & 14.63 & 99 & 18 & 4.86 & -5.52 & 2010.523 \\
S2-16 & 12.34 & 15.76 & 100 & 19 & -1.07 & 2.06 & 2010.201 \\
S4-71 & 12.38 & 14.57 & 100 & 19 & 0.77 & -4.09 & 2010.194 \\
S3-26 & 12.38 & 14.48 & 100 & 19 & -2.57 & -2.07 & 2010.133 \\
S3-30 & 12.46 & 14.74 & 69 & 19 & 1.66 & -2.94 & 2010.064 \\
S3-374 & 12.55 & 15.23 & 100 & 19 & -2.76 & -2.85 & 2010.080 \\
S1-4 & 12.57 & 14.77 & 100 & 19 & 0.88 & -0.66 & 2010.057 \\
S1-22 & 12.62 & 14.73 & 51 & 19 & -1.57 & -0.52 & 2010.135 \\
S5-233 & 12.64 & 14.73 & 33 & 10 & 5.59 & 0.67 & 2010.312 \\
S2-19 & 12.71 & 14.83 & 100 & 19 & 0.38 & 2.31 & 2010.075 \\
S1-14 & 12.80 & 14.78 & 84 & 18 & -1.32 & -0.37 & 2009.666 \\
S4-36 & 12.82 & 15.29 & 83 & 11 & -3.69 & 1.78 & 2009.810 \\
S5-191 & 12.87 & 14.91 & 100 & 19 & 3.18 & -4.89 & 2010.357 \\
S2-22 & 12.96 & 15.01 & 100 & 19 & 2.30 & -0.21 & 2010.168 \\
S1-1 & 13.10 & 15.11 & 100 & 19 & 1.04 & 0.03 & 2009.895 \\
S4-258 & 13.11 & 16.37 & 91 & 18 & -4.40 & -1.63 & 2009.761 \\
S4-180 & 13.19 & 15.87 & 41 & 18 & -4.29 & -1.33 & 2011.094 \\
IRS 7SE & 13.24 & 15.91 & 99 & 18 & 2.99 & 3.46 & 2009.595 \\
S2-5 & 13.27 & 15.39 & 99 & 19 & 1.94 & -0.79 & 2010.021 \\
S5-237 & 13.28 & 15.33 & 63 & 14 & 5.49 & 1.02 & 2012.046 \\
S5-187 & 13.29 & 15.27 & 99 & 19 & -1.71 & -5.55 & 2009.655 \\
S2-74 & 13.30 & 15.59 & 100 & 18 & 0.11 & 2.78 & 2010.109 \\
S1-21 & 13.31 & 15.32 & 100 & 19 & -1.64 & 0.09 & 2010.177 \\
S2-21 & 13.33 & 15.42 & 100 & 19 & -1.62 & -1.67 & 2010.222 \\
S2-36 & 13.43 & 15.61 & 98 & 19 & 1.99 & 0.44 & 2009.947 \\
S1-12 & 13.53 & 15.58 & 100 & 19 & -0.75 & -1.03 & 2010.234 \\
S1-19 & 13.58 & 15.68 & 83 & 13 & 0.43 & -1.64 & 2010.024 \\
S4-169 & 13.64 & 15.57 & 100 & 12 & 4.41 & 0.28 & 2009.872 \\
S0-14 & 13.69 & 15.73 & 96 & 18 & -0.76 & -0.29 & 2010.316 \\
S0-15 & 13.70 & 15.86 & 56 & 17 & -0.97 & 0.18 & 2009.722 \\
IRS 34NW & 13.71 & 16.90 & 90 & 18 & -3.78 & 2.83 & 2010.283 \\
S4-287 & 13.73 & 15.87 & 100 & 19 & 0.13 & -4.77 & 2010.188 \\
S5-34 & 13.73 & 16.18 & 89 & 18 & -4.33 & -2.74 & 2010.071 \\
S3-331 & 13.75 & 15.85 & 95 & 18 & -1.22 & 3.65 & 2010.150 \\
S3-17 & 13.81 & 16.43 & 34 & 18 & -1.41 & 2.85 & 2009.185 \\
S2-7 & 13.91 & 16.17 & 100 & 18 & 0.93 & 1.85 & 2010.257 \\
S3-25 & 14.04 & 16.13 & 100 & 18 & 1.41 & 2.95 & 2009.923 \\
S2-58 & 14.05 & 16.27 & 94 & 18 & 2.14 & -1.13 & 2010.157 \\
S0-2 & 14.12 & 16.13 & 89 & 18 & -0.01 & 0.17 & 2007.893 \\
S1-8 & 14.22 & 16.48 & 99 & 19 & -0.58 & -0.92 & 2010.185 \\
S0-4 & 14.23 & 16.25 & 39 & 13 & 0.45 & -0.33 & 2010.008 \\
S0-9 & 14.28 & 16.27 & 82 & 18 & 0.22 & -0.60 & 2009.769 \\
S4-196 & 14.32 & 16.56 & 100 & 19 & 2.24 & -3.93 & 2010.135 \\
S3-190 & 14.33 & 17.12 & 100 & 19 & -3.19 & 1.41 & 2010.223 \\
S3-208 & 14.47 & 16.36 & 100 & 19 & -0.98 & -3.41 & 2010.198 \\
S6-64 & 14.53 & 16.34 & 74 & 18 & -3.05 & -5.86 & 2010.405 \\
S3-96 & 14.56 & 17.14 & 68 & 12 & -3.13 & -0.63 & 2010.217 \\
S0-3 & 14.65 & 16.74 & 98 & 19 & 0.34 & 0.12 & 2008.393 \\
S1-2 & 14.78 & 16.82 & 100 & 19 & 0.08 & -1.02 & 2009.985 \\
S0-1 & 14.78 & 16.92 & 67 & 18 & 0.04 & -0.26 & 2006.300 \\
S4-12 & 14.83 & 17.45 & 70 & 18 & -2.86 & 2.84 & 2010.034 \\
S1-18 & 14.87 & 17.13 & 69 & 19 & -0.79 & 1.50 & 2009.682 \\
S0-5 & 15.03 & 17.13 & 72 & 12 & 0.17 & -0.36 & 2009.636 \\
S1-33 & 15.04 & 17.13 & 100 & 19 & -1.25 & -0.00 & 2009.977 \\
S0-31 & 15.10 & 17.10 & 88 & 11 & 0.57 & 0.45 & 2009.906 \\
S3-3 & 15.13 & 17.20 & 100 & 19 & 3.09 & -0.64 & 2010.081 \\
S0-7 & 15.17 & 17.26 & 97 & 17 & 0.51 & 0.10 & 2010.379 \\
S3-155 & 15.18 & 17.03 & 99 & 18 & -1.84 & -2.83 & 2010.185 \\
S0-11 & 15.19 & 17.27 & 98 & 18 & 0.49 & -0.06 & 2010.011 \\
S2-50 & 15.20 & 17.34 & 85 & 16 & 1.70 & -1.51 & 2009.533 \\
S6-44 & 15.22 & 17.43 & 92 & 19 & -2.15 & -6.00 & 2010.113 \\
S2-306 & 15.28 & 17.89 & 52 & 11 & -0.49 & -2.89 & 2010.382 \\
S2-29 & 15.29 & 17.43 & 87 & 17 & 1.95 & -2.16 & 2010.009 \\
S3-268 & 15.29 & 17.56 & 87 & 19 & -2.15 & -3.03 & 2010.422 \\
S1-29 & 15.34 & 17.63 & 82 & 11 & 1.07 & 0.16 & 2010.071 \\
S4-8 & 15.35 & 17.54 & 99 & 19 & -0.45 & -3.97 & 2010.186 \\
S4-314 & 15.41 & 17.60 & 99 & 19 & 4.41 & -2.03 & 2010.075 \\
S0-19 & 15.43 & 17.52 & 79 & 17 & -0.01 & 0.40 & 2009.642 \\
S2-40 & 15.44 & 17.62 & 34 & 18 & 1.74 & 1.27 & 2009.828 \\
S3-314 & 15.46 & 17.56 & 79 & 19 & 3.84 & -0.09 & 2010.262 \\
S5-106 & 15.53 & 17.96 & 86 & 18 & -4.35 & -3.20 & 2010.067 \\
S2-76 & 15.62 & 18.18 & 68 & 17 & -0.22 & 2.81 & 2011.550 \\
S0-16 & 15.64 & 17.75 & 44 & 11 & 0.23 & 0.17 & 2007.593 \\
S4-262 & 15.71 & 17.77 & 98 & 18 & 4.28 & -1.96 & 2010.562 \\
S3-65 & 15.71 & 18.01 & 98 & 16 & -1.24 & -2.80 & 2009.687 \\
S0-8 & 15.79 & 17.81 & 88 & 16 & -0.23 & 0.16 & 2008.370 \\
S0-20 & 15.90 & 17.90 & 67 & 16 & 0.05 & 0.14 & 2008.089 \\
S1-27 & 16.07 & 18.51 & 56 & 12 & -1.03 & 0.19 & 2008.871 \\
S0-36 & 16.11 & 17.97 & 39 & 18 & -0.54 & -0.73 & 2007.994 \\
S0-61 & 16.43 & 18.46 & 63 & 7 & -0.20 & 0.44 & 2007.638 \\
S0-23 & 16.44 & \text{---} & 39 & 0 & 0.25 & -0.29 & 2008.044 \\
S0-40 & 17.16 & 19.17 & 82 & 15 & -0.23 & -0.10 & 2007.193 \\
\enddata
\end{deluxetable*}


\section{Methodology and Results} 
\label{sec:methodology_and_results}

\subsection{Photometric Periodicity Search} 
\label{sub:photometric_periodicity_search}

To detect binary star systems in our sample, we searched for periodic variability in our sample's stellar light curves.
Besides binary systems, periodic flux variability in stellar light curves can also originate from intrinsic sources of variability, such as stellar pulsations or rotation of stars with persistent star spots. Radial velocity follow-up with spectroscopy can confirm if the flux variability is indeed due to a stellar binary.

Periodic variability in binary systems commonly originates from stellar eclipses. Furthermore, we expect to be sensitive to variability originating from ellipsoidal effects and irradiation effects, both of which arise strongly in close binary systems.
Ellipsoidal variability originates when a close companion induces tidal distortions on a member of the binary system, with variable flux observed over the orbital period of the binary system due to orientation of the distorted star to the observer \citep[e.g.,][]{Morris:1985, Mazeh:2008}.
Irradiation effects are caused by a difference in surface temperature of the stars in the binary system. Incident radiation from the hotter component star in the binary can differentially heat one side of the cooler companion star. The heated side of the cooler companion will exhibit higher flux, and over the course of the binary period, the rotation of the differentially heated star in the binary system leads to observed periodic flux variability \citep[e.g.,][]{Peraiah:1982, Wilson:1990, Davey:1992, Prsa:2016}.

We employed the Lomb-Scargle periodicity search in our experiment  \citep{Lomb:1976, Scargle:1982}, which is particularly sensitive to quasi-sinusoidal variability and works effectively with the sparse and irregular sampling of our experiment. However, the Lomb-Scargle search is not as sensitive to binary systems with eclipses narrow in phase. For such light curves, the Box Least Squares (BLS) periodogram \citep{Kovacs:2002} can be used. However, when we tested the BLS method on mock binary systems injected into our sample  with eclipses narrow in phase, we were not able to successfully retrieve these systems (detailed further in \S~\ref{sec:BLS_pdgram_tests}). We would need more observations with finer sampling for the BLS method to effectively detect binary systems with narrow eclipses. In our experiment, we therefore employ only the Lomb-Scargle periodicity search which is well suited to our experiment's irregular observation cadence.

A description of our experiment's Lomb-Scargle periodicity search is provided in \S~\ref{ssub:trended_multiband_periodicity_search}, and we list the criteria for detections in \S~\ref{ssub:detection_criteria_for_periodic_variability}. \S~\ref{ssub:results_detection_of_periodic_signals} details the results from the periodicity search: a detection of 3 stars exhibiting significant periodic variability.

\subsubsection{Trended, multiband periodicity search} 
\label{ssub:trended_multiband_periodicity_search}

In this experiment, we performed a \textit{trended}, \textit{multiband} periodicity search. The trended component of the search allowed accounting for flux variability found by \citetalias{Gautam:2019} in several GC stellar light curves on timescales $\gtrsim 10$ years. The multiband component of the search allowed incorporating both $K'$- and $H$-band photometry data to more robustly detect periodic signals than single-band observations alone. We implemented the trended multiband Lomb-Scargle periodicity search framework outlined by \citet{VanderPlas:2015}, which extends the Lomb-Scargle periodicity search method. The multiband, trended framework \emph{simultaneously} allows observations in multiple passbands to be searched for periodic signals and long-term flux trends in the average stellar flux. Such long-term flux trends may be expected due to intrinsic stellar variability over long time baselines or due to the proper motion of stars behind spatial inhomogeneities in the foreground extinction screen (as suggested by \citetalias{Gautam:2019}).

Close binary systems often exhibit tidal distortions that yield quasi-sinusoidal flux variability in magnitude space \citep{Morris:1985, Mowlavi:2017}. Close binaries are the ones to which this experiment's observation cadence is most sensitive to due to their typically short periods and broad eclipses in phase. Due to the quasi-sinusoidal flux variability in magnitude space, we carried out our periodicity search in magnitude space rather than in flux space. The Lomb-Scargle periodic variability search is most sensitive to sinusoidal signals, making it particularly suited to search for such binary systems in magnitude space.

The periodicity search model consisted of two different components for every period $P$ searched:
\begin{enumerate}
    \item A trended sinusoid \emph{base model} to model the long-term trend and periodic signal shared across all bands. We allowed the long-term trend ($A_{\text{base,0}} \cdots A_{\text{base,}n}$) to extend from a first-order polynomial (i.e., linear) up to a fourth-order polynomial: $1 \leq n \leq 4$. $B_{\text{base}}$ and $C_{\text{base}}$ allowed modeling the amplitude and phase of the sinusoidal periodic signal:
    \begin{eqnarray}
        m_{\text{base}} = A_{\text{base,0}} &+& \sum_{k=1}^{n} A_{\text{base,}k}(t - t_0)^k \nonumber \\
        &+& B_{\text{base}}\sin{\left[\frac{2 \pi}{P} (t - t_0)\right]} \nonumber \\
        &+& C_{\text{base}}\cos{\left[\frac{2 \pi}{P} (t - t_0)\right]}.    \label{eq:trend_per_base_model}
    \end{eqnarray}
    In our periodicity search, $t_0$ is a fixed parameter that we set as our first observation date: $t_0 \equiv 53858.512$. To fit for an arbitrary phase shift in the search, we include both the sin and cos components in the models.
    \item Two \emph{band-specific models} to model the residual periodic signal in each band not captured by the base model. In the trend, $A_{K',0}$ and $A_{H,0}$ allowed capturing the color difference between the bands, while $A_{K',1}$ and $A_{H,1}$ allow different linear slopes for the long-term trend (e.g., as expected by reddening due to changing extinction). The band-specific periodicity components ($B_{K'}$, $C_{K'}$, $B_{H}$, and $C_{H}$) can allow capturing differences in the variability amplitude or phase shift across the different bands:
    \begin{eqnarray}
        m_{K'\text{ model}} = A_{K',0} &+& A_{K',1}(t - t_0)        \nonumber \\
        &+& B_{K'}\sin{\left[\frac{2 \pi}{P} (t - t_0)\right]}  \nonumber \\
        &+& C_{K'}\cos{\left[\frac{2 \pi}{P} (t - t_0)\right]}, \label{eq:trend_per_Kp_model}
    \end{eqnarray}
    \begin{eqnarray}
        m_{H\text{ model}} = A_{H,0} &+& A_{H,1}(t - t_0)           \nonumber \\
        &+& B_{H}\sin{\left[\frac{2 \pi}{P} (t - t_0)\right]}   \nonumber \\
        &+& C_{H}\cos{\left[\frac{2 \pi}{P} (t - t_0)\right]}.  \label{eq:trend_per_H_model}
    \end{eqnarray}
\end{enumerate}
The base and band-specific models were added to the weighted mean magnitude of observations in each band ($\overline{m}_{K'}$ and $\overline{m}_{H}$) to obtain the final model magnitudes for the respective band:
\begin{eqnarray}
    m_{K'} &=& \overline{m}_{K'} + m_{\text{base}} + m_{K'\text{ model}},\\
    m_{H} &=& \overline{m}_{H} + m_{\text{base}} + m_{H\text{ model}}.
\end{eqnarray}

Before running the periodicity search, we determined the polynomial order of the long-term flux trend to use in the periodicity search. The polynomial trend order ($n$) was determined by initial polynomial fits to only the $K'$-band flux values for each star, since our $K'$-band data spanned a longer time baseline than our $H$-band data. We used Wilks Theorem \citep{Wilks:1938} to compare the polynomial model likelihoods for a given order ($L_n$):
    \begin{eqnarray}
        D &=& -2 \ln \left( \frac{L_{n}}{{L_{n+1}}} \right)   \\
        &=& 2 (\ln L_{n+1} - \ln L_n).
    \end{eqnarray}

We favored higher order polynomial fits for a star's flux only if $p(D)$, the probability of obtaining a better fit with a higher order, exceeded $5\sigma$.

For every test period in the periodogram, a total of 12--15 parameters were fit: $A$, $B$, and $C$ for the base model and the two band-specific models (with 12 total parameters for $n=1$, going up to 15 total parameters for $n=4$). Our implementation of a multiband, trended Lomb-Scargle periodogram is available at an online software repository\footnote{\url{https://github.com/abhimat/gatspy}}, forked from the software package \textsc{gatspy} \citep{gatspy, VanderPlas:2015}.

We computed two metrics to establish the significance of detected candidate signals in our periodicity search:

\begin{enumerate}
    \item We calculated a \emph{false-alarm probability} to assign significance to powers in the computed Lomb-Scargle periodogram. The false-alarm probability was implemented by extending the bootstrap methods outlined by \citet{Ivezic:2014} and \citet{VanderPlas:2018}, using 10,000 mock light curves in each observed waveband for the star.
When generating a mock light curve in a given waveband, we first de-trended the observations by subtracting the best fit long-term polynomial trend for the respective band calculated during the Lomb-Scargle analysis. In each waveband, these de-trended observations were then randomly drawn, with replacement, at each observation time. After drawing the random observations, we added the long-term polynomial trend for each band back to the observations. This procedure allowed the mock light curves to include any observed long-term trends in order to properly account for the increase in the trended Lomb-Scargle periodogram power when there is a significant long-term trend.
Lower values for the false-alarm probability indicated that the detection was less likely to be a false signal generated by flux measurement uncertainties or by our experiment's observation cadence.
	
	\item We computed the \emph{significance of the sinusoid amplitude} in the detected signal as a measure of the signal strength. This was done by fitting a trended sinusoid model to the observed flux at the most significant period determined from the Lomb-Scargle periodicity search using an MCMC fitting routine implemented with \textsc{emcee} \citep{Foreman-Mackey:2013}.
To simplify the model fit for this sinusoid amplitude calculation, we fit the trended sinusoid model (equation~\ref{eq:trend_per_base_model}), no $K'$-band-specific model ($A_{K',0} = A_{K',1} = B_{K'} = C_{K'} = 0$ in equation~\ref{eq:trend_per_Kp_model}), and only a linear term in the $H$-band-specific model ($B_{H} = C_{H} = 0$ in equation~\ref{eq:trend_per_H_model}). We compared the amplitude of the sinusoid as
\begin{eqnarray}
    \text{Sin Amplitude} = \sqrt{B_{\text{base}^2} + C_{\text{base}^2}},
\end{eqnarray}
and the uncertainty in the sinusoid amplitude as
\begin{eqnarray}
    \sigma_{\text{Sin Amplitude}} = \sqrt{\sigma_{B\text{, base}}^2 + \sigma_{C\text{, base}}^2},
\end{eqnarray}
in order to determine the strength of the sinusoid variability:
\begin{eqnarray}
    \frac{\text{Sin Amplitude}}{\sigma_{\text{Sin Amplitude}}}.
\end{eqnarray}
Periodic flux variability from an astrophysical origin with amplitude much greater than our experiment's photometric uncertainty will have higher amplitude significance in our experiment.
\end{enumerate}


\begin{figure}[b]
  \centering
    \includegraphics[width=.49\textwidth]{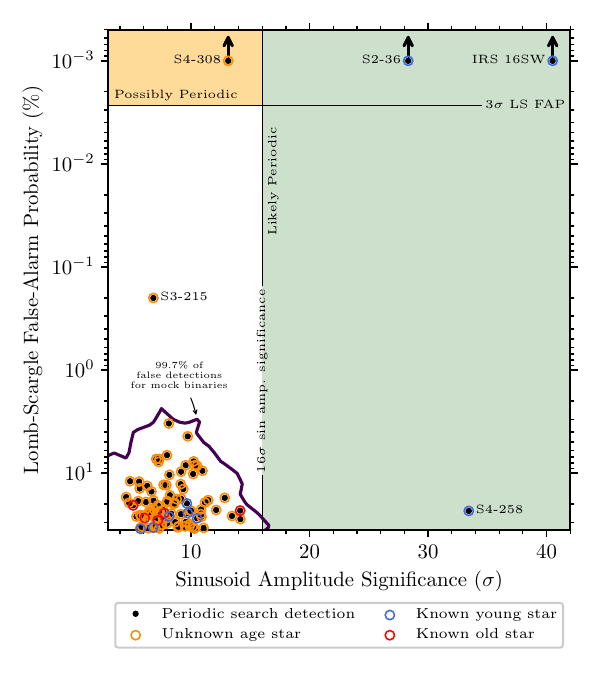}
  \caption{
    Periodic flux variability detected in our experiment is plotted as a function significance of signal amplitude (horizontal axis) and bootstrap false-alarm probability (vertical axis). Background colors indicate detection confidence regions from our analysis of injected binary signals, with green indicating \textit{likely} periodic variability and orange indicating \textit{possibly} periodic variability (see \S~\ref{ssub:detection_criteria_for_periodic_variability} for details of these regions.).
    For comparison, we additionally plot the $99.7\%$ contour of \textit{false} detections from the mock binary injection and recovery analysis: periodic signals inconsistent with the period and amplitude of the mock binary signals (see \S~\ref{ssub:recovery_of_binary_signals}). Therefore, the contour outlines the region of the parameter space susceptible to false detections.
  }
  \label{fig:LS_sig_amp}
\end{figure}

\subsubsection{Detection criteria for periodic variability} 
\label{ssub:detection_criteria_for_periodic_variability}
We implemented a set of criteria to identify likely periodic variable signals in our experiment similar to those used by \citetalias{Gautam:2019}. Since our search was primarily intended to search for young eclipsing binaries in this experiment, we only searched for periodically variable candidates in the range of 1--100 days. Distinguishing real signals much longer than 100 days from false positive aliases caused by our observation cadence can be difficult. Furthermore, this experiment is not very sensitive to binary signals at long periods much larger than 100 days, since such binaries typically have narrower eclipses in phase \citep[e.g.,][]{Mowlavi:2017}. The period cutoff was validated in our later binary signal recovery procedure, where fewer than 1\% of mock binaries with periods $\approx 100$ d were recovered with our periodicity search (see Figure~\ref{fig:mock_binary_param_dists}).

We additionally removed possible periodically variable candidates that exhibited high power in the periodogram at timescales greater than 1 yr. Long-term flux variability trends that were picked up as periodically variable at periods longer than $\sim 1$ yr often led to aliases in our binary period search range.

Finally we examined the distribution of detections picked up in our mock binary variability search, and used the distribution of false detections to inform where in the parameter space of our significance measures our search is unlikely to falsely detect a periodic signal (see \S~\ref{ssub:recovery_of_binary_signals} for more details).
Detections with sinusoid amplitude significance $\geq 16 \sigma$ and false-alarm probability $\leq 40\%$ were unlikely to be false, so we consider any detections in this region where the sinusoid amplitude significance is large to be \textit{likely} periodic.

We considered the region with sinusoid amplitude significance $< 16 \sigma$ but False-Alarm probability~$\leq 0.270\%$ (i.e. $3 \sigma$ in a normal distribution) to host \textit{possibly} periodic variables. The low amplitude significance in this region makes it difficult to verify likely variability, but the low False-Alarm probability suggests a possible periodic signal.

The regions of likely and possible periodic variability are indicated in the parameter space in Figure~\ref{fig:LS_sig_amp} as shaded colored areas: the green region corresponding to signals that are \emph{likely} periodic and the orange region corresponding to signals that are \emph{possibly} periodic.
It is difficult with this experiment's sensitivity to determine if detections in the parameter space outside of these two regions are originating from true periodic flux variability.


\begin{figure*}[t]
  \centering
    \includegraphics[width=.95\textwidth]{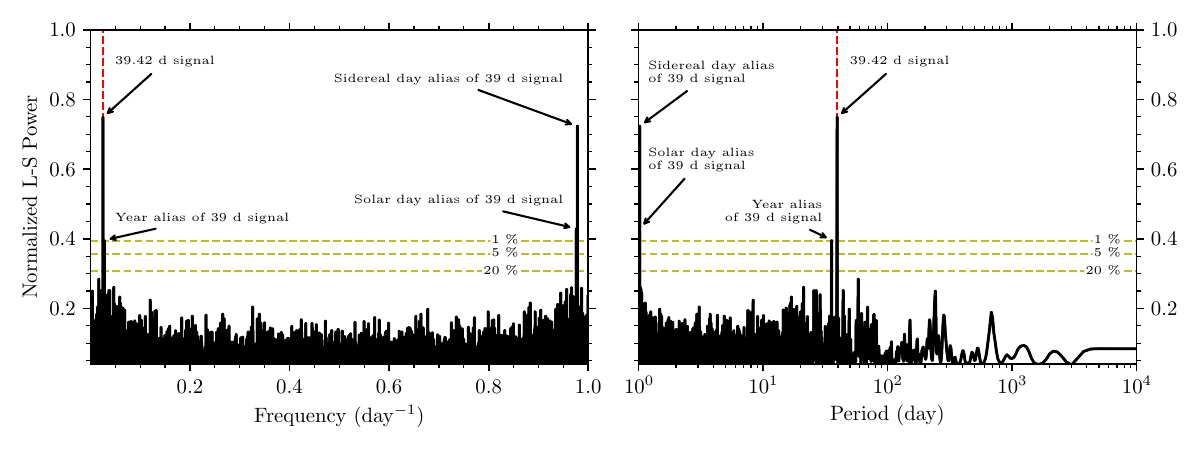}
    \includegraphics[width=.95\textwidth]{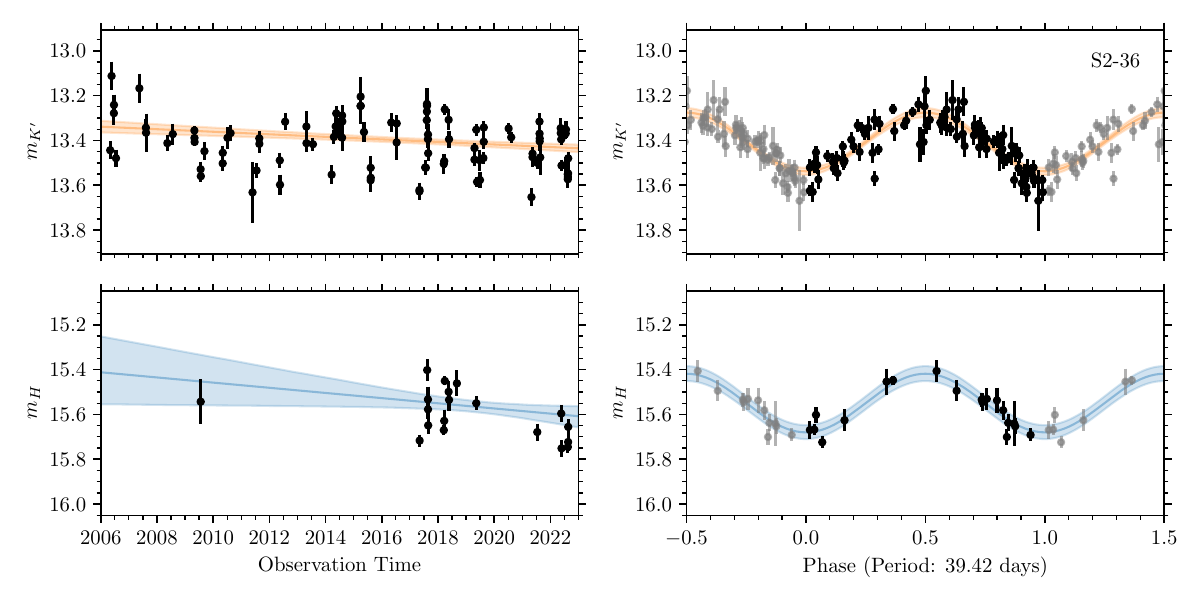}
  \caption{Detection of \textbf{S2-36}'s periodic variability in our experiment. S2-36's periodic flux variability was previously reported by \citet{Gautam:2019}. Spectroscopic measurements reported in this work demonstrate that S2-36 is an early-type star and has significant RV variability (see \S~\ref{sec:S2-36_spec_binary_confirmation}), making it a likely young binary star.
  \textbf{Top row:} S2-36's flux periodograms in frequency (left) and period (right) space. The vertical dashed line indicates the most significant peak in the periodogram, corresponding to the detected periodic signal. The horizontal dashed lines indicate bootstrap false-alarm probability levels. Also labelled are the three most prominent aliases to the periodic signal. These aliases originate from, in order of power, the sidereal day, solar day, and yearly typical observing cadences of our experiment, and are detected for all strongly periodic signals in our periodograms.
  \textbf{Middle and bottom rows:} S2-36's flux measurements in our experiment in $K'$-band (middle row) and $H$-band (bottom row). The bottom two left panels show every observed flux measurement. The colored solid line and band show the best fit and $1\sigma$ confidence interval, respectively, of the long-term 1\textsuperscript{st}-order polynomial trend fit to S2-36's flux in each band. The bottom two right panels show the de-trended flux phased to the detected photometric variability period of 39.42 d. The colored solid line and band show the best fit and $1\sigma$ confidence interval, respectively, of the sinusoidal component of the periodicity model fit in each band.
  }
  \label{fig:det_plot_S2-36}
\end{figure*}

\begin{figure*}[t]
  \centering
    \includegraphics[width=.95\textwidth]{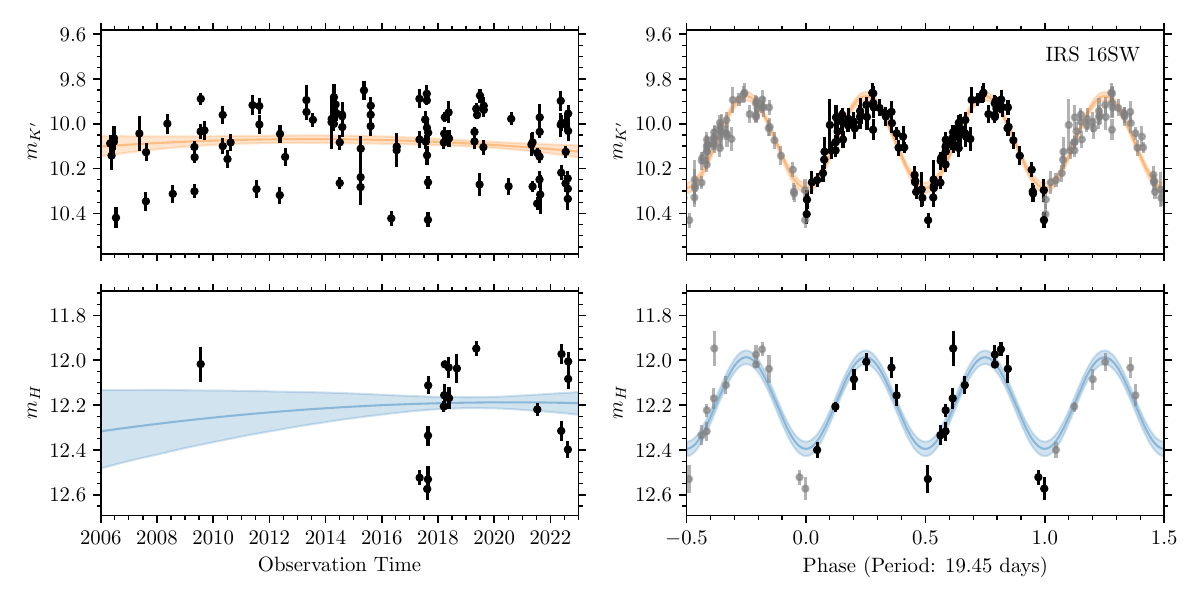}
    \hrule
    \includegraphics[width=.95\textwidth]{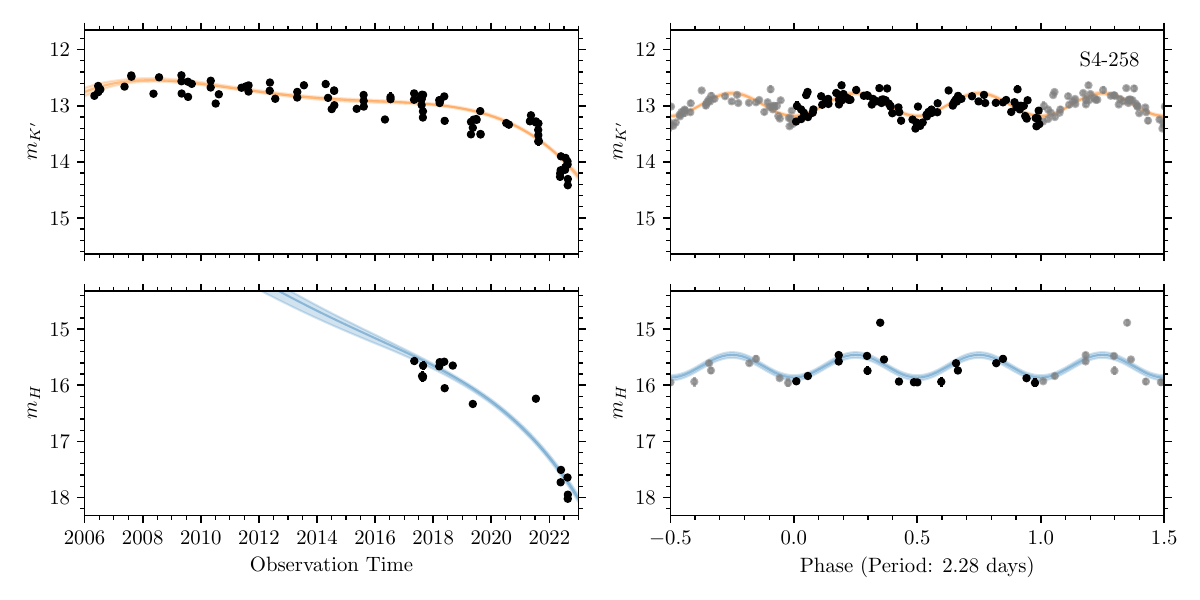}
  \caption{Detection of the eclipsing binaries \textbf{IRS 16SW} (top two rows) and \textbf{S4-258} (bottom two rows) in our experiment. Left panels show observed flux measurements with the long-term polynomial trend, and right panels plot de-trended flux measurements phased to known binary orbital periods.
  The long-term trend fit to IRS~16SW is a 2\textsuperscript{nd}-order polynomial, while the long-term trend fit to S4-258 is a 4\textsuperscript{th}-order polynomial. The long-term dip in flux for S4-258 has an associated color change that makes it consistent with extinction, suggesting that S4-258 is passing behind foreground extinguishing features at the GC over time (Haggard et al. in prep.).
  }
  \label{fig:det_plot_knownbinaries}
\end{figure*}

\begin{figure*}[t]
  \centering
    \includegraphics[width=.95\textwidth]{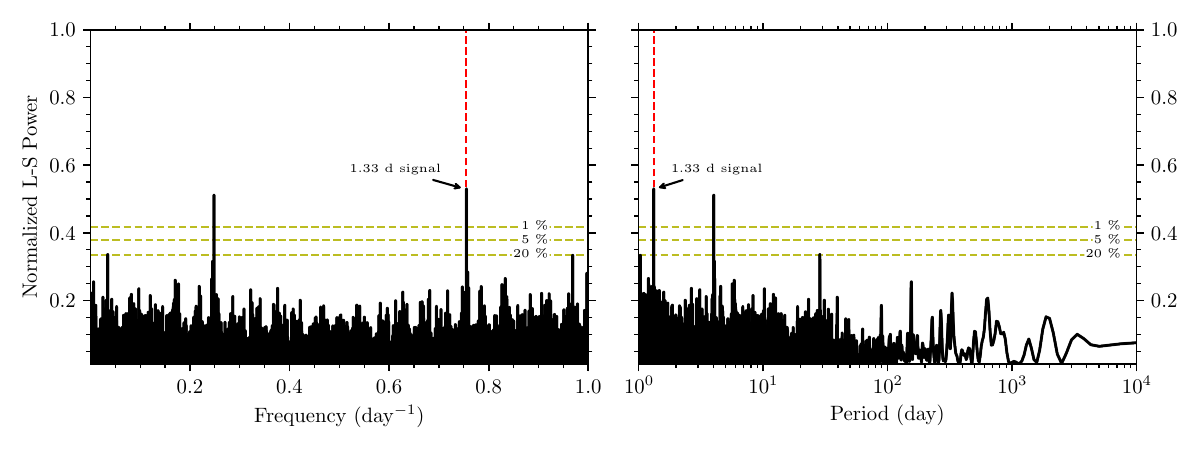}
    \includegraphics[width=.95\textwidth]{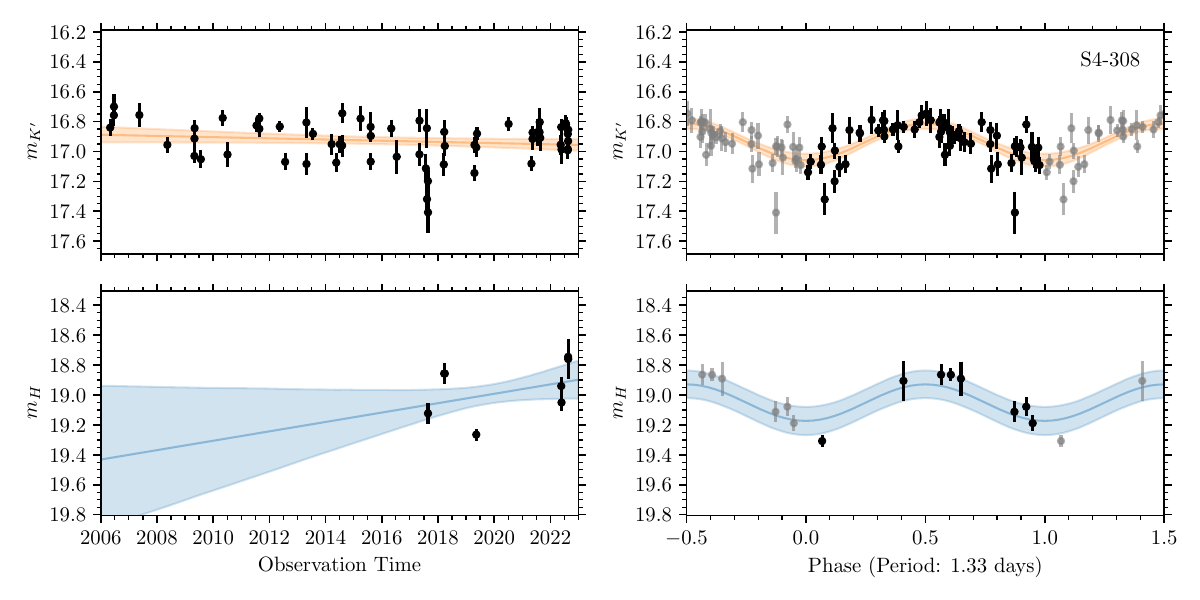}
  \caption{Same as Figure~\ref{fig:det_plot_S2-36}, but for \textbf{S4-308}. The long-term trend fit to S4-308 is a 1\textsuperscript{st}-order polynomial.
  The periodic flux variability in S4-308 has not been reported previously. We report it as a \textit{possibly} periodic variable due to its small amplitude of variability ($\approx 0.24$ mag) and the photometric precision of our experiment. Future, more precise experiments, or additional observations beyond those presented in this work may be able to verify the possible periodically variable flux of this star. Spectroscopic measurements of S4-308 are difficult to obtain with 8--10 m ground-based telescopes due to its faint flux, but measurements taken with JWST or with future ELTs can determine if S4-308 is a young star, and if it is a stellar binary with detections of RV variability.
}
  \label{fig:det_plot_S4-308}
\end{figure*}

\subsubsection{Results: Detection of Periodic Signals} 
\label{ssub:results_detection_of_periodic_signals}

Our experiment's periodic signal detections are plotted in Figure~\ref{fig:LS_sig_amp}. Variability that we can identify as periodic has large flux amplitudes compared to our experiment's photometric precision and low false-alarm probability.
We found three known young stars in our search sample that passed our criteria for detection as \emph{likely} periodic variables: S2-36 (photometric period = 39.42 d; Figure~\ref{fig:det_plot_S2-36}), IRS 16SW (photometric period = 9.72 d), and S4-258 (photometric period = 1.14 d Figure~\ref{fig:det_plot_knownbinaries}). IRS 16SW and S4-258 are known eclipsing binary stars, each with binary orbital periods approximately double the photometric variability period detected in our experiment (19.45 d and 2.28 d, respectively) due to their approximately equal eclipse depths \citep{Ott:1999, Peeples:2007, Rafelski:2007, Pfuhl:2014, Gautam:2019}. The periodic variability in S2-36 has been reported previously by \citet{Gautam:2019}. With two new Keck~I OSIRIS spectra reported in \S~\ref{sec:S2-36_spec_binary_confirmation}, S2-36 exhibits a spectrum typical of early-type stars and shows significant radial velocity (RV) variability, likely making it a young binary system. All three of these sources had periodic variability that was very significant in the parameter space of significance measures, where we are confident in our experiment's detections. With the detections in this and previous experiments, there are four known binaries within a half parsec surrounding the GC SMBH (see Figure~\ref{fig:gc_binaries_fov}).

The light curves of all three binary shapes appear to be quasi-sinusoidal, however it is important to note that due to the construction of our Lomb-Scargle periodicity search (\S~\ref{ssub:trended_multiband_periodicity_search}), the best-fit period here is where each light curve best fits a sinusoid. Small deviations to the measured period are possible from our reported period since the light curves of these systems phased to the true binary orbital period may deviate slightly from a sinusoid, such as sharper declines in flux from eclipses that can be slightly smeared when fitting a sinusoid. Due to their quasi-sinusoidal shape, all three binaries detected in our experiment are likely to be contact binaries or irradiation binaries (see mock binary example light curves in Figures~\ref{fig:mock_binary_examples_detached} -- \ref{fig:mock_binary_examples_irrad}). Previous astrophysical models of the binaries IRS~16SW \citep{Peeples:2007} and S4-258 \citep{Pfuhl:2014} have demonstrated that both these systems are contact binaries with one or both stellar components overflowing the Roche lobe. While our two spectra reported in \S~\ref{sec:S2-36_spec_binary_confirmation} confirm significant RV variability for S2-36, indicative of a binary system, they are not able to confirm if the $\approx 39$ day periodic flux signal corresponds to approximately half or the full binary orbital period. Additional spectroscopic observations in addition to the flux observations reported here are required to determine if the flux variability our experiment detects originates from a contact system or from irradiation.

One unknown age star had a bootstrap false-alarm test significance higher than $3 \sigma$ but did not have a sinusoid significance high enough to pass our likely periodic bounds: S4-308 (photometric period = 1.33 d ; Figure~\ref{fig:det_plot_S4-308}). The periodic variability in S4-308 has not been reported previously, and is difficult to confirm with the photometric precision of our experiment due to its small amplitude of variability ($\approx 0.24$ mag). Future, more precise photometric experiments, or additional observations beyond those presented in this work may be able to verify the possible periodically variable flux of this star. Spectroscopic measurements of S4-308 are difficult to obtain with 8--10 m ground-based telescopes due to its faint flux, $\overline{m}_{K'} = 16.9$ \citep[e.g.,][provide a discussion of spectroscopy limits for GC stars with current and future telescopes]{Do:2019}. Spectroscopic measurements taken with the James Webb Space Telescope (JWST) and ``extremely large telescope'' facilities (ELTs) can determine if S4-308 is a young star, and if it is a stellar binary with detections of RV variability.



\subsection{Binary Star Fraction Determination} 
\label{sub:binary_star_fraction_determination}
To use this experiment's detections of binary systems to constrain the underlying intrinsic binary fraction of GC young stars, we estimated our experiment's sensitivity to young binary systems. This is particularly challenging due to the long time baseline of our experiment and the large proper motion of GC stars behind a differential extinction screen, which results in approximately half of all GC stars to exhibit flux variability \citep{Gautam:2019}. To combat this challenge, we first simulated a mock population of young binary systems using population characteristics expected for young stars in the GC. This procedure is described in more detail in \S~\ref{ssub:mock_binary_population}. We next simulated light curves for these mock binary systems in the $K'$- and $H$-bands, described in \S~\ref{ssub:mock_binary_light_curves}.
From these, 100 mock binary system light curves were then injected into each of our sample's observed light curves. We determined the fraction of the injected binary signals that can be recovered by our periodicity search analysis.
The procedure allowed determining how likely a binary periodic signal could be detected in each of our sample stars' variable light curves.
Injection of binary signals into our sample light curves is described further in \S~\ref{ssub:injection_of_binary_signals}, while \S~\ref{ssub:recovery_of_binary_signals} describes the recovery of injected signals using our periodicity search techniques.
Finally, \S~\ref{ssub:results_bin_frac} provides the results of our binary fraction measurement by combining our periodicity search detections with the sensitivity analysis.

\subsubsection{Mock binary population} 
\label{ssub:mock_binary_population}

We generated a mock population of 20,000 young binary systems to evaluate our experiment's sensitivity to photometric binary signals. In each mock binary system we hereafter denote the initially more massive component as \emph{Star 1}, and the initially less massive component as \emph{Star 2}. Five parameters were drawn for each mock binary system:
\begin{enumerate}
    \item $m_{i, 1}$: Initial mass for Star 1. This quantity was drawn from a top heavy initial stellar mass function observed for the GC: $p(m_{i, 1}) \propto m_{i, 1}^{- \alpha}$ with $\alpha = 1.7$ \citep{Lu:2013}. We selected $m_{i, 1}$ in the range of 10 -- 100 $M_{\odot}$.
    \item $q_i$: Initial mass ratio of the binary system, i.e. $q_i = m_{i, 2} / m,_{i, 1}$. We drew $q$ from a power law distribution, with a power law slope constrained by observations of local massive stars: $p(q_i) \propto q_i^{- \kappa}$ with $\kappa = -0.1$ \citep{Sana:2012}. Values of $q_i$ were selected in the range of 0.1 -- 1.0.
    \item $P$: Orbital period of the binary system. $P$ was drawn from a power law distribution, with a power law slope constrained by observations of local massive stars: $p( \log_{10}(P/\text{day}) ) \propto \log_{10}(P/\text{day})^{\pi}$ with $\pi = -0.55$ \citep{Sana:2012}. We sampled orbital periods in the range of $10^0$ -- $10^2$ days.
    \item $e$: Orbital eccentricity of the binary system. We drew $e$ from a power law distribution constrained by observations of local massive stars: $p(e) \propto e^{\eta}$ with $\eta = -0.45$ \citep{Sana:2012}. We selected $e$ in the range of 0 -- 1.
    \item $i$: Inclination of the binary system relative to the observer. We drew $i$ from a distribution flat in $\cos i$, with $i$ spanning from $0 \degree$ -- $180 \degree$.
\end{enumerate}

The distribution of all parameters we used for the binary star population, except for inclination, is informed by reasonable expectations for the GC young star population. The mass function we use to select $m_{i, 1}$ follows observational constraints for the GC young stars by \citet{Lu:2013}.
However, the distribution of binary parameters of the GC young stars has not been observationally constrained. Therefore, since our GC young star population consists of young OB-type stars, we drew $q_i$, $P$, and $e$ from distributions for massive stars in the solar neighborhood as measured by \citet{Sana:2012}. We discuss in more detail possible biases to our results from the choice of parameter distributions in \S~\ref{sub:discussion_bin_frac_dependence_mock_population}.
Finally, the distribution of binary star inclinations ($i$) originates from a uniform distribution of inclination in 3D space.


\subsubsection{Generation of mock binary light curves} 
\label{ssub:mock_binary_light_curves}

\begin{figure*}[t]
  \centering
    \includegraphics[width=.6\textwidth]{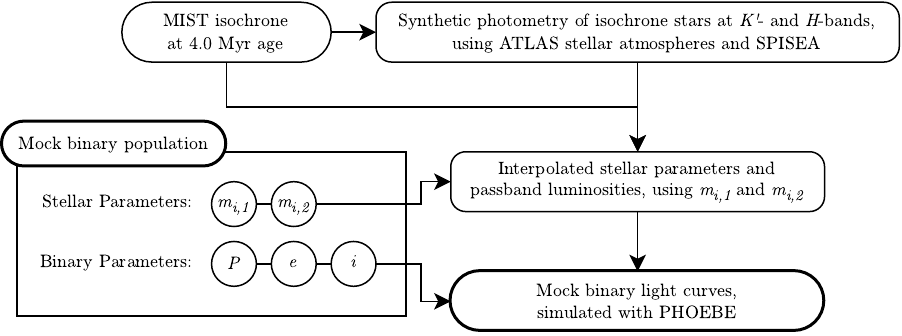}
  \caption{Procedure to generate the simulated light curves for this experiment's mock binary population. Mock binary light curves generated from this procedure were later injected into sample stellar light curves, outlined in Figure~\ref{fig:binary_injection_procedure}.}
  \label{fig:mock_light_curve_procedure}
\end{figure*}

We generated mock light curves in our experiment's observation passbands for the mock binary systems generated in \S~\ref{ssub:mock_binary_population}. This procedure is outlined in Figure~\ref{fig:mock_light_curve_procedure}. We first used the initial mass of the component stars ($m_{i, 1}$, $m_{i, 2}$) in the binary system to interpolate the remaining stellar parameters that are needed to calculate the observed binary light curves. We used isochrones constructed from stellar evolutionary models of young, massive stars for this procedure, detailed further in \S~\ref{par:stellar_params}. We then derived simulated light curves for each mock binary with an astrophysical binary modeling code (\S~\ref{par:binary_light_curve_models}).
The software we developed to carry out both of these steps, \textsc{Phitter}, is available online \citep{phitter:1_0_0}.

\paragraph{Stellar parameters from evolutionary models} 
\label{par:stellar_params}

The stellar parameters of the component stars in each mock binary system were derived using theoretical stellar isochrones calculated with the stellar population synthesis code \textsc{SPISEA} \citep{Hosek:2020}. We used MIST isochrones \citep{Choi:2016, Dotter:2016} computed at an age of 4.0 Myr \citep[consistent with the age of the young GC stars as estimated by][]{Lu:2013} and at solar metallicity to obtain stellar parameters. We then calculated synthetic photometry for the isochrones at our experiment's observation bandpasses ($K'$ and $H$) with the \textsc{SPISEA} code, using ATLAS atmosphere models \citep{Castelli:2003}. To derive synthetic photometry for each isochrone, we used the NIR extinction law derived from wide-field GC observations by \citet{Nogueras-Lara:2018}. We assumed a line of sight $K_{S}$-band extinction towards the GC of $A_{K_{S}} = 2.54$ \citep{Schodel:2010} and a distance to the GC of $R_0 = 7.971 \text{ kpc}$ \citep{Do:2019a} to obtain apparent magnitudes from the synthetic photometry.

From the isochrones, we linearly interpolated all stellar properties required for the stars for generating mock light curve. Using the initial mass of each star (i.e., $m_{i, 1}$, $m_{i, 2}$), we interpolated the five other stellar quantities needed for each star for light curve calculation: stellar mass at age 4.0 Myr ($m_1$, $m_2$), stellar radius ($R_1$, $R_2$), effective surface temperature ($T_{\text{eff}, 1}$, $T_{\text{eff}, 2}$), and the passband luminosities ($L_{K', 1}$, $L_{H, 1}$, $L_{K', 2}$, $L_{H, 2}$). The \textsc{Phitter} software interfaced with \textsc{SPISEA} to derive the binary population's stellar parameters and synthetic photometry.


\begin{figure*}[p]
  \centering
    \includegraphics[width=.95\textwidth]{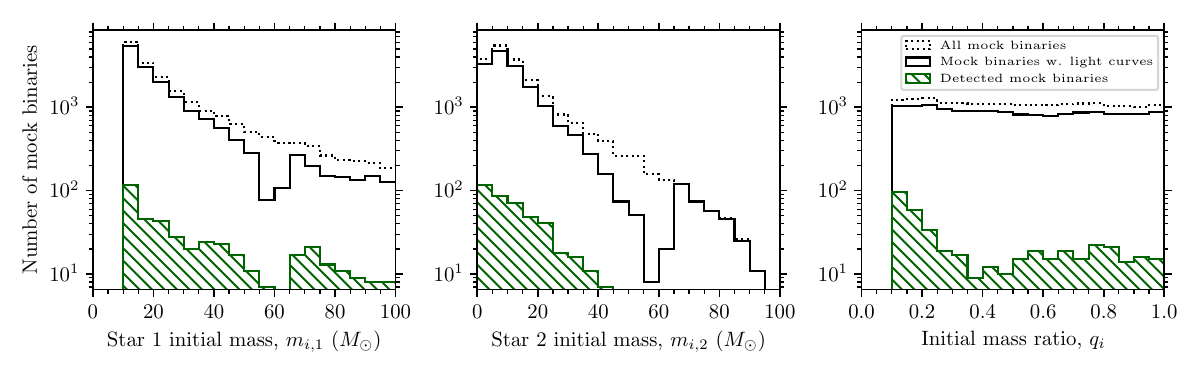}
    \includegraphics[width=.95\textwidth]{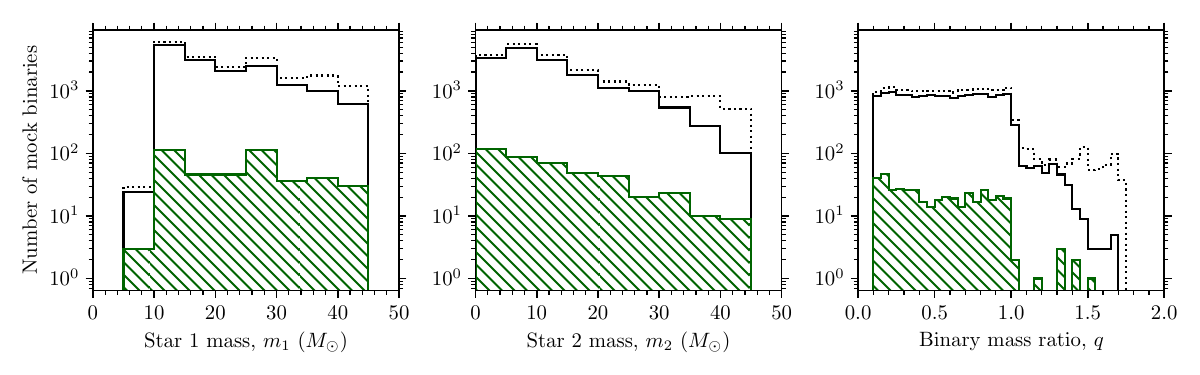}
    \includegraphics[width=.95\textwidth]{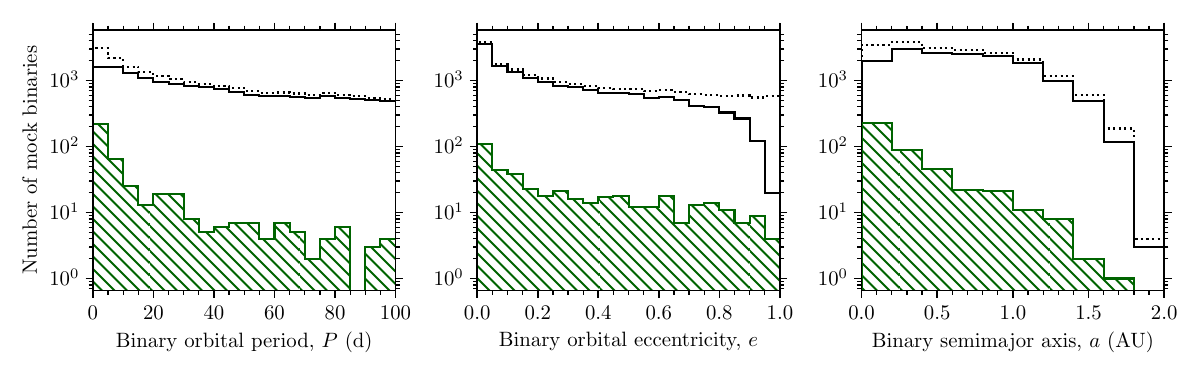}
    \includegraphics[width=.95\textwidth]{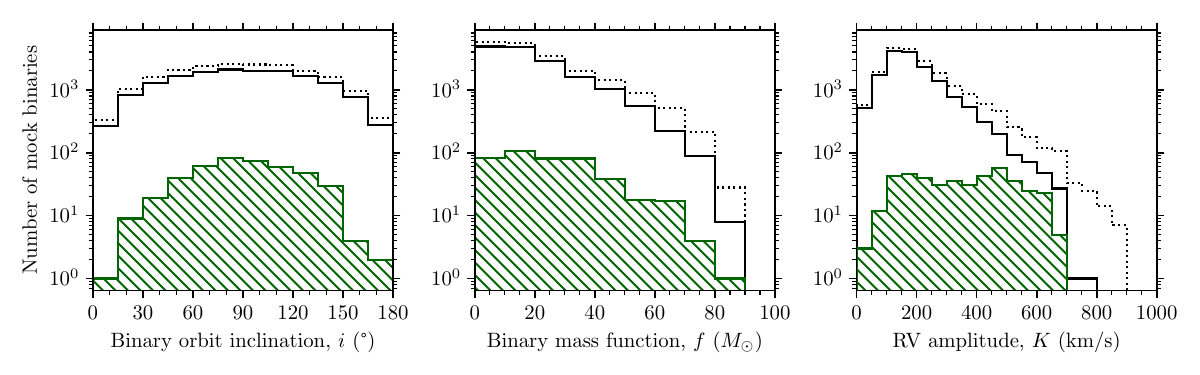}
  \caption{Distributions for the binary parameters of the mock binary population. Black histograms plot the entire mock binary library, while the green histogram plots those systems that were detected as a \emph{likely} detection in an injected light curve. The first row plots distributions for the initial masses of the component stars in the binary systems, while the second row plots the component mass distributions after the 4 Myr lifetimes. Although $0 \leq q_i < 1$, due to higher mass loss in star 1, some systems have $q > 1$.}
  \label{fig:mock_binary_param_dists}
\end{figure*}

\begin{figure*}[bt]
  \centering
    \includegraphics[width=.5\textwidth]{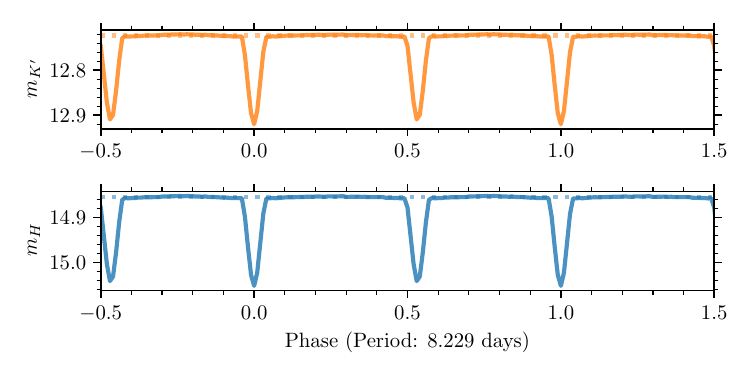}\includegraphics[width=.5\textwidth]{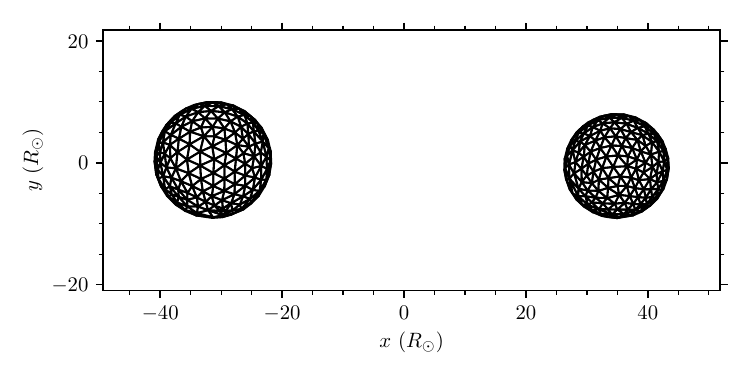}
    \includegraphics[width=.5\textwidth]{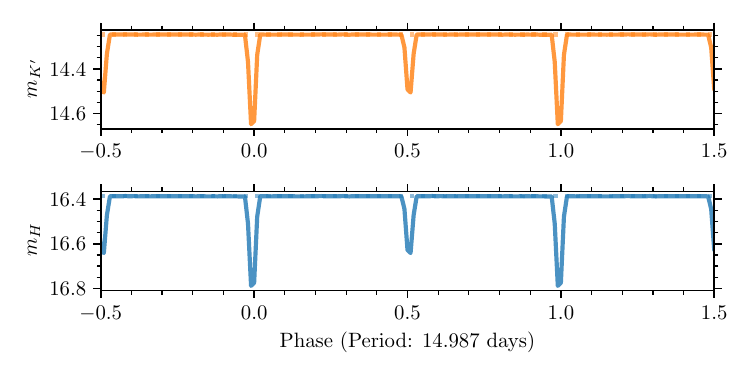}\includegraphics[width=.5\textwidth]{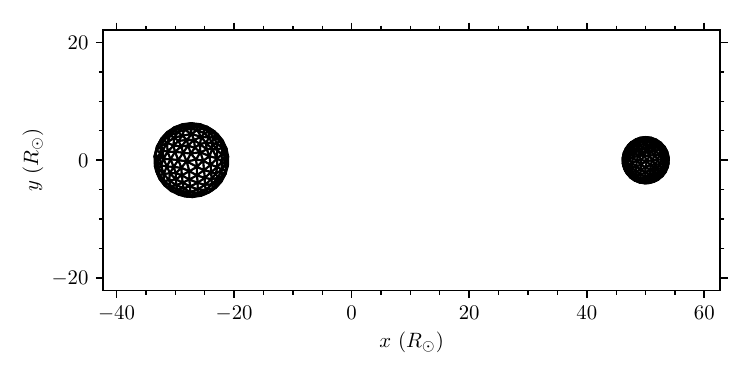}
    \includegraphics[width=.5\textwidth]{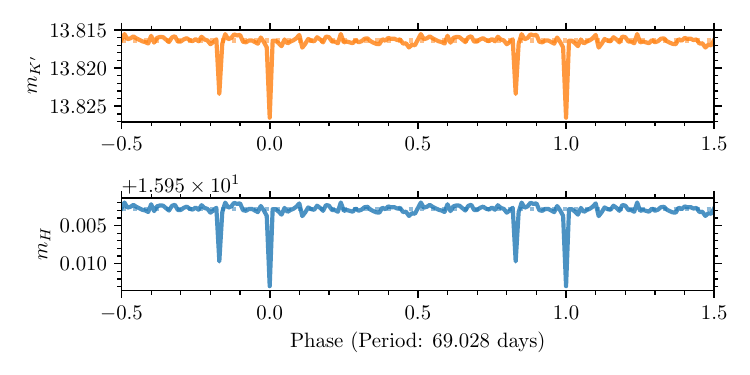}\includegraphics[width=.5\textwidth]{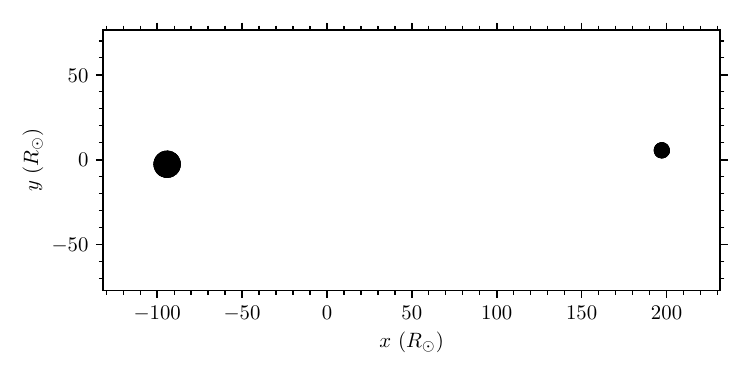}
  \caption{Examples of detached binary systems in our mock binary samples. Simulated light curves in $K'$-band and $H$-band are shown in the left column, with dotted horizontal lines indicating the median magnitude of each binary system used during the light curve injection procedure. The right column shows a mesh surface plot of each binary system. Mesh surface plots are plotted on the plane of the sky (here, $x$ and $y$) at 0.25 phase in each binary system.
  These systems have nearly $90\degree$, edge-on inclination, leading to eclipses. The bottom example system plotted has $e \gg 0$, leading to eclipses close together in phase. The narrow eclipses in phase of far separated, detached eclipsing binary systems makes them difficult to detect in our experiment due to our sparse observation cadence and the Lomb-Scargle periodogram being optimized for sinusoidal signals.
  The noise visible in the out-of-eclipse portion of the light curve in the last example indicates noise in our models originating from our mesh set-up of stellar atmospheres in PHOEBE. The noise is 0.002 mag in most observations, much smaller than our observations' photometric uncertainty, and is obvious in this example due to the small eclipse depth.
  }
  \label{fig:mock_binary_examples_detached}
\end{figure*}

\begin{figure*}[t]
  \centering

 \includegraphics[width=.5\textwidth]{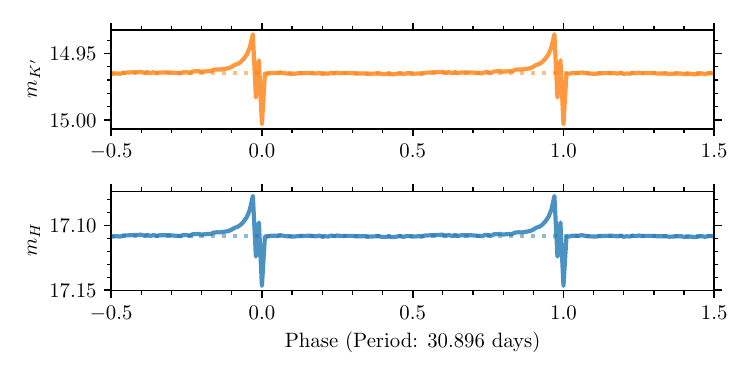}\includegraphics[width=.5\textwidth]{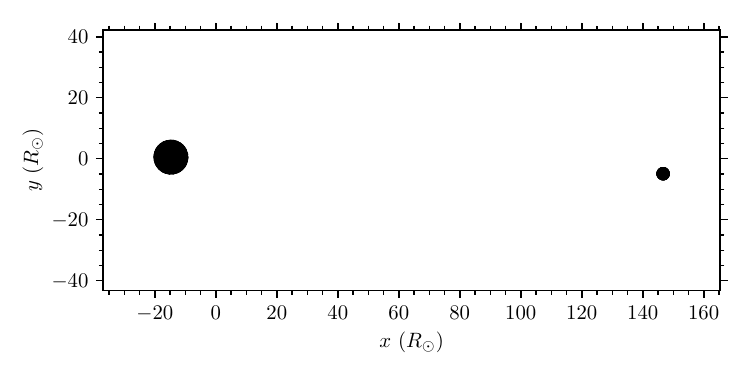}

  \caption{Same as Figure~\ref{fig:mock_binary_examples_detached}, but with an example of a \emph{heartbeat} binary star. In this type of system, the orbit is highly eccentric. During periapse, tidal deformations of the component stars lead to the characteristic light curves shown here. The narrow-in-phase nature of the flux variability makes these systems difficult to detect in our experiment.}
  \label{fig:mock_binary_examples_hb}
\end{figure*}

\begin{figure*}[p]
  \centering
    \includegraphics[width=.5\textwidth]{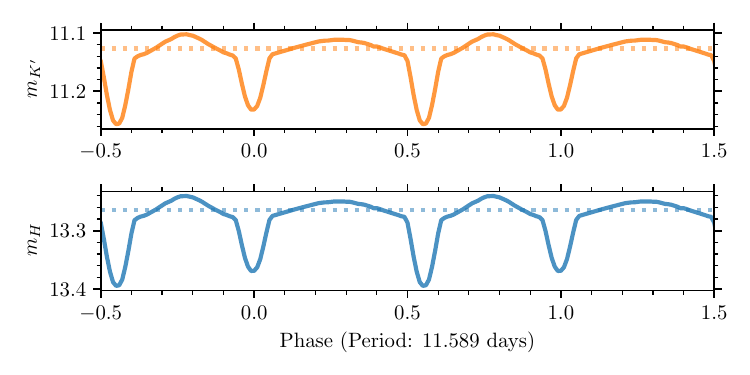}\includegraphics[width=.5\textwidth]{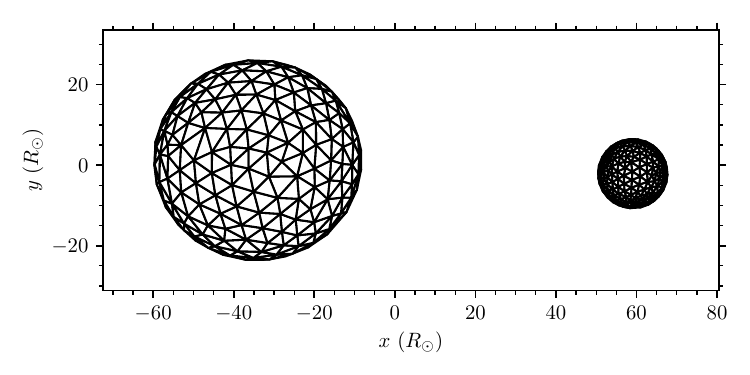}
   \includegraphics[width=.5\textwidth]{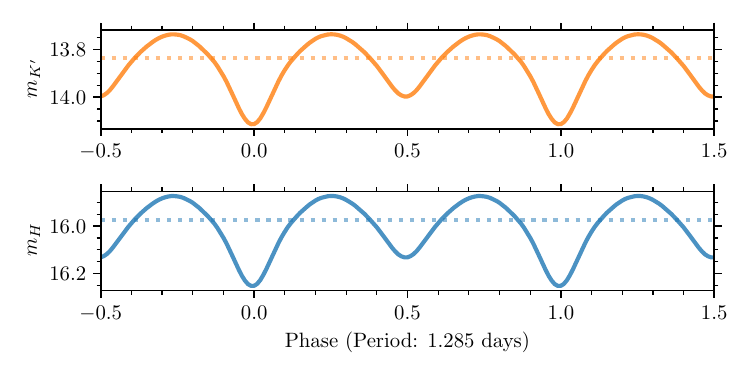}\includegraphics[width=.5\textwidth]{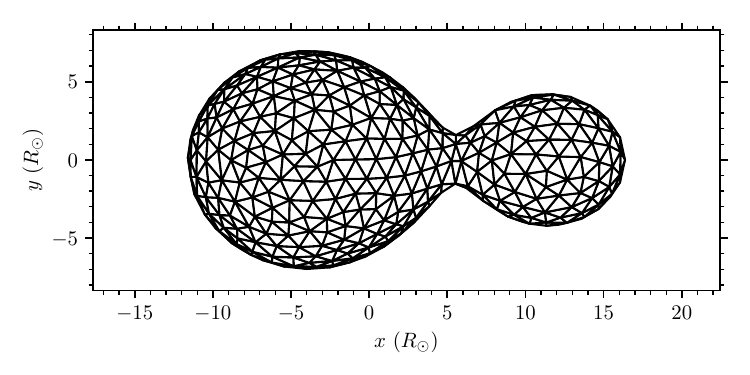}
    \includegraphics[width=.5\textwidth]{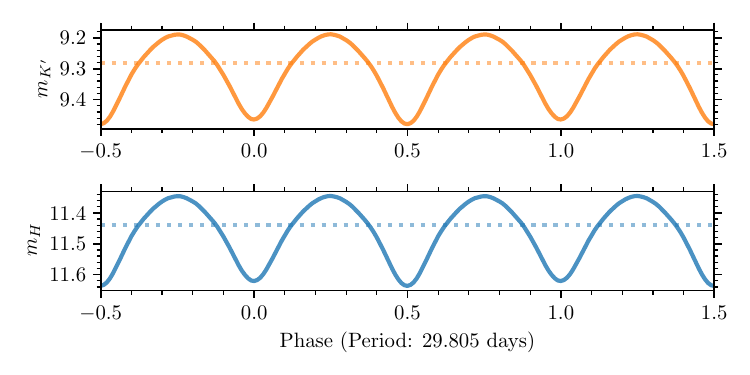}\includegraphics[width=.5\textwidth]{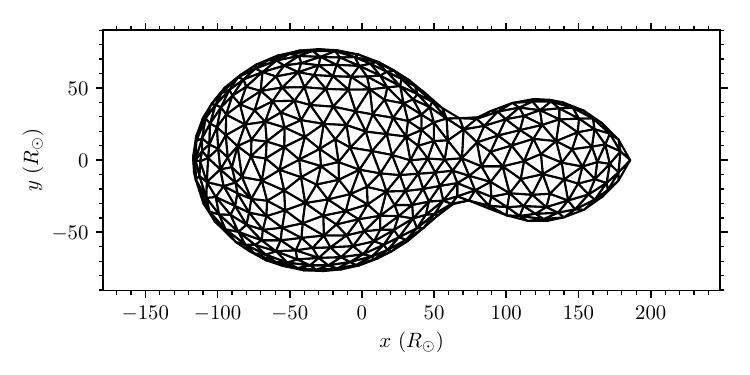}
    \includegraphics[width=.5\textwidth]{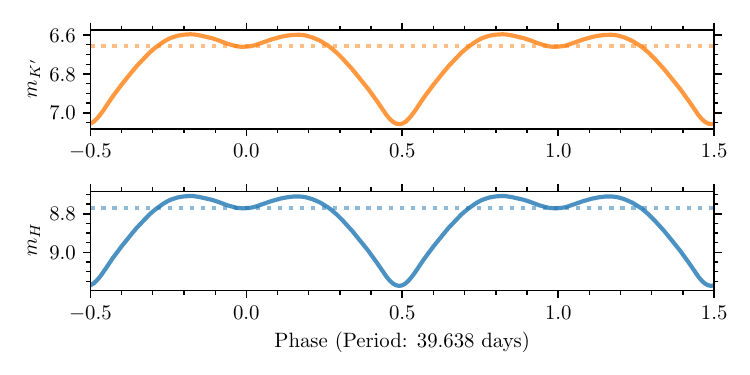}\includegraphics[width=.5\textwidth]{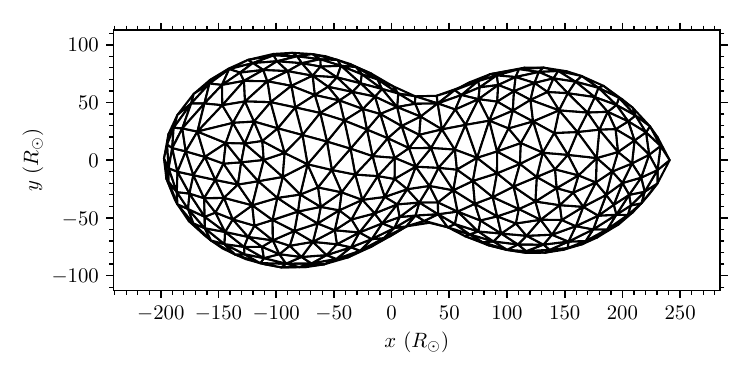}

  \caption{Same as Figure~\ref{fig:mock_binary_examples_detached}, but with examples of close binary systems in our mock binary samples. 
  The broad eclipses in phase of these systems, and the quasi-sinusoidal light curve shape of contact systems in magnitude space lead to easier detection in our Lomb-Scargle periodicity search. The out of eclipse flux variability in the light curves of the first example demonstrates \emph{ellipsoidal} variability, where tidal deformations on the surfaces of the component star from the mass of their respective companions leads to quasi-sinusoidal light curves in magnitude space.
  Note that in all but the last example systems, there are two dips in flux with approximately equal depth per orbital period. This will cause these first three example systems to be detected at approximately half their binary orbital period in our Lomb-Scargle periodogram. The last system, due to unequal eclipse depths, will be detected at the binary orbital period in the Lomb-Scargle periodogram.
  }
  \label{fig:mock_binary_examples_close}
\end{figure*}

\begin{figure*}[p]
  \centering
      \includegraphics[width=.5\textwidth]{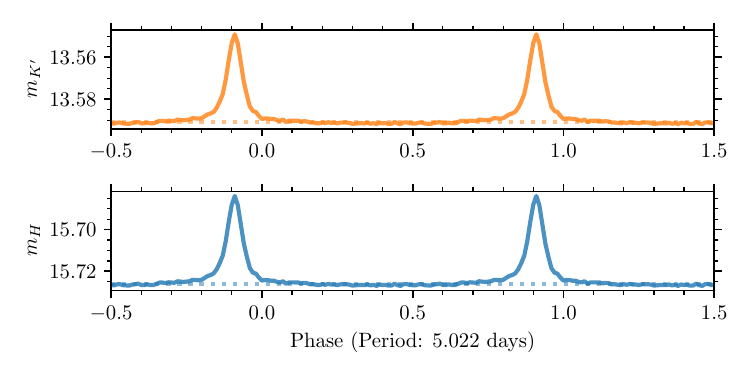}\includegraphics[width=.5\textwidth]{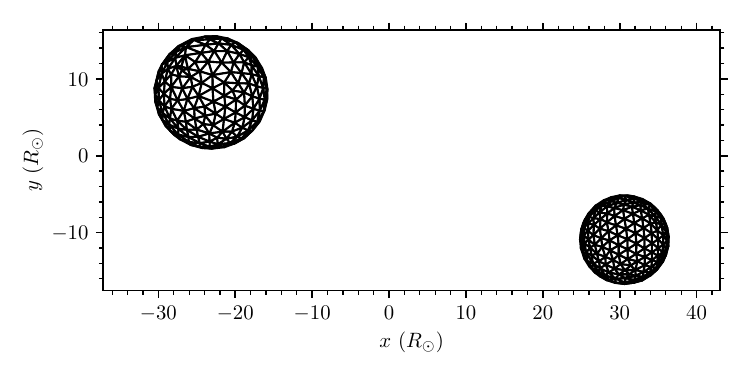}
      \includegraphics[width=.5\textwidth]{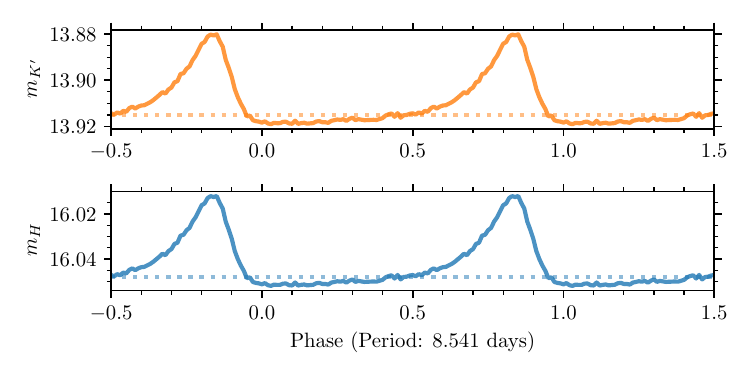}\includegraphics[width=.5\textwidth]{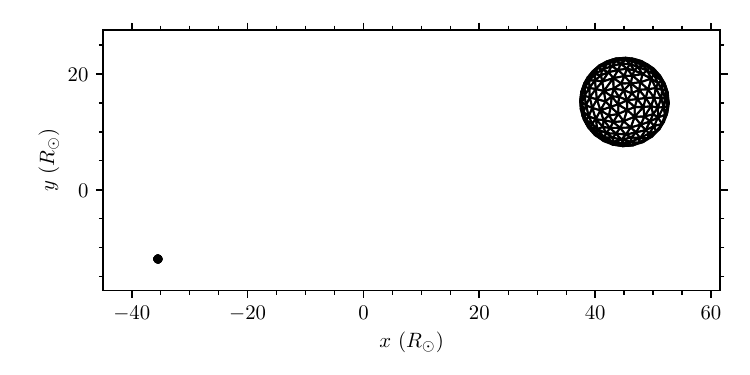}
   \includegraphics[width=.5\textwidth]{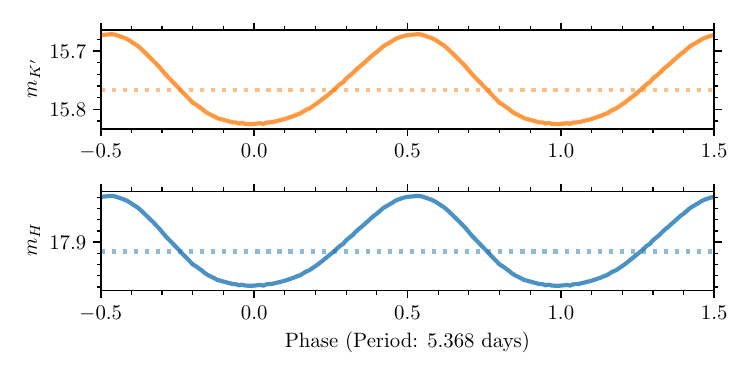}\includegraphics[width=.5\textwidth]{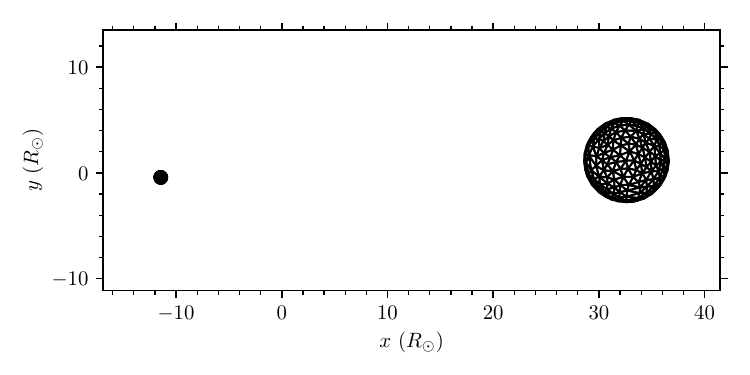}
   \includegraphics[width=.5\textwidth]{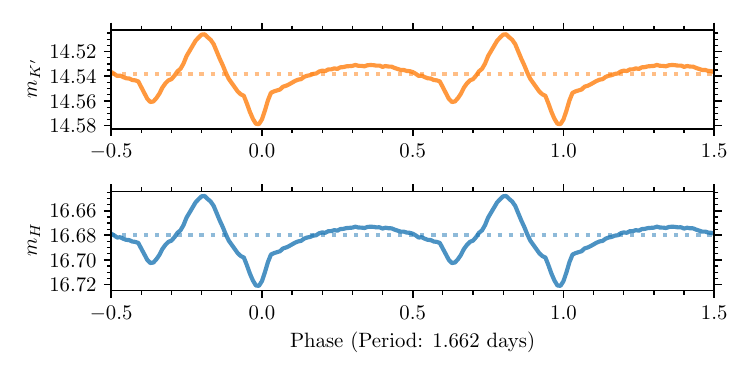}\includegraphics[width=.5\textwidth]{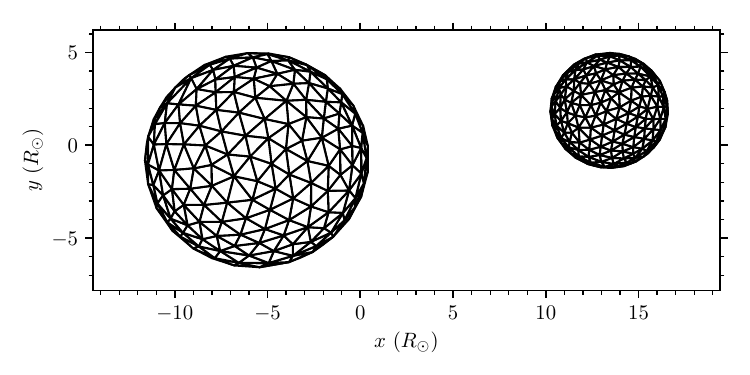}
   \includegraphics[width=.5\textwidth]{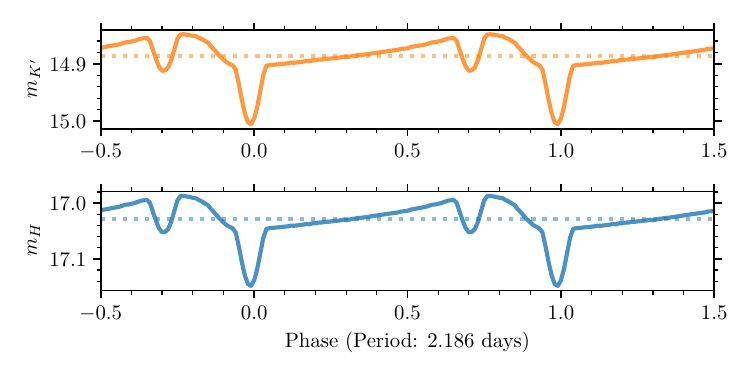}\includegraphics[width=.5\textwidth]{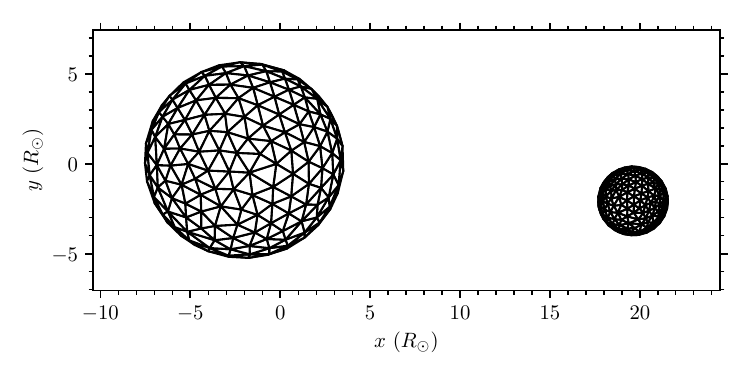}

  \caption{Same as Figure~\ref{fig:mock_binary_examples_detached}, but with examples of binary systems exhibiting irradiation variability, where surface temperature differences between the component stars leads to the hotter star heating one side of its cooler companion. This leads to brightening variability that can appear quasi-sinusoidal in close cases (like example 3 here).
  Note that unlike most close binary systems or contact systems (Figure~\ref{fig:mock_binary_examples_close}), these systems have one large dip or peak in flux per orbital period. This will cause these systems to be detected at approximately their binary orbital period in our Lomb-Scargle periodogram.
  Examples 4 and 5 demonstrate nearly edge-on systems, where the irradiation effect is coupled with eclipses.}
  \label{fig:mock_binary_examples_irrad}
\end{figure*}

\begin{figure*}[t]
  \centering
    \includegraphics[width=\textwidth]{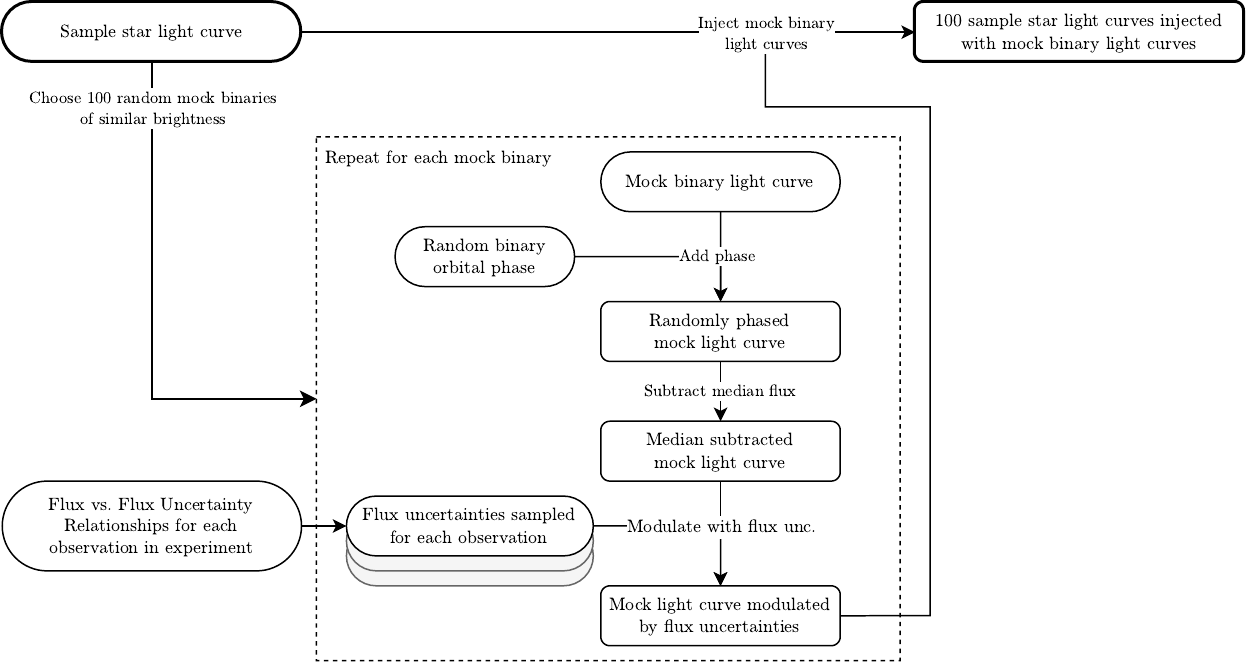}
  \caption{Outline of the procedure to inject mock binary light curves into every sample stellar light curve in our experiment.}
  \label{fig:binary_injection_procedure}
\end{figure*}

\paragraph{Binary light curve models} 
\label{par:binary_light_curve_models}
We generated binary models using the \textsc{PHOEBE 2.4} binary light curve simulation software \citep{Prsa:2016, Horvat:2018, Jones:2020, Conroy:2020}.
With \textsc{PHOEBE}, we constructed mock binary systems and simulated fluxes at 100 times uniformly across phase. In addition to the interpolated stellar parameters, we passed additional parameters of the binary system to use for constructing the model binary system. These quantities include the binary orbital period ($P$), orbital eccentricity ($e$), and inclination of the binary orbital plane ($i$).

PHOEBE uses a mesh geometry to simulate stellar atmospheres. For mock binaries in detached or semi-detached configurations, we constructed the mesh for each stellar component with 500 triangles. In contact binary configurations, the contact envelope was constructed with 1000 triangles. Due to the use of a mesh, our photometric models had photometric noise in the simulated light curves of $\approx 0.001$ mag (visible in the third example binary light curve in Figure~\ref{fig:mock_binary_examples_detached} due to its small eclipse depths). The noise originating from our choice for the number of triangles for simulating stellar atmospheres is much smaller than our experiment's photometric precision (reaching down to a median of $\approx 0.04$ mags across all observations; Figure~\ref{fig:mag_magerr}).

The determination of whether a model binary system is a detached, semidetached, or contact binary system was made before the simulated fluxes were calculated, with the help of the \textsc{Phitter} software. \textsc{PHOEBE} currently requires setting up each of these configurations of model binary systems in a separate manner. When starting to set up a model binary system, we calculated the Roche overflow limit. If the maximum radius of the physically larger of the two stellar components in the model binary system was between 98.5\% and 101.5\% of the Roche overflow limit, the detached binary configuration in \textsc{PHOEBE} would not work. In these cases, the model was set up as a semidetached binary system. For smaller or larger cases, detached or contact binary models were used, respectively.
In semidetached or contact binary systems, we fixed $e$ to 0, since \textsc{PHOEBE} does not support eccentric models in such systems. We expect such close binaries would be expected to have circularized orbits due to tidal circularization \citep[e.g.,][]{Bluhm:2016}, so this is a reasonable approximation.

Some of our 20,000 mock systems had non-physical or unstable parameters (e.g., large eccentricities that would lead to stellar mergers, or stellar radii much larger than the size of the binary orbit). We identified and removed such mock binary systems at this stage.
Our final mock binary population consists of 16,027 binary systems for which we successfully generated mock light curves. The distribution of parameters for the mock binary populations for which we simulated light curves are shown in Figure~\ref{fig:mock_binary_param_dists}. We also present examples of various types of light curve variability present in our mock binary library:
\begin{itemize}
    \item Figure~\ref{fig:mock_binary_examples_detached} shows examples of detached binary systems with eclipses far separated in phase. Eclipses narrow in phase are difficult to detect with the sinusoidal fits in our Lomb-Scargle periodicity search. Furthermore, narrow eclipses in phase are easy to be missed in our experiment's sparse observation cadence.
    \item Figure~\ref{fig:mock_binary_examples_hb} shows an example of a \emph{heartbeat} binary star. These types of binaries have very eccentric orbits. During periapse, tidal deformations in the component stars of the binary lead to flux variability appearing like their eponymous ``heartbeats'' as seen in electrocardiograms \citep[e.g.,][]{Welsh:2011, Thompson:2012, Shporer:2016}. The flux variability is typically narrow in phase, so detection of such signals in our experiment is difficult for similar reasons to those of detached binaries with narrow eclipses in phase.
    \item Figure~\ref{fig:mock_binary_examples_close} highlights examples of close binary systems. In such systems, eclipses are typically broad in phase, and many exhibit quasi-sinusoidal flux variability in magnitude space outside of the eclipses, originating from tidal distortions \citep[ellipsoidal variability, e.g.,][]{Morris:1985, Mazeh:2008}. In contact binaries, where one or both components are overflowing the Roche lobe, the flux variability approaches smooth quasi-sinusoidal variability across the entire phase. Due to their wide eclipses and quasi-sinusoidal flux variability in magnitude space, such systems are easily detectable by our experiment. If the eclipses are approximately equal in depth, such systems will be detected at half the binary orbital period in our Lomb-Scargle periodogram.
    \item Figure~\ref{fig:mock_binary_examples_irrad} shows various examples of binary systems with irradiation variability (i.e., reflection variability). The variability in such stars originates from a large difference in surface temperature between the component stars, leading to differential heating of the cooler star by the hotter star. The temperature difference leads to variability that can be seen as increases in flux \citep[e.g.,][]{Peraiah:1982, Wilson:1990, Davey:1992, Prsa:2016}.
    In further separated systems, this variability is typically narrow in phase and therefore difficult to detect with our experiment. However, in closer binary systems, the variability appears quasi-sinusoidal in magnitude space. Such variability can be effectively detected at the binary orbital period in the Lomb-Scargle periodogram.

    We also include examples of systems where eclipses are present with irradiation variability. Such signals may be challenging to detect with our periodicity search unless the variability flux amplitudes are large.
\end{itemize}


\subsubsection{Injection of mock binary signals into observed stellar light curves} 
\label{ssub:injection_of_binary_signals}
We injected 100 mock binary light curves into each of our sample's stellar light curves. Figure~\ref{fig:binary_injection_procedure} provides an overview of our binary light curve injection procedure that is described in this subsection below:

We first selected binaries of similar brightness from our mock binary library to inject into each of our stellar sample's light curves. For each sample star, we selected all mock binary systems that had combined system median $K'$-band magnitudes within $\pm 0.25$ mag of the median sample star $K'$-band magnitude. This allowed compensating for the variable extinction screen towards the GC \citep[as observed by][]{Schodel:2010}. If a given sample star had fewer than 10 mock binaries of similar $K'$ magnitude, we increased the search range in steps of 0.25 mag until at least 10 similar brightness mock binaries were selected. From these similar brightness stars, we then drew 100 random mock binaries, with replacement. For each of the 100 drawn mock binaries, we also picked a random phase shift in the interval $[0, 1)$.
After applying the random phase shift, we then sampled from the mock binary simulated light curves all observation times for the sample star. After sampling the light curve, we subtracted the median flux from the mock binary light curve. This procedure resulted in obtaining 100 mock light curves of binaries similar in brightness to the sample star, at each of the sample star's observation dates with random phasing.

A particular detail to note in our mock binary selection process is that since we allowed replacement when randomly selecting similar brightness mock binaries, the same mock binary could be selected for injection multiple times into a given sample star's light curve. However, each time a mock binary was drawn, a different random phase was selected. This meant that even for repeatedly injected mock binaries, phase shifts in the injected light curves would be different.

We next injected the mock binary light curves into our sample star's observation after modulating the mock light curve with the observations' photometric uncertainties. For every observation date, we constructed a flux vs. flux uncertainty relationship, in bins of half magnitude. In each bin, we calculated the median magnitude uncertainty (median $\sigma_{m}$) and the median absolute deviation in the magnitude uncertainty. This flux vs. flux uncertainty relationship for six example observations from our dataset is shown in Figure~\ref{fig:obs_unc_examples} in Appendix~\ref{sec:additional_deets_injection_recovery}.
At the given star's flux for the observation date, we sampled the observation date's flux vs. flux uncertainty relationship to choose a flux uncertainty for the injected binary sample. The sampled flux uncertainty was picked from a normal distribution (i.e., Gaussian distribution) with mean set at the sampled median flux uncertainty and standard deviation set at the median absolute deviation. The sampled flux uncertainties were then used to apply a flux modulation to the mock binary light curve to simulate our experiment's photometric noise.
Finally, the resulting modulated mock light curve was added to the observations of the sample star's light curve. This entire procedure was repeated for each mock binary light curve injected for a sample star.
At the end of this procedure, we obtained 112,900 sample light curve injected with a mock binary signal (1129 sample stars $\times$ 100 mock binary light curves per sample star). Three examples from our mock binary variability injection procedure are shown in Figure~\ref{fig:sbv_examples}, in Appendix~\ref{sec:additional_deets_injection_recovery}.


\subsubsection{Recovery of injected binary signals with periodicity search} 
\label{ssub:recovery_of_binary_signals}
This experiment's periodicity search was next run on the injected mock binary light curves in order to determine the \textit{recovery fraction} of mock binary signals in each sample star's light curve. The recovery fraction estimates the answer to the question: if a true signal like that found in our mock binary library is present in a given sample star's light curve, what is the probability that our experiment can detect it strongly enough to classify it as a \emph{likely periodic} detection? For each of the 112,900 sample light curves injected with a mock binary signal we ran the trended, multiband periodicity search described in \S~\ref{ssub:trended_multiband_periodicity_search}. We then obtained the recovery fraction of mock binary signals for every star as the fraction of mock binary signals detected by the periodicity search.

In order to consider a candidate periodic star detection to be consistent with the injected mock binary signal, the candidate detection had to pass the following criteria:
\begin{enumerate}
    \item At least one of the period match criteria:
    \begin{itemize}
        \item The period of the most significant detection from the periodicity search matched either the orbital period or the half orbital period of the injected mock binary to within 2\% of the period. Detections at the half orbital period are expected for eclipsing binary systems where the eclipses are of similar depths, or in ellipsoidal binary systems.
        \item The period of the most significant detection from the periodicity search matched the sidereal day alias of either the orbital period or the half orbital period of the injected mock binary to within 2\% of the alias period. Due to our experiment's observation cadence, the sidereal day alias is the strongest alias detected in our periodicity search for real periodic signals (see \S~\ref{ssub:results_detection_of_periodic_signals}). This alias can frequently dominate over the true signal in our periodicity search
        
        The sidereal day alias of a true signal's period is \citep[following][]{VanderPlas:2018}:
        \begin{eqnarray}
            P_{\text{alias}} = \left|\frac{1}{P_{\text{true}}} - \frac{1}{P_{\text{sidereal}}}\right|^{-1},
        \end{eqnarray}
        where $P_{\text{sidereal}} \approx 0.99727 \text{ day}$, in the time unit of solar days in MJD that we use in our experiment.
    \end{itemize}
    The 2\% period matching criteria originated from attempting to match the injected binaries' orbital periods with periodicity detections in the Lomb-Scargle periodogram. Since the Lomb-Scargle method uses a sinusoid model to search for periodic signals, the most significant period  is where the phased light curve best matches a sinusoid, which can be slightly different than the true orbital period of the injected binary star. The difference largely arises from deviations to the light curve from a sinusoid shape: e.g., from eclipses not being evenly distributed in phase (when the binary orbits have non-zero eccentricity), or the deviations of the flux from the stellar eclipses, tidal distortions, and irradiation leading to non-sinusoidal shapes. Therefore, we needed to allow for a difference between the measured period from the periodicity search and the true orbital period of the binary. We tested allowing for a 1\%, 2\%, or 5\% difference between the periodogram measured period and the true injected period on injected mock binary stars. The 1\% criteria missed many true detections that were otherwise matched with the 2\% or 5\% criteria. However the 5\% criteria matched false positives, largely for short period binary systems where the true injected period was close to the $\approx$ daily observation cadence. We therefore opted for the 2\% criteria for period matching.
    
    \item Amplitude of the candidate detection signal within $\pm 0.1 \text{ mag}_{K'}$ of the mock binary signal amplitude.
    
    \item Mock binary signal amplitude $\geq 0.01$ mag. Smaller magnitude signals could lead to spurious detections when conducting the periodicity search. 
\end{enumerate}

We separated all candidate periodic detections from the periodicity search run on our injected signal light curves by whether or not they are \emph{inconsistent} or \emph{consistent} with injected binary signals (i.e., \textit{false} or \textit{true} detections, respectively). We constructed a 2D histogram of the false candidate detections in the space of the two parameters we use for significance: the bootstrap false-alarm probability and the sinusoid amplitude significance. The 99.7\% (i.e., $3 \sigma$) contours from the histogram are plotted in Figure~\ref{fig:LS_sig_amp}. The distribution of false detections in the parameter space informed the \textit{likely} and \textit{possibly} periodic detection bounds in the periodicity search performed for our stellar sample (see \S~\ref{ssub:results_detection_of_periodic_signals} for details and illustrations of these ranges in Figure~\ref{fig:LS_sig_amp}).

We used the periodic binary signals recovered in the \emph{likely periodic} region of the significance parameter space to calculate the \emph{recovery fraction} for each star in our sample. The recovery fraction of injected binaries is plotted for all stars in Figure~\ref{fig:recovery_frac} in Appendix~\ref{sec:additional_deets_injection_recovery} and listed in the sample table (Table~\ref{tab:stellar_sample} in Appendix~\ref{sec:stellar_sample}). We used the recovery fraction to estimate the underlying true binary fraction of young stars at the GC, detailed in \S~\ref{ssub:results_bin_frac}.


\begin{figure}[t]
  \centering
        \includegraphics[width=.48\textwidth]{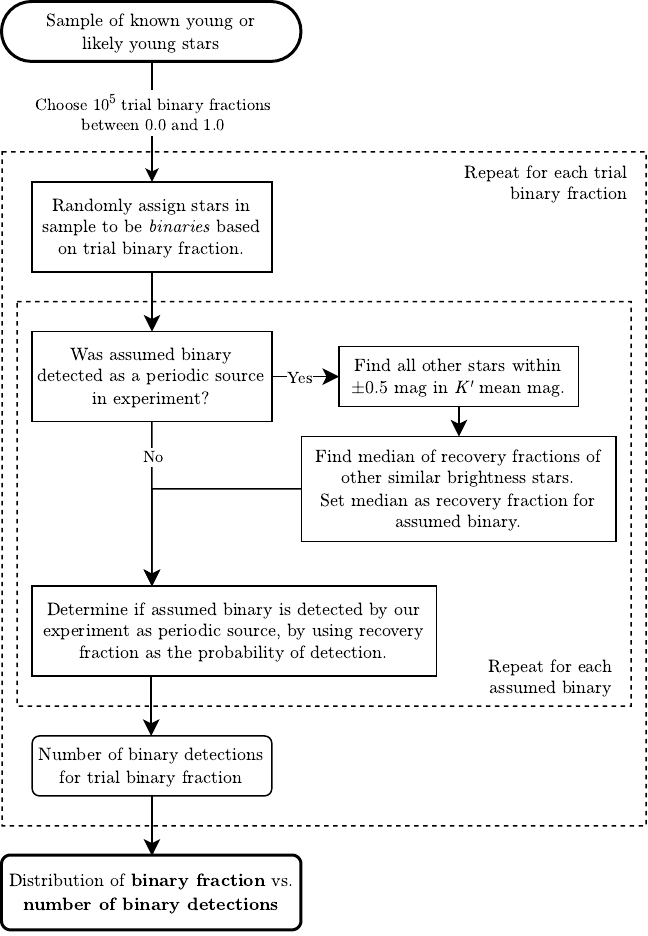}
  \caption{Procedure to estimate the GC young star binary fraction from our experiment's detections and the recovery fractions we estimated from our young mock binary light curve injections and recovery.}
  \label{fig:bin_frac_sim_procedure}
\end{figure}

\begin{figure}[t]
  \centering
        \includegraphics[width=.5\textwidth]{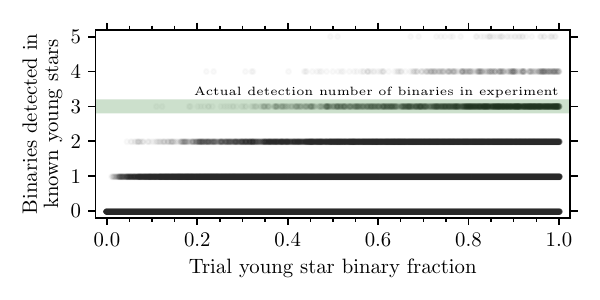}
  \caption{Distribution of trial binary fraction vs. number of binary detections for this experiment's known young star sample. This distribution is obtained at the end of the procedure outlined in Figure~\ref{fig:bin_frac_sim_procedure}. The green band indicates the true number of binary detections in our experiment's sample of known young stars: 3 binaries.}
  \label{fig:bin_frac_numdetect}
\end{figure}

\begin{figure*}[t]
  \centering
    \includegraphics[width=\textwidth]{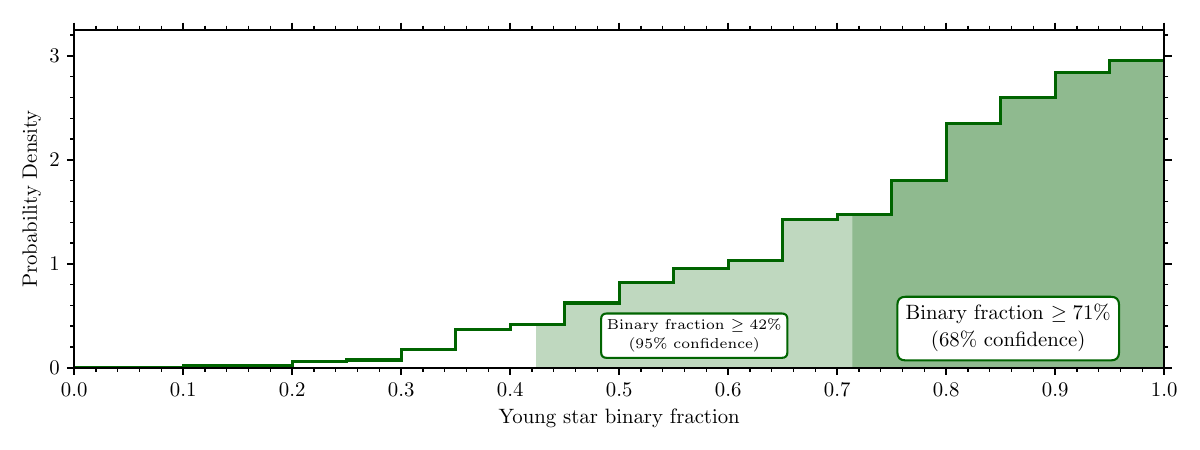}
    \includegraphics[width=\textwidth]{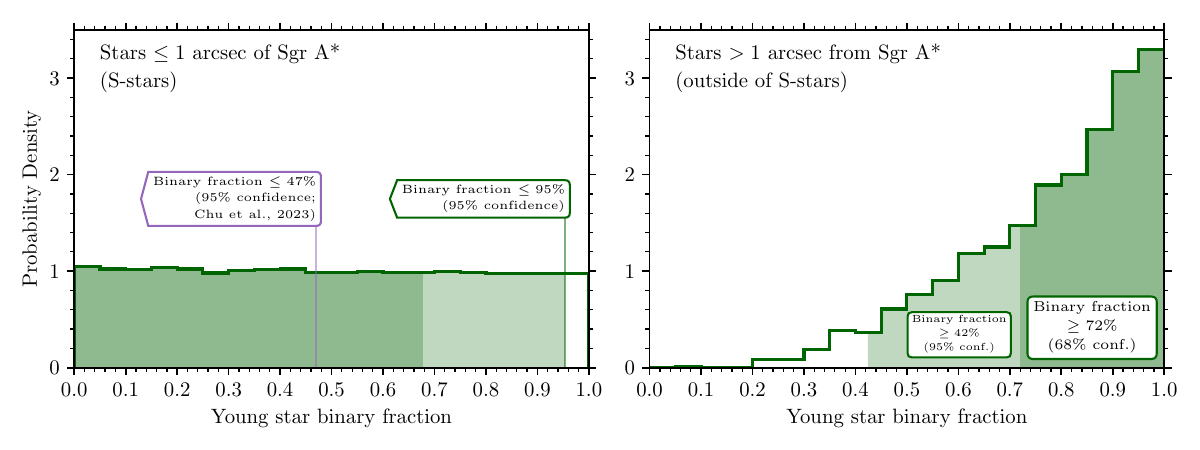}

  \caption{\textbf{Top: }Estimate of the GC young star binary fraction, based on the 3 detections among 102 spectroscopically confirmed young stars in this experiment. The young, massive star binary fraction is at least 71\% (68\% confidence), or at least 42\% (95\% confidence) within the central $\approx 0.4$ pc surrounding the GC SMBH.
  \textbf{Bottom: } Same as top panel, but considering stars inside and outside the central arcsecond separately. The left panel shows our constraints for stars within the central arcsecond (i.e., the S-stars within 0.04 pc of the central SMBH), while the right panel shows our constraints for stars outside the central arcsecond. We also plot for reference the 95\% confidence upper limit on the young S-star binary fraction from the spectroscopic binary search conducted by \citet{Chu:2023} in the left panel. Photometric experiments are less sensitive to binaries, and with no detections in the central arcsecond region our experiment's young star binary fraction constraints in the region are not as precise as those obtained via spectroscopic measurements by \citet{Chu:2023}. The stars in the two populations inside and outside the central arcsecond have likely had dynamically different lives, and dynamical models predict a higher binary fraction at larger distances from the central SMBH (see Figure~\ref{fig:dyn_prediction_comparisons}).
  }
  \label{fig:binfrac_dist_young_all}
\end{figure*}

\subsubsection{Results: Constraints on the GC Young Star Binary Fraction} 
\label{ssub:results_bin_frac}

Using the recovery fraction calculated in \S~\ref{ssub:recovery_of_binary_signals}, we were able to estimate the sensitivity of our experiment to binary systems and obtain a measurement of the GC young star binary fraction. The procedure to do this is presented as a diagram in Figure~\ref{fig:bin_frac_sim_procedure} and detailed below.

Among the sample of 102 known young stars, we calculated a distribution of expected number of binary systems our experiment would be able to detect based on different assumed binary fractions. To do so, we conducted $10^5$ Monte Carlo simulations of trial binary fractions in the range of 0.0 -- 1.0. For each trial binary fraction in each Monte Carlo simulation, we randomly assigned a fraction of the sample stars to be binaries based on the trial binary fraction.
We then went through every star assumed to be a binary in the Monte Carlo simulation, and determined if our experiment could detect the binary as a photometrically periodic source.
This determination was based on the star's recovery fraction to define the probability of detection. If a given star was determined as likely periodic in our actual experiment (i.e., IRS~16SW, S2-36, and S4-258), we set its recovery fraction to be the median of the recovery fraction of all non-periodic stars with $\overline{m}_{K'}$ within $\pm 0.5$ mag.
At the end of each trial Monte Carlo simulation, we then obtained the total number of photometric binary detections for the assumed trial binary fraction.
From all $10^5$ simulations, we obtained a distribution of trial binary fraction vs. number of binary detection for the sample of known young stars. This distribution is illustrated in Figure~\ref{fig:bin_frac_numdetect}.

In order to obtain a binary fraction constraint, we cut the trial binary fraction vs. binary detection number distribution where the number of detections is the same as that from our actual experiment: 3 binary stars detected (see \S~\ref{ssub:results_detection_of_periodic_signals}). The resulting binary fraction distribution is plotted in the top row of Figure~\ref{fig:binfrac_dist_young_all}.
At 68\% confidence, the binary fraction of young stars at the GC is at least 71\%. At 95\% confidence, we place a lower limit of 42\% on the binary fraction of GC young stars.

We additionally obtained a binary fraction measurement on smaller subsamples in our experiment: young stars located inside and outside the central arcsecond. The central arcsecond approximately contains the S-star cluster, and the young stars inside and outside the S-star cluster may have had different different dynamical origins and star formation scenarios. A significant difference in the stellar binary fraction between the two regions could provide support to these differences. This experiment's binary fraction measurement for the two regions is plotted in the bottom row of Figure~\ref{fig:binfrac_dist_young_all}.
Inside the central arcsecond, we detect 0 binaries. Due to the lack of sensitivity to binaries at the brightness of the young S-stars and the small size of the subsample, the detection of 0 binaries in the subsample is not able to place a significant constraint on the underlying binary fraction in the region.
Outside the central arcsecond, however, we detect 3 binaries. In this region, the detections allow obtaining a slightly tighter constraint on the underlying young star binary fraction than that obtained with our complete sample: at least 72\% (with 68\% confidence), or at least 42\% (with 95\% confidence).




\begin{deluxetable}{lL}[t]
    \tablecolumns{2}
    \tablecaption{Young Star Stellar Binary Fractions\label{tab:young_bin_fracs}}
    \tablehead{
        \colhead{Population}        &
        \colhead{Binary Fraction}
    }
    \startdata
    WR and OB stars in Galactic Center     &\\
    (68\% conf., this work)  & \geq 71\% \\
    (95\% conf., this work)  & \geq 42\% \\
    \tableline
    B stars in S-stars ($\leq 1''$ of SMBH)  & \leq 95\% \\
    (95\% conf., this work)          &   \\
    B stars in S-stars ($\lesssim 1''$ of SMBH)  & \leq 47\% \\
    \citep[95\% conf.,][]{Chu:2023}  &   \\
    \\
    WR and OB stars outside S-stars   &  \\
    ($> 1''$ of SMBH)   &   \\
    (68\% conf., this work)  & \geq 72\% \\
    (95\% conf., this work)  & \geq 42\% \\
    \tableline
    O stars in solar neighborhood   & (69 \pm 9) \%  \\
    \citep[68\% conf.,][]{Sana:2012}   &   \\
    O stars in solar neighborhood   & (31 \pm 7) \%  \\
    close binaries: $P = 2$--20 d   &   \\
    \citep[68\% conf.,][]{Moe:2013} &   \\
    B stars in Milky Way    & \gtrsim 60 \%  \\
    \citep{Duchene:2013}    &   \\
    B stars in Milky Way    & (21 \pm 5) \%  \\
    close binaries: $P = 2$--20 d   &   \\
    \citep[68\% conf.,][]{Moe:2013}    &   \\
    B stars in LMC  & (7 \pm 2) \%  \\
    intermediate-period binaries: $P = 20$--50 d    &   \\
    \citep[68\% conf.,][]{Moe:2015b}   &   \\
    \enddata
\end{deluxetable}

\section{Discussion} 
\label{sec:discussion}
This experiment's results show a high stellar binary fraction for the known young, massive stars at the GC, of at least 71\% (68\% confidence). A comparison of our binary fraction measurements with other studies of the GC binary fraction and other young star populations are listed in Table~\ref{tab:young_bin_fracs}. The GC binary fraction is consistent with that of O stars \citep[$69\% \pm 9\%$;][]{Sana:2012} and of B stars \citep[$\gtrsim 60 \%$;][]{Duchene:2013} in the solar neighborhood.
As Figure~\ref{fig:bin_frac_numdetect} illustrates, the biggest source of uncertainties in our result is the small number of detections: 3 detections from a sample of 102 young stars.
Besides the number of detections, the binary fraction measurement results are also dependent on the mock population chosen to represent the binary population. We discuss in \S~\ref{sub:discussion_bin_frac_dependence_mock_population} how our choices for our mock binary population affect the final implied binary fraction final measurement.
The young star stellar binary fraction is of particular importance for constraints on star formation in the GC, and the implications of the high binary fraction are considered in more detail in \S~\ref{sub:discussion_in_situ_star_formation}. We consider our measurements of the GC binary fraction with the recent measurement of a low binary fraction for the GC young S-stars by \citet{Chu:2023}, and what the combined results can inform about the GC dynamical environment in \S~\ref{sub:discussion_dynamical_interactions}. Finally, other implications for the GC environment from our results are considered in \S~\ref{sub:discussion_gc_environment}.

\subsection{Dependence of binary fraction measurement on mock population characteristics} 
\label{sub:discussion_bin_frac_dependence_mock_population}
A crucial component to measure the intrinsic binary fraction from the observed binary systems in our experiment was the population of mock binaries we generated to model the underlying population of binaries at the GC (described in \S~\ref{ssub:mock_binary_population}). We generated the binary star population with parameters expected for the young, massive stars in the area. The age and masses of the stars were chosen based on estimates from \citet{Lu:2013}, who found that all young, early-type stars at the GC are consistent with a single star formation event, and estimated a most probable age of $\approx 4$ Myr for the young stars and a top-heavy mass function. While the mass function is determined from prior observations of GC stars \citep{Lu:2013}, the binary systems parameters of mass ratio, period, or eccentricity are informed by constraints from local massive star populations \citep{Sana:2012} since these distributions have not yet been determined for the GC.
The GC young stellar population may have differences in the binary star parameters from local populations, and the differences may thus bias the resulting binary fraction estimate we inferred in our work. With a photometric experiment and just 3 binary detections, we are unable to use our observations to significantly constrain the population's characteristics.

Our experiment's difficulty in placing constraints on the underlying binary population parameters of the young stars in the central half parsec of the Galaxy stems primarily from photometric surveys being limited in their sensitivity to off-edge binary systems. That limitation, however, can be addressed by future GC binary surveys.
In particular, spectroscopic surveys of the GC stellar population will be more deeply sensitive to less edge-on inclination binary systems, allowing a more complete census of the binary population. With ground-based AO datasets, sensitivity to binary stars at $m_{K'} \lesssim 16$ (corresponding to B-type and more luminous stars for the GC young star population) can be reached \citep{Chu:2023, Do:2019}. Ongoing spectroscopic surveys with JWST will allow deeper spectroscopy to $m_{K'} \lesssim 19$ (corresponding to A-type and more luminous stars for the GC young star population).
The binary population census allowed by these spectroscopic surveys will be able to address if the binary star population parameters are indeed consistent with those of local young star populations.

We consider below in more detail three possible differences to the underlying mock binary population that would imply differences to the inferred binary fraction:
the assumed age of the young star population (\S~\ref{ssub:differences_in_assumed_age_of_young_star_population}),
the stellar isochrones and models used for the young star parameters (\S~\ref{ssub:differences_in_young_star_models}),
and the distribution of binary orbital periods (\S~\ref{ssub:differences_in_binary_orbital_period_distribution}).

\subsubsection{Differences in assumed age of young star population} 
\label{ssub:differences_in_assumed_age_of_young_star_population}
The choice of age of the mock binary population was informed by previous observational constraints \citep{Lu:2013}, which determined that all the massive, young stars in the GC making up the YNC are consistent with having formed in a single star formation event. The star formation event likely took place between 2.5 and 5.8 Myr ago (95\% confidence).
The adopted age of 4 Myr for the population of mock binary stars used to derive our results in \S~\ref{sub:binary_star_fraction_determination} was chosen to match the most probable age solution inferred by the prior work's observational constraints.
To examine the effect of the assumed population age on the resulting binary fraction estimate, we repeated our analysis in \S~\ref{sub:binary_star_fraction_determination} assuming several different young star population ages between 2.5 Myr and 5.8 Myr, the 95\% confidence bounds estimated by \citet{Lu:2013}.

Assuming a stellar population age for the GC young stars younger than 4 Myr resulted in no difference to the resulting binary fraction estimate. Stellar ages younger than 4 Myr were preferred in the \citet{Lu:2013} estimate: comprising 73\% confidence in their posterior distribution for population age. In this young star population age region, our binary fraction estimate results from \S~\ref{ssub:results_bin_frac} still hold.
However, assuming young star ages older than 4 Myr does lead to slightly lower young star binary fraction estimates: the 95\% confidence lower limits for the implied young star binary fraction are reduced to $\geq 33\%$ (5.0 Myr age) and $\geq 30\%$ (5.8 Myr age), compared with $\geq 42\%$ for the 4 Myr and younger ages. Stellar population ages older than 4 Myr have slightly less support by the \citet{Lu:2013} age estimate: with 19\% confidence for 4.0 -- 5.0 Myr and 25\% confidence for 4.0 -- 5.8 Myr.

The source of the difference in the inferred binary fraction at older population ages is due to the expected presence of more stars that have evolved off the main sequence. In particular, MIST stellar evolution models for ages $> 4$ Myr predict that stars at the bright end of our sample ($10 \lesssim m_{K'} \lesssim 12$) have entered the red giant phase with physically much larger stellar atmospheres. Since larger stellar radii corresponds to wider eclipses in phase, these binaries are expected to be more easily detected in our experiment. Our experiment's detection of 3 binaries at the GC then implies a slightly lower binary fraction.

This experiment's results motivate future studies to more precisely constrain the age of the GC young star population.
So far, previous estimates of the young star population's age favor ages $< 4$ Myr, where our estimated binary fraction constraint measurements hold. However, an older stellar population age extending to $\approx 6$ Myr cannot yet be ruled out, where our experiment estimates a lower stellar binary fraction.
Furthermore, it is important to note that the \citet{Lu:2013} estimates are derived with single star evolution models. Close interactions in binary systems, such as mass exchange and mergers, lead to the observation of post-main-sequence stars, like Wolf Rayet stars, at older stellar ages, which may in turn imply an older age for GC young stars.
Clearly, additional work is needed to better constrain the GC young star population age with considerations of binarity.


\subsubsection{Differences between stellar parameters in young star models} 
\label{ssub:differences_in_young_star_models}
The binary fraction estimates relied on realistic models of young, massive stars in order to determine our experiment sensitivity. However, many uncertainties still remain in the understanding massive star evolution, and this uncertainty is reflected in the different stellar parameter predictions offered by stellar evolution models \citep[e.g.,][]{Agrawal:2022}.
For our photometric study with sparse time sampling, one of the key determinants of detection was stellar radius. Larger stellar sizes resulted in binary light curves with wider eclipses in phases. Such systems are more easily detectable in our experiment.

We analyzed differences in the predictions of stellar radii for stellar ages spanning 2.5 -- 5.8 Myr and stellar phases spanning the range of our stellar sample for the MIST isochrones \citep{Choi:2016}, derived from MESA stellar evolution models \citep{Paxton:2011}, and the Geneva stellar isochrones and models \citep{Ekstrom:2012, Yusof:2013}.
MIST and Geneva stellar models are commonly used for models of young, early-type stars.
For stars near the main sequence phase, both MIST and Geneva predicted comparable stellar radii for a given initial stellar mass. However, MIST expects stellar radii $\approx 2 \times$ to $3 \times$ larger for stars near the end of the main sequence phase (corresponding to $10 \lesssim m_{K'} \lesssim 12$ for our sample at the GC). For star populations $> 4$ Myr in age, MIST predicts the bright end of our sample to be populated by red giant stars, with stellar radii approaching $\approx 10 \times$ larger than those predicted by Geneva for the same initial star mass.
Our analysis demonstrates that if our sensitivity analysis was based off of Geneva stellar evolution models, \emph{fewer} mock binaries would be recovered due to the typically smaller stellar radii and narrower eclipses, and our 3 binary detections would suggest a \emph{higher underlying binary fraction}.
Uncertainties remain in massive star evolution models, but the typically larger stellar radii predicted by MIST for the stellar evolution phases used in our work suggest that our binary fraction result may be an \textit{underestimate} if GC young stars are physically smaller than predicted by MIST.

Lastly, the stellar models we considered assume isolated, single star evolution only, but close interactions between component stars in binary systems are frequent, especially in young, massive stars \citep{Sana:2012}. These interactions will leave differences in the implied stellar parameters. BPASS stellar models \citep{Eldridge:2017} consider binary evolution in their detailed stellar evolution code. However, many aspects of binary evolution, such as detailed mass transfer, are still not included in models due to the difficulty in computation. Such close binary interactions are expected to be frequent in the GC's dense stellar environment surrounding the central SMBH \citep[e.g.,][]{Stephan:2019}, and therefore our use of only single star evolutionary models in the mock binaries is a limitation. Future work with detailed consideration of binary interactions in stellar models will assist in obtaining more accurate constraints of the GC binary population.

\subsubsection{Differences in binary orbital period distribution} 
\label{ssub:differences_in_binary_orbital_period_distribution}
Notably, two of the three stellar binaries detected in our experiment have orbital periods longer than 10 days, while the fourth known binary in the central half parsec region surrounding Sgr A*, IRS~16NE, has a much longer orbital period of 224.1 days.
The distribution of observed binary periods may suggest a deviation from the orbital period distribution of massive binaries in the solar neighborhood as measured by \citet{Sana:2012}, demonstrating that massive binaries are typically found with shorter orbital periods.
We performed a KS test to determine the probability that the observed periods can be consistent with the local distribution of observed periods: $p = 0.074$ when only considering GC binaries detected photometrically, $p = 0.008$ when also including the long-period GC binary IRS~16NE. Here, $p$ indicates the probability of the observed sample of binary periods are consistent with being drawn from the distribution of observed binary periods for local massive stars. The small values of $p$ from the KS test hint towards a possible deviation from the local orbital period distribution, especially when considering IRS~16NE. However, more GC binary detections are needed with longer periods to detect a significant deviation from the local population.

If the GC young binaries are indeed found more often in longer period systems (i.e., a flatter period distribution with $\pi > -0.55$ as defined for the period distribution in \S~\ref{ssub:mock_binary_population}), the implied intrinsic GC young star binary fraction would be even higher than our results reported in \S~\ref{ssub:results_bin_frac}: for larger values of $\pi$, the mock binary population would contain more binaries at longer periods, while our experiment is more sensitive to shorter period binaries (see Figure~\ref{fig:mock_binary_param_dists}). Therefore, our recovery analysis would detect \textit{fewer} binaries for every trial binary fraction, resulting in our experiment's detection of 3 binaries implying a \textit{higher binary fraction}. For this reason, the binary fraction estimate in our results would serve as an \textit{underestimate} if GC young stars are more likely to be in longer period binaries.



\begin{figure*}[t]
  \centering
    \includegraphics[width=\textwidth]{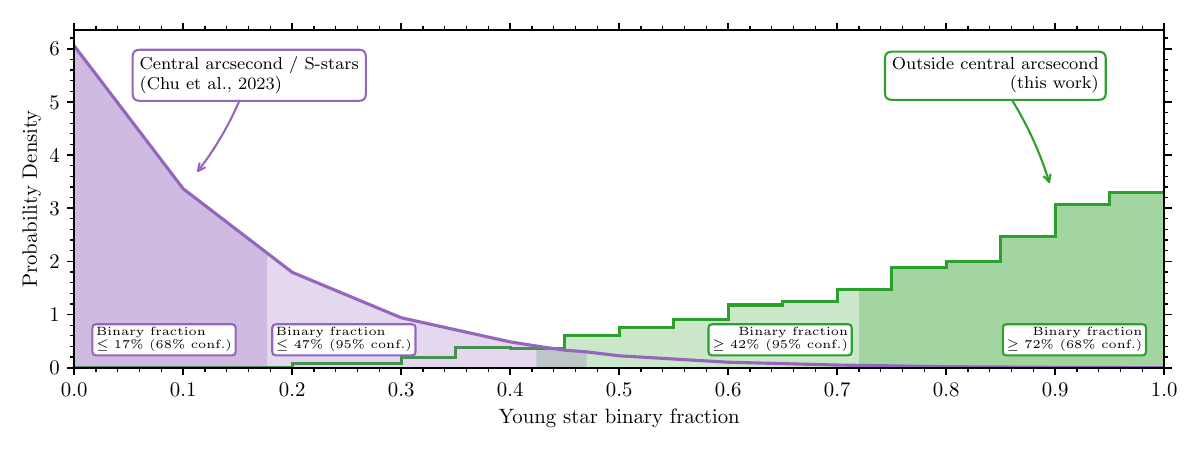}
  \caption{Comparison of the young star binary fraction estimated for the young stars outside the central arcsecond (this work, green line) with the estimate for the young S-stars in the central arcsecond \citep[][purple line]{Chu:2023}. The probability distributions for the young star binary fraction for the two populations are significantly different: they are unlikely to share the same underlying stellar binary fraction ($p < 1.4\%$). Lighter and darker shading inside each probability density function indicate $1\sigma$ (68\%) and $2\sigma$ (95\%) confidence intervals, respectively.}
  \label{fig:binfrac_cen_arcsec_compare}
\end{figure*}

\begin{figure*}[t]
  \centering
  \includegraphics[width=\textwidth]{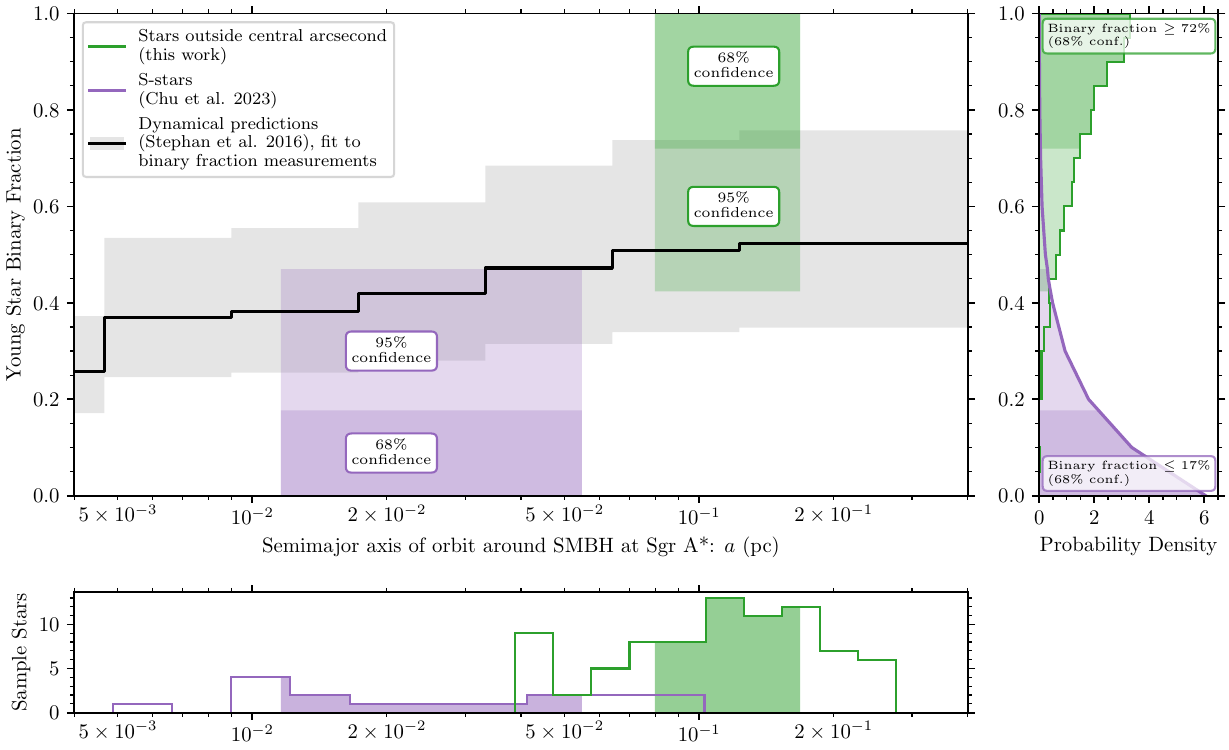}
  \caption{
      \textbf{Main panel:} The young star binary fraction constraints from this work for the stars outside the central arcsecond (green) and from \citet{Chu:2023} for the S-stars (purple) are plotted against the semimajor axis of the respective samples' stellar orbits around the central SMBH at Sgr A*.
      The black line shows binary star depletion at the GC via dynamical interactions by \citet{Stephan:2016}, scaled to an initial binary fraction best fit to binary fraction measurements from this work and for the S-stars: $(52^{+23}_{-17})\%$ ($1 \sigma$ uncertainties on the fit, indicated on the plot as the grey band).
      The SMA range plotted for both observational measures spans the interquartile range of the SMAs in the respective samples: 25\textsuperscript{th} -- 75\textsuperscript{th} percentile.
      \textbf{Right panel:} Same as Figure~\ref{fig:binfrac_cen_arcsec_compare}, plotted here for comparison to the dynamical prediction fit.
      \textbf{Bottom panel:} The semimajor axis distributions of the binary fraction measurements' respective stellar samples are indicated. This experiment's ``outside central arcsecond" sample was selected by choosing stars that have a projected distance $> 1''$ from Sgr A* ($> 0.04$ pc from the SMBH). The interquartile range of SMAs is shaded in the histograms of both samples.
  }
  \label{fig:dyn_prediction_comparisons}
\end{figure*}

\subsection{Constraints on in situ star formation} 
\label{sub:discussion_in_situ_star_formation}

The degree of fragmentation during star formation is imprinted on a young star population's stellar multiplicity \citep{Duchene:2013}, so this experiment's high young star binary fraction estimate allows a constraint on the possible in situ star formation processes that have occurred at the GC. One of the most compelling routes proposed for in situ star formation at the GC is via fragmentation of an accretion disk that may have previously surrounded the SMBH \citep[e.g.,][]{Levin:2003, Milosavljevic:2004, Nayakshin:2005a}.
\citet{Nayakshin:2007} conducted numerical simulations demonstrating the formation of a stellar disk from a gravitationally unstable gaseous accretion disk surrounding the central SMBH.
Their simulations with long cooling timescales resulted in a top-heavy initial mass function, as already observationally measured for the GC young stars by \citet{Lu:2013}, and high stellar binary fractions, which now this work's measurements support.

Observational evidence of a high stellar binary fraction motivates additional theoretical study of disk formation for the GC young stars. Since the top-heavy initial mass function and high stellar binary fraction suggest a slow cooling timescale during star formation, additional work is necessary to understand the slow cooling's origin. Slower cooling may be expected for the disk in scenarios where the strong accretion activity onto the central SMBH could heat the disk \citep[there is some observational evidence for past strong accretion activity onto the GC SMBH; e.g.,][]{Li:2013}. However, previous simulations of disk star formation have lacked effects like black hole feedback \citep[e.g.,][]{Morris:2023}. Our results suggest that the inclusion of such advanced phenomena and other advanced physics like magnetic fields in disk formation simulations are warranted.

\subsection{High binary fraction in the context of GC dynamical interactions} 
\label{sub:discussion_dynamical_interactions}
When comparing our findings with previous measurements of the binary fraction close to the SMBH, we find a significant difference in the young star binary fraction with distance from the SMBH.
Figure~\ref{fig:binfrac_dist_young_all} shows our constraints on the GC young star binary fraction separated by the population found inside $1''$ ($\approx 0.04$ pc in projected distance) of the SMBH (i.e. the S-star population) and outside $1''$ of the SMBH. We detected no binary systems among the known 20 young S-stars in our experiment. However the improbability of detecting binaries among these 20 stars means that our null detection isn't very constraining of the underlying binary fraction. Outside of the S-stars, our sample contains 82 known young stars, of which 3 are detected in our experiment as binaries. This allows us to constrain the underlying young star binary fraction in this region further away from Sgr A* to be at least 72\% (with $1 \sigma$ / 68\% confidence), or at least 42\% (with $2 \sigma$ / 95\% confidence). These results are summarized in Table~\ref{tab:young_bin_fracs}. Combining our results with those of \citet{Chu:2023}, who place a $2 \sigma$ / 95\% confidence upper limit of 47\% on the binary fraction of the young S-stars, the GC young star binary fraction appears to have a radial dependence based on the distance from the SMBH. In fact, when comparing the probability density function implied by our measurement of the binary fraction outside the central arcsecond to that of the S-stars by \citet{Chu:2023} (Figure~\ref{fig:binfrac_cen_arcsec_compare}), the probability that the two regions share the same underlying binary fraction is $< 1.4\%$.

A depletion of stellar binaries closer towards the SMBH as evidenced by combining the results of our study and those of \citet{Chu:2023} is predicted by dynamical models for the GC. Binary evaporation by frequent stellar interactions in the dense GC environment is one avenue of depletion \citep[e.g.][]{Alexander:2014, Rose:2020}. Furthermore, binary mergers are expected to be common in the GC. These mergers are expected to be induced by the EKL mechanism, where stellar binaries form a hierarchical triple with the central SMBH. In this scenario, the orbit of the binary components around each other can become highly eccentric, enough to occasionally cause stellar mergers \citep{Naoz:2016, Stephan:2016, Stephan:2019}, and possibly resulting in the production of stellar merger objects seen as the G-objects \citep[e.g.,][]{Ciurlo:2020}. Both processes are predicted to be strongest closer towards the central SMBH. In order to test our measurements with these predictions, we fit the \citet{Stephan:2016} prediction of binary depletion after 6 Myr for GC young stars with the binary fraction measurement from this experiment (\S~\ref{ssub:results_bin_frac}) and from \citet{Chu:2023}. This fit is shown in Figure~\ref{fig:dyn_prediction_comparisons}, showing that the dynamical predictions can adequately describe the observed depletion in binary fraction towards the central SMBH, with a best fit initial binary fraction of $(52^{+24}_{-17})\%$. In this fit and in Figure~\ref{fig:dyn_prediction_comparisons}, the estimation of the semimajor axis (SMA) of the orbit of our sample star's orbits around the SMBH was done using the following assumptions: stars are located at their respective orbit's apoapse, the eccentricity of the orbit around the SMBH is 0.3, and that the $z$ distance from the SMBH, the distance along the line of sight from the SMBH, is obtained by $z = r_{\text{2D}} / \sqrt{2}$.

Our study's measurement of a high stellar binary fraction at the 68\% confidence level may suggest a steeper depletion in stellar binaries than predicted by \citet{Stephan:2016}. A steeper depletion can hint towards an imprint of the star formation process, leaving an initially inhomogeneous spatial distribution of stars, like a radial edge inside which stars cannot form. Such an inner edge to star formation is present in several models of in situ models \citep[e.g., disk formation][]{Nayakshin:2007}, and is observationally motivated by the inner edge observed for the clockwise disk of GC young stars. The steeper depletion could also be caused by collisions between stars in the dense GC environment, as predicted by \citet{Rose:2020, Rose:2023}. However, our constraint is not precise enough to significantly demonstrate that current dynamical predictions are lacking. If such a steep decline in stellar binaries beyond what is expected from dynamical processes is indeed present at the GC, future, more precise measurements of the GC stellar binary fraction would be needed to detect it.

While our work demonstrates a high stellar binary fraction for GC young stars outside the central arcsecond, our constraints are not very precise, which future studies can address. The primary limitation in precision of this experiment's results are due to the photometric search for stellar binary systems. As our binary recovery analysis demonstrates (see \S~\ref{ssub:recovery_of_binary_signals} and Figure~\ref{fig:recovery_frac}), photometric surveys with flux precision and observation cadence like our experiment's are not very sensitive to binary systems. This fact, coupled with the typically high stellar variability seen for GC stars \citep{Gautam:2019} makes detecting signatures of binarity difficult photometrically. A future spectroscopic survey of the young stars at our sample's projected distances from the SMBH can constrain the stellar binary fraction much more precisely. Additionally, such a spectroscopic survey can allow characterizing the binary parameters of the GC binary population \citep[such as using the methods of][]{Price-Whelan:2017}, which is not yet possible with the detection of just 3 binaries in our experiment.


\subsection{Other implications for GC environment\\from high stellar binary fraction} 
\label{sub:discussion_gc_environment}
The high binary fraction of the young stars can have broader implications for the galactic center stellar population. The currently high binary fraction among the young, massive stars may suggest that previous in situ star formation episodes at the GC similarly resulted in the massive stars having higher binary fractions than massive stars in the solar neighborhood.

One outcome of typically high stellar binary fractions from past episodes of GC star formation would be a higher  fraction of millisecond pulsars among the GC pulsar population. Typical binary evolution scenarios predict millisecond pulsars form due to accretion onto a neutron star in a binary system \citep[e.g.,][]{Wijnands:1998, Tauris:2006}. The accreting material imparts angular momentum onto the neutron star, leading to the formation of a millisecond pulsar. Due to their lower radio flux \citep[e.g.,][]{Lorimer:2004}, millisecond pulsars are typically more difficult to detect than other types of pulsars in the GC, made especially difficult by the high dispersion and scattering towards the GC environment \citep[e.g.,][]{Bower:2015, Wharton:2019}. This may contribute to the challenge of few detections of pulsars in the GC environment (i.e., the \textit{missing pulsar problem}).

The high stellar binary fraction additionally suggests that the GC (and galactic nuclei in general) are promising sources of gravitational wave signals, such as those observed by the LIGO-Virgo-KAGRA collaboration. Previous dynamical simulations have demonstrated that gravitational interactions with the SMBH may merge two stellar-mass black holes \citep[e.g.,][]{Antonini:2012, Hoang:2018, Wang:2021b}. In these simulations, the merger rate highly depends on the assumed stellar binary fraction. With high stellar binary fractions in the galactic nuclei close to the SMBH, as our experiment has found for the GC, the merger rate can be comparable to other dynamical channels that have been suggested in the literature \citep{Hoang:2018, Stephan:2019}.
Mergers via the EKL mechanism in particular may leave a signature in future gravitational wave detectors planned, such as LISA \citep{Amaro-Seoane:2017}. For example, eccentricity oscillations expected by the EKL mechanism may be detected in future detectors \citep{Hoang:2019}, and the extreme proximity to the GC SMBH can be extracted from the signal \citep{Xuan:2023}.
Thus, the high stellar binary fraction measured for the GC in our experiment implies that future LISA observations may detect a gravitational wave signal from our own Galactic center.


\section{Conclusions} 
\label{sec:conclusions}
In this work, we presented the first measurements of the young star binary fraction for the young, massive stars located in the central $\approx 0.4$ pc of the young nuclear cluster in the GC. This experiment detected 3 stellar binary systems out of 102 known young stars in the region via a search for periodic flux variability: IRS~16SW, S4-258, and S2-36. Out of these three stars, the binary nature of IRS~16SW and S4-258 has been previously reported, while the binary nature of S2-36 is first reported in this work.

We performed an analysis to estimate the sensitivity of our experiment to young binary systems. We simulated light curves for a young star binary population consisting of $\approx 16,000$ mock binary systems. By injecting and attempting to recover these mock binary signals into and from our experiment's observed light curves, we were able to estimate our experiment's sensitivity to a young binary system if present in each of our sample's light curves. This allowed us to then estimate the underlying young star binary fraction as determined by our detection of 3 binaries in this experiment's sample of known young stars.

The GC young, massive star binary fraction is at least $71\%$, at 68\% confidence, or at least $42\%$, at 95\% confidence, in the central $\approx 0.4$ pc of the GC. When specifically considering the stellar population outside the S-star cluster (i.e., outside of the central arcsecond), the young, massive star binary fraction is at least $72\%$, at 68\% confidence, or at least $42\%$, at 95\% confidence. Such a high stellar binary fraction is consistent with or higher than the typically high stellar binary fractions observed in local OB star populations ($\approx 60\%$ -- 70\%).
In addition, the young star binary fraction outside the S-star cluster is significantly higher than that of the young stars in the S-star cluster. It is unlikely that the young B stars in the S-star cluster and the WR and OB stars outside the S-star cluster share the same stellar binary fraction ($p < 1.4\%$).

The observed radial dependence of the binary fraction at the GC is consistent with dynamical predictions for the depletion of binary stars close to the central SMBH via dynamical processes that lead to binary evaporation or binary mergers. Furthermore, the high stellar binary fraction at the GC may also suggest consistency with star formation scenarios with long cooling timescales, but more detailed simulations are required to understand the formation of binaries in in situ models of GC star formation. Future studies of the GC binary population, particularly spectroscopic studies that are more sensitive to stellar binaries, will allow a more precise measurement of the GC young star binary fraction and the underlying population's component star and binary parameters.

\begin{acknowledgements}
We thank the reviewers for their helpful comments.
Support for this work was provided by the W. M. Keck Foundation, the Heising Simons Foundation, the Gordon and Betty Moore Foundation, and the National Science Foundation award numbers 1909554 and 1836016.
M.W.H. is supported by the Brinson Prize Fellowship.
S.N. acknowledges the partial support from NASA ATP 80NSSC20K0505 and from NSF-AST 2206428 grant as well as thanks Howard and Astrid Preston for their generous support. 
We thank Eric E. Becklin for helpful feedback on this experiment's analysis, and Ellen~Van~Wyk for helpful comments on this manuscript's text and plots.
This work used computational and storage services associated with the Hoffman2 Shared Cluster provided by UCLA Office of Advanced Research Computing's Research Technology Group.
The data presented herein were obtained at the W. M. Keck Observatory, which is operated as a scientific partnership among the California Institute of Technology, the University of California, and the National Aeronautics and Space Administration. We thank the staff of the Keck Observatory for their help in obtaining the observation data. The Observatory was made possible by the generous financial support of the W. M. Keck Foundation.
The authors wish to recognize and acknowledge the very significant cultural role and reverence that the summit of Maunakea has always had within the indigenous Hawaiian community. We are most fortunate to have the opportunity to conduct observations from this mountain. 
\end{acknowledgements}

\facilities{Keck:II (NIRC2), Keck:I (OSIRIS)}

\software{
    \textsc{NumPy} \citep{oliphant2006guide, van2011numpy},
    \textsc{Astropy} \citep{Astropy-Collaboration:2013, Astropy-Collaboration:2018, Astropy-Collaboration:2022},
    \textsc{SciPy} \citep{2020SciPy-NMeth},
    \textsc{KAI} \citep{KAI:1_0_0},
    \textsc{gatspy} \citep{VanderPlas:2015, gatspy},
    \textsc{PHOEBE} \citep{Prsa:2016, Horvat:2018, Jones:2020, Conroy:2020},
    \textsc{SPISEA} \citep{Hosek:2020},
    \textsc{OSIRIS} Toolbox \citep{OSIRISdrp, Lockhart:2019},
    \textsc{emcee} \citep{Foreman-Mackey:2013},
    \textsc{Matplotlib} \citep{Hunter:2007, Matplotlib:3_6_2},
    \textsc{Phitter} \citep{phitter:1_0_0},
    \textsc{binary\_fraction} \citep[software written to conduct this experiment's analysis;][]{binary_fraction:1_0_0},
    \textsc{StarKit} \citep{starkit}.
}

\bibliographystyle{aasjournal}
\bibliography{PaperBib,Software}

\appendix

\begin{figure}[t]
  \centering
    \includegraphics[width=0.49\textwidth]{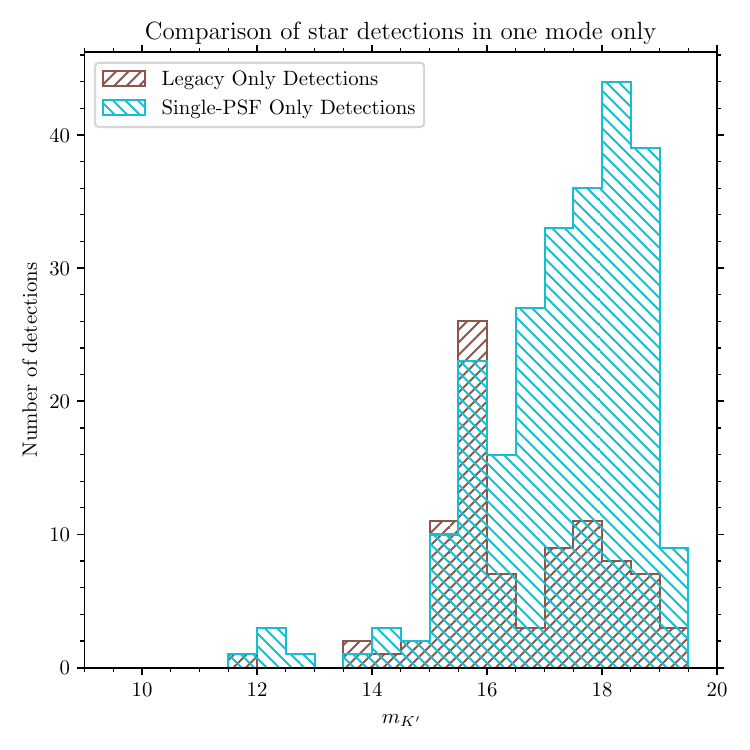}
  \caption{Sources only detected in either Legacy mode or Single-PSF mode, and missed by the other StarFinder mode. Single-PSF mode tends to detect faint stars, particularly in the PSF haloes of bright stars, that are often missed by Legacy mode.
  }
  \label{fig:airopa_dets}
\end{figure}

\begin{figure}[t]
  \centering
    \includegraphics[width=0.49\textwidth]{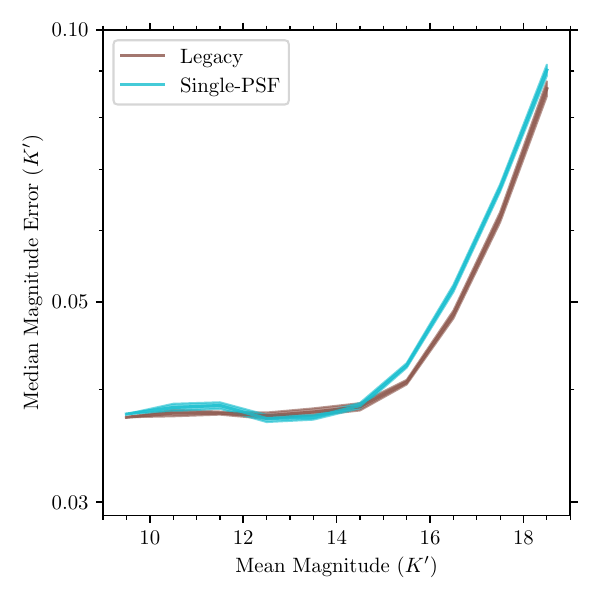}
  \caption{Same as Figure~\ref{fig:mag_magerr}, but comparing photometric flux uncertainties estimated by Legacy mode and by Single-PSF mode. Both modes have comparable flux uncertainties for stars brighter than $m_{K'} \approx 15$.
  }
  \label{fig:airopa_artifact_num_phot_unc_comparison}
\end{figure}

\begin{figure*}[t]
  \centering
    \includegraphics[align=c, width=0.31\textwidth]{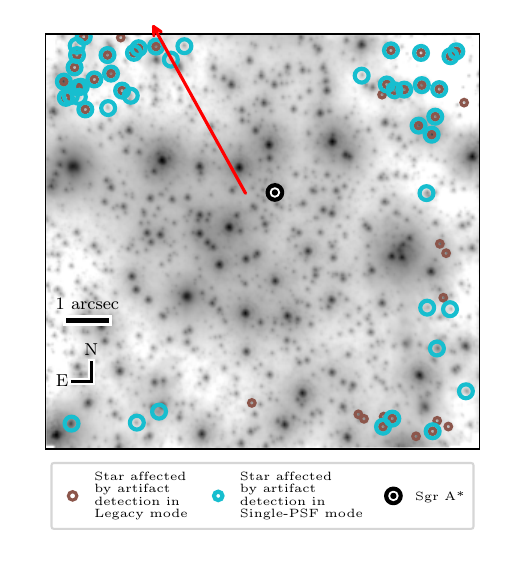}
    \includegraphics[align=c, width=0.375\textwidth]{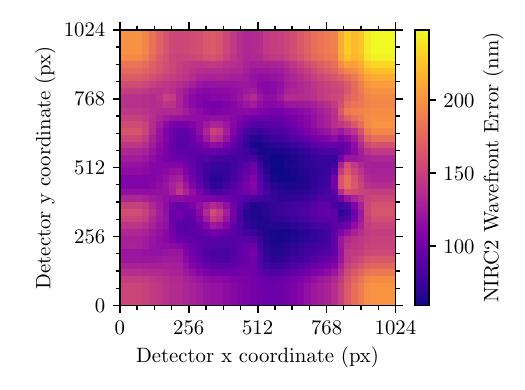}
    \includegraphics[align=c, width=0.30\textwidth]{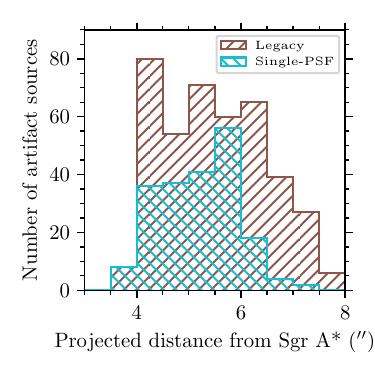}
  \caption{
    \textbf{Left:} Stars affected by the artifact source detections plotted on the field of view. Color of circles around stars indicate in which mode each labeled star had artifact sources associated with it. The base of the red arrow indicates the mean position of the 9 stars used to construct the PSF model used for PSF fitting, weighted by the flux of each PSF star. The red arrow points towards the direction of the tip-tilt star used during AO observations, located approximately $16.9''$~north, $10.1''$~east of Sgr A*.
    \textbf{Middle:} Example map of the NIRC2 instrumental wavefront error. The orientation of all images used for this experiment are the same, and the example image's orientation in the middle panel matches the orientation of the instrumental wavefront error map in the right panel.
    Overall, the combination of the instrumental wavefront error and anisoplanatism due to the position of the tip-tilt star from the science field of view results in the presence of artifact stars and their location on the field of view.
    \textbf{Right:} A comparison of the number of artifact sources detected as a function of distance of Sgr A*. Legacy mode is more often affected by artifact sources.
  }
  \label{fig:artifact_srcs_wfe_map}
\end{figure*}

\section{Use of AIROPA Single-PSF mode} 
\label{sec:use_of_airopa_single_psf_mode}
This experiment implemented AIROPA ``Single-PSF'' mode StarFinder \citep{Witzel:2016} in order to detect stellar sources in our imaging data and estimate their flux. Single-PSF mode StarFinder was designed to offer several improvements for point source detection compared to the ``Legacy" mode StarFinder used in \citetalias{Gautam:2019}. We evaluated the changes due to Single-PSF mode in the context of our experiment's science goals, and we present an overview of the analysis in this section. In this experiment, we used Single-PSF mode because of more detections of faint stars, reduction in artifact sources detected, and consistent photometric uncertainties with Legacy mode.

Single-PSF mode detects more stars across our experiment's field of view. When considering stars detected in at least 30 nights of this experiment's observations and stars confirmed to be true detections via visual inspection, Legacy mode detects 972 stars while single-PSF mode detects 1130 stars. Figure~\ref{fig:airopa_dets} demonstrates the difference in these detections: the discrepancy in the number of detections is largely originating from faint stars, $m_{K'} \gtrsim 16$, where Single-PSF detects many more stars often missed by Legacy mode.

We additionally evaluated the photometric uncertainties estimated from both modes, shown in Figure~\ref{fig:airopa_artifact_num_phot_unc_comparison}. Both Legacy and Single-PSF mode have comparable photometric uncertainties for bright stars (i.e., $m_{K'} \lesssim 15$). Higher uncertainties for fainter stars can be accounted by the higher number of detections of fainter stars.

Single-PSF mode detects fewer \textit{artifact sources} near the edge of the field of view than Legacy mode. Artifact sources are caused by anisoplanatism and instrumental wavefront error, leading to the PSF shape near field edges to be elongated \citep[described by][]{Jia:2019, Gautam:2019}. As a consequence, some elongated single sources can be fit as multiple sources during the PSF fitting routine. Importantly for photometry experiments such as this work, this effect results in a too low estimate for stellar flux since the flux is split across multiple source detections. We implemented the astrometric matching criteria developed by \citet{Jia:2019} to identify artifact sources. The left panel of Figure~\ref{fig:artifact_srcs_wfe_map} shows the location of the artifact sources detected by this experiment's PSF fitting. It also indicates the direction from the experiment field towards the tip-tilt star used for AO observations. The middle panel of Figure~\ref{fig:artifact_srcs_wfe_map} shows an example of the instrumental wavefront error for the NIRC2 imager at Keck Observatory. Artifact sources in the northeast and southwest corners of the field are partly from anisoplanatism due to the off-axis location of the tip-tilt star relative to the science field of view. Artifact sources are also often found in regions of high instrumental wavefront error particularly in the corners and the west edge of the field. 
In Legacy mode, 46 total stars are identified to be affected by artifact sources, across a total of 421 observations for these stars. In Single-PSF mode, a similar number of stars, 48 stars, are affected by artifact sources. However, this issue only affects these stars a total of 213 observations, a reduction by almost a factor of 2. This reduction in artifact sources is illustrated in the right panel of Figure~\ref{fig:artifact_srcs_wfe_map}. 

In the experiment presented in this work, we decided to use Single-PSF detections for the following reasons: Single-PSF mode detects many more stars than Legacy mode, particularly fainter stars in the PSF haloes of bright stars. Single-PSF mode detects fewer artifact sources, which affect the flux estimates for sources near the edge of our experiment's field of view. Finally, Single-PSF mode has comparable photometric precision to Legacy mode.

\begin{deluxetable*}{ll|lll|lll}[h]
    \tablewidth{0pt}
    \tablecolumns{8}
    \tablecaption{Photometric Calibrator Star Bandpass Corrected Reference Fluxes\label{tab:bandpass_corr}}
    \tablehead{
        \colhead{Star Name}     &
        \colhead{Spectral Type} &
        \colhead{$K$s$_{\text{S10}}$}                        &
		\colhead{$K$s$_{\text{S10}} - K'_{\text{NIRC2}}$}    &
        \colhead{$K'_{\text{NIRC2}}$}                                   &
        \colhead{$H_{\text{S10}}$}                        &
		\colhead{$H_{\text{S10}} - H_{\text{NIRC2}}$}    &
        \colhead{$H_{\text{NIRC2}}$}
    }
    \startdata
    IRS 16NW    & Wolf-Rayet    & $10.14 \pm 0.06$   & $-0.05$          & $10.19 \pm 0.06$  & $12.03 \pm 0.06$  & $-0.15 \pm 0.01$  & $12.18 \pm 0.07$  \\
    S3-22       & Late-Type     & $11.03 \pm 0.06$   & $-0.05$          & $11.08 \pm 0.06$  & $13.16 \pm 0.07$  & $-0.18$           & $13.34 \pm 0.07$  \\
    S2-22       & Early-Type    & $12.98 \pm 0.09$   & $-0.05$          & $13.03 \pm 0.09$  & $14.75 \pm 0.08$  & $-0.14 \pm 0.01$  & $14.89 \pm 0.09$  \\
    S4-3        & Late-Type     & $12.91 \pm 0.07$   & $-0.05$          & $12.96 \pm 0.07$  & $14.86 \pm 0.07$  & $-0.16 \pm 0.01$  & $15.02 \pm 0.08$  \\
    S1-1        & Early-Type    & $13.09 \pm 0.07$   & $-0.05$          & $13.04 \pm 0.07$  & $14.95 \pm 0.07$  & $-0.15 \pm 0.01$  & $15.10 \pm 0.08$  \\
    S1-21       & Early-Type    & $13.26 \pm 0.06$   & $-0.07 \pm 0.01$ & $13.33 \pm 0.07$  & $15.60 \pm 0.07$  & $-0.20 \pm 0.01$  & $15.80 \pm 0.08$  \\
    S1-12       & Early-Type    & $13.52 \pm 0.07$   & $-0.05 \pm 0.01$ & $13.57 \pm 0.07$  & $15.41 \pm 0.08$  & $-0.15 \pm 0.01$  & $15.56 \pm 0.09$  \\
    S2-2        & Late-Type     & $13.97 \pm 0.06$   & $-0.04$          & $14.01 \pm 0.06$  & $15.51 \pm 0.06$  & $-0.13 \pm 0.01$  & $15.64 \pm 0.07$  \\
    S3-88       & Late-Type     & $14.23 \pm 0.06$   & $-0.05$          & $14.28 \pm 0.06$  & $16.24 \pm 0.07$  & $-0.17 \pm 0.01$  & $16.41 \pm 0.08$  \\
    S2-75       & Late-Type     & $14.38 \pm 0.08$   & $-0.05$          & $14.43 \pm 0.08$  & $16.46 \pm 0.08$  & $-0.18 \pm 0.01$  & $16.64 \pm 0.09$  \\
    S3-36       & Late-Type     & $14.61 \pm 0.09$   & $-0.05$          & $14.66 \pm 0.09$  & $16.49 \pm 0.09$  & $-0.16 \pm 0.01$  & $16.65 \pm 0.10$  \\
    S1-33       & Early-Type    & $14.96 \pm 0.07$   & $-0.05$          & $15.01 \pm 0.07$  & $16.84 \pm 0.08$  & $-0.15 \pm 0.01$  & $15.01 \pm 0.09$  \\
    \enddata   
    \tablerefs{S10: \cite{Schodel:2010}}
\end{deluxetable*}

\section{Photometric Calibration Details} 
\label{sec:phot_calibration_details}
Table~\ref{tab:bandpass_corr} collects the bandpass corrections for the calibrator stars from the \citet{Schodel:2010} catalog measurements to the respective NIRC2 bandpasses. The bandpass corrected $K'_{\text{NIRC2}}$ and $H_{\text{NIRC2}}$ are the reference calibrator magnitudes used for the absolute photometric calibration step used in this work.

Since the calibrator stars and the bandpass calibration were re-derived for this work compared to \citetalias{Gautam:2019}, we analyzed the differences in flux between the two calibrations.
The updated calibration tends to result in flux estimates for stars being fainter from the calibration presented in \citetalias{Gautam:2019} by $\approx 0.09$ mags. The median of the difference in stellar flux estimates derived from the two photometric calibrations is $m_{K'\text{,2022calib}} - m_{K'\text{,2018calib}} = 0.090 \pm 0.005$.


\begin{figure*}[p]
  \centering
    \includegraphics[width=\textwidth]{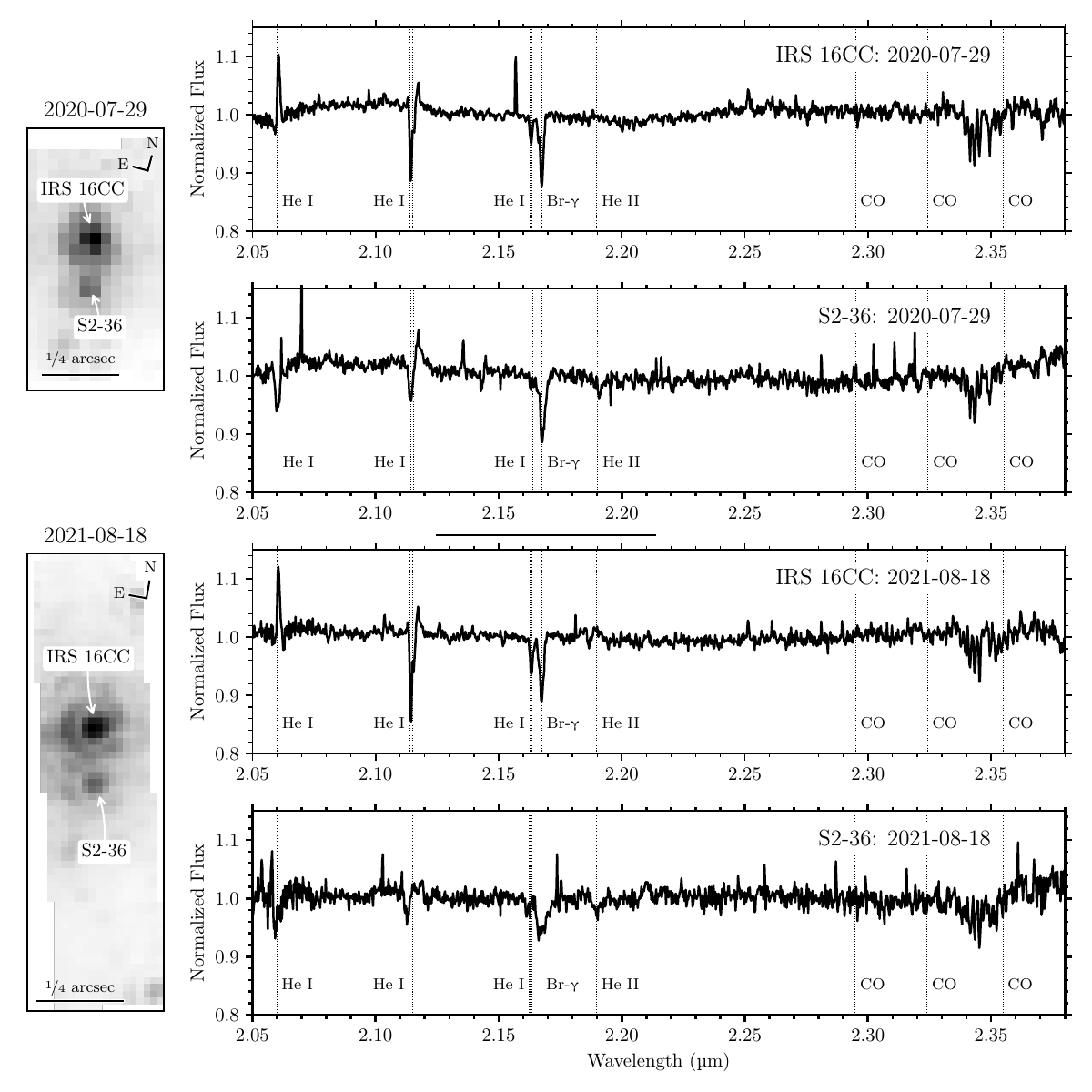}
  \caption{Overview of the two OSIRIS IFS observations taken of S2-36 to verify that S2-36 is an early-type binary star. The top half of the plot shows the 2020-07-29 observation, taken with the 35 \nicefrac{mas}{pixel} IFS configuration, while the bottom half shows the 2021-08-18 observation, taken with the 20 \nicefrac{mas}{pixel} IFS configuration. For each observation, the left panel displays the IFS data cube collapsed along wavelength to show the field of view surrounding S2-36 in the observation. The right panels show the extracted spectra of both IRS~16CC and S2-36 (top and bottom panels for each observation, respectively). In each spectrum, reference spectral line wavelengths are indicated, shifted to the best fit RV from a one-star spectral model fit to the given spectrum. From the observed spectra IRS~16CC and S2-36 are both early-type stars, primarily due to the presence of the Hydrogen \brgamma absorption line at 2.16 \um~and the lack of CO band head absorption lines at $\approx 2.3$ \um~that would be present for late-type stars with cooler atmospheres. Furthermore, S2-36's observed spectra in both observations has some differences from those of IRS~16CC in both observations, notably in the He I lines, suggesting lack of contamination from IRS~16CC's flux due to our background subtraction procedure.}
  \label{fig:S2-36_spec_overview}
\end{figure*}

\begin{figure*}[htb!]
  \centering
    \includegraphics[width=\textwidth]{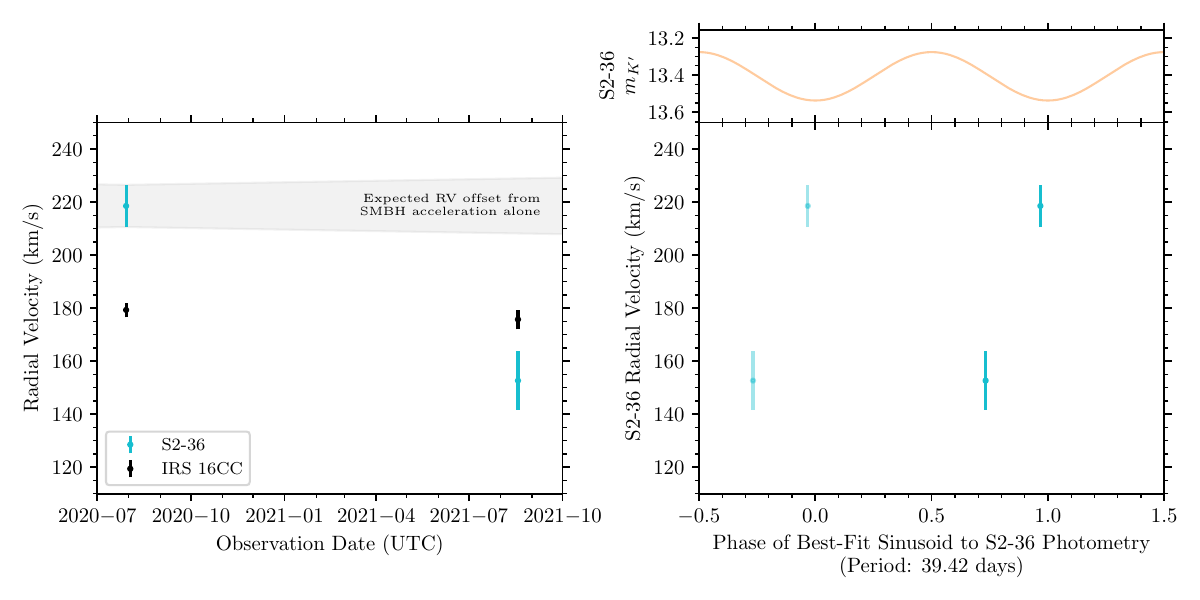}
  \caption{\textbf{Left:} Radial Velocity (RV) measurements of the stars S2-36 and IRS~16CC in our spectroscopic observations. IRS~16CC does not have a significant RV variation between the 2020-07 and 2021-08 observations. However, S2-36 does have significant RV variation between the two observations, regardless of the single-star or the double-star spectral fit to derive RV for the 2021-08 observation. The grey band shows expected RV offset from the 2020-07 observation's RV if S2-36's RV variation originated from the gravitational acceleration of the SMBH alone. S2-36's RV variation can not be completely accounted for by the SMBH's acceleration, demonstrating that it is a spectroscopic binary.
  \textbf{Right:} RV measurements of S2-36 phased to the best-fit sinusoid to S2-36's photometric variability. For reference, the best-fit sinusoid model to S2-36's photometric variability is shown in the top right panel. The 2021-08 observation was conducted at phase $\approx 0.75$ and the 2020-07 observation was conducted at phase $\approx 1.0$.
  }
  \label{fig:S2-36_rv_points}
\end{figure*}

\begin{deluxetable*}{lDccccc|C}[t]
    \decimals
    \tablecolumns{11}
    \tablecaption{OSIRIS spectroscopic observations and RV measurements of S2-36\label{tab:S2-36_Observations_Table_Spectra}}
    \tablehead{
        \colhead{Date}                      &
        \multicolumn2c{MJD}                 &
        \colhead{Wavelength Range}          &
        \colhead{Frames}                    &
        \colhead{Int. Time}                 &
        \colhead{Scale}                     &
        \colhead{SNR}                       &
        \colhead{RV}                        
        \\
        \colhead{(UT)}                      &
        \multicolumn2c{}                    &
        \colhead{(\textmu m)}               &
        \colhead{}                          &
        \colhead{(s)}                       &
        \colhead{(mas/pixel)}               &
        \colhead{}                          &
        \colhead{(\nicefrac{km}{s})}
    }
    \startdata
    2020-07-29                  & 59059.300   & 1.965 -- 2.381 ($K_{\text{bb}}$)  & 6    & 900   & 35    & 73   & 218.1 \pm 8.1     \\
    2021-08-18\tablenotemark{T} & 59444.249   & 1.965 -- 2.381 ($K_{\text{bb}}$)  & 4    & 900   & 20    & 118  & 152.9 \pm 13.4    \\
    \enddata
    \tablenotetext{T}{denotes observation conducted with TRICK, the NIR tip-tilt sensor on Keck I, allowing for improved AO correction.}
\end{deluxetable*}

\begin{deluxetable}{lDc|C}[h]
    \decimals
    \tablecolumns{5}
    \tablecaption{RV measurements of IRS~16CC\label{tab:16CC_Observations_Table_Spectra}}
    \tablehead{
        \colhead{Date}                      &
        \multicolumn2c{MJD}                 &
        \colhead{SNR}                       &
        \colhead{RV}
        \\
        \colhead{(UT)}                      &
        \multicolumn2c{}                    &
        \colhead{}                          &
        \colhead{(\nicefrac{km}{s})}
    }
    \startdata
    2020-07-29                  & 59059.300   & 182   & 179.3 \pm 2.8 \\
    2021-08-18\tablenotemark{T} & 59444.249   & 165   & 175.7 \pm 3.7 \\
    \enddata
    \tablenotetext{T}{denotes observation conducted with TRICK.}
\end{deluxetable}


\section{Spectroscopic confirmation of S2-36 as a young, binary star system}
\label{sec:S2-36_spec_binary_confirmation}
We performed spectroscopy of S2-36 to determine if its observed periodic photometric variability is due to a young stellar binary system. With two spectroscopic observations, we verified that S2-36 is an early-type star belonging to the young, massive star population at the GC and that it is a binary system with significant radial velocity (RV) variations that cannot be accounted by its orbit around the central SMBH at Sgr A*.

\subsection{Observations and spectral extraction} 
\label{sub:observations_and_spectral_extraction}
We performed two laser-guide-star adaptive optics (LGSAO) NIR spectroscopic observations of S2-36 at the 10 m W. M. Keck I telescope using the OSIRIS integral field spectrograph \citep[IFS; ][]{Larkin:2006}: on 2020 July 29 and on 2021 August 18. Both observations were performed in the $K_{\text{bb}}$ bandpass (wavelength range: 1.965~\textmu m -- 2.381~\textmu m).
The 2020 observation had a pixel scale of 35 \nicefrac{mas}{pixel}, while the 2021 observation was performed with a higher spatial resolution of 20 \nicefrac{mas}{pixel} and used Keck I's NIR tip-tilt sensor \citep[TRICK,][]{Wizinowich:2014} allowing better AO correction and separation of S2-36's flux from the brighter nearby star IRS~16CC. 
We reduced the OSIRIS IFS data using the latest version of the OSIRIS Data Reduction Pipeline \citep{OSIRISdrp, Lockhart:2019}.
An overview of the two spectroscopic observations is provided in Table~\ref{tab:S2-36_Observations_Table_Spectra}.

The stars S2-36 ($\overline{m}_{K'} = 13.4$) and IRS~16CC ($\overline{m}_{K'} = 11.0$) are only separated by $\approx 160$ mas on sky, presenting a challenge for the extraction of S2-36's spectra from the IFS observation data. Due to the proximity, additional care was required to properly subtract the background when extracting S2-36's spectrum in order to reduce the impact on it from the brighter star IRS~16CC's flux. Our spectrum extraction procedures followed those outlined by \citet{Do:2013a} with the following parameters: for the 2020-07 spectral observation, we used an extraction aperture with a radius of 1.5 pixels (52.5 mas). For sky and background subtraction in the 2020-07 observation, we used the median flux values from a $3 \times 3$ pixel ($105 \times 105$ mas) box in an empty region, located $\approx 2$ pixels ($\approx 70$ mas) south and $\approx 7$ pixels ($\approx 245$ mas) west of S2-36. For the 2021-08 observation, the extraction aperture was a radius of 1.5 pixels (30 mas). Sky and background subtraction for the 2021-08 observation was performed using a 2 pixel annulus (40 mas) around the extraction aperture, as was done in works such as \citet{Do:2013a}. Using other box regions for background subtraction rather than the annulus subtraction on the 2021-08 observation made a negligible difference to the extracted spectrum, likely due to the high angular resolution provided by the 20 \nicefrac{mas}{pixel} plate scale and improved AO correction from TRICK. The OSIRIS IFS field of view surrounding S2-36 in each of our spectroscopic observations and the extracted spectra of the stars IRS~16CC and S2-36 in both observations are shown in Figure~\ref{fig:S2-36_spec_overview}.


\subsection{Spectral typing of S2-36 as an early-type star} 
\label{sub:spec_early_type}
From the flux alone, S2-36 is either an early-type star (spectral type O or B), indicating that it is a member of the young star population at the GC, or a late-type giant star (spectral type M or K), belonging to the old star population making up the nuclear star cluster \citep[see e.g.,][]{Do:2013a, Do:2019}. In the $K_{\text{bb}}$ bandpass, the primary distinction between the two spectral types are the Hydrogen \brgamma line (2.16 \um) and the He I lines (2.06 \um, 2.11 \um, and 2.16 \um), present in early-type stars with hot atmospheres and not present in late-type giants with cooler atmospheres, and the CO rovibrational band head absorption lines (2.294 \um), present in late-type stars with cool atmospheres and not present in early-type stars \citep[see e.g.,][]{Buchholz:2009, Do:2013a}. As Figure~\ref{fig:S2-36_spec_overview} demonstrates, both spectra of S2-36 indicate S2-36 is an early-type star, like the nearby IRS~16CC \citep[typed by][]{Do:2009b}, due to the presence of \brgamma and He absorption lines in its spectra and lack of the CO band head. Furthermore, S2-36 appears to have an ionized Helium (He II) absorption line at 2.189 \um, typically only present in hot stars with $\teff \gtrsim 32,000$ K \citep{Hanson:2005}. These spectra demonstrate that S2-36 is likely a member of the young, massive star population in the GC.


\subsection{Measurement of a significant radial velocity difference for S2-36} 
\label{sub:sig_RV_diffs}
In order to measure radial velocities from S2-36's spectral observations, we performed spectral template fitting, expanding on a similar method to that described by \citet{Chu:2023}. We fit the observed spectra to model stellar atmosphere spectra in the BOSZ stellar atmosphere grid \citep{Bohlin:2017}. Interpolation of the stellar atmosphere grid was performed using the \textsc{StarKit} package \citep{starkit} and we implemented an MCMC fitting routine with \textsc{emcee} \citep{Foreman-Mackey:2013} in order to derive the best-fit and uncertainties to the 6 physical parameters fit for each star: surface temperature ($T_{\text{eff}}$), surface gravity ($\log g$), overall metallicity ($\left[\frac{\text{M}}{\text{H}}\right]$), alpha-element abundance ($\left[\frac{\alpha}{\text{Fe}}\right]$), rotational velocity ($v_{\text{rot}}$), and the line-of-sight radial velocity ($v_{\text{rad}}$ or RV). As noted by \citet{Do:2019a}, when measuring RV, complete spectral fitting (rather than a Gaussian fit to the Hydrogen \brgamma line as is sometimes done) leads to smaller uncertainties and less bias for early-type stars in the NIR $K$-band.
The RV measurements that we derived from the two observations of S2-36 are listed in Table~\ref{tab:S2-36_Observations_Table_Spectra}, while Table~\ref{tab:16CC_Observations_Table_Spectra} lists RV measurements of the nearby star IRS~16CC. A more detailed evaluation of the radial velocities in the context of possible binary models for S2-36 will be presented in a future publication (Gautam et al., in prep.).

We measured a significant RV difference for S2-36 between the two observations, indicative of a spectroscopic binary system.
Observed RVs for S2-36 and IRS~16CC are shown in Figure~\ref{fig:S2-36_rv_points} and listed in Tables~\ref{tab:S2-36_Observations_Table_Spectra} and \ref{tab:16CC_Observations_Table_Spectra}. The measurement of S2-36's RV in both observations is significantly different than that of IRS~16CC, suggesting lack of contamination in the spectral extraction of S2-36.
At the GC, another source of RV variation is the orbital motion of stars around the central SMBH. Using the acceleration on the plane of the sky measured with the proper motion of S2-36 \citep{Sakai:2019} and the SMBH mass and distance \citep{Do:2019a}, we calculated an estimate of the line-of-sight acceleration on S2-36 due to the SMBH's gravitational acceleration: $a_z = 1.1^{+1.2}_{-0.5} \frac{\text{km}}{\text{s yr}}$ ($3 \sigma$ or 99.7\% confidence). Therefore, between the two spectroscopic observations taken, $\Delta \text{RV} = 1.2^{+1.3}_{-0.5}$ \nicefrac{km}{s} is expected from the SMBH on S2-36, which is much smaller than the observed RV difference.
Therefore, our two spectroscopic observations confirm that S2-36 is indeed an early-type stellar binary, belonging to the GC young star population.


\section{Tests on mock binaries with the Box Least Squares periodogram} 
\label{sec:BLS_pdgram_tests}

We performed a series of tests of the Box Least Squares periodogram (BLS) method \citep{Kovacs:2002} on model binary light curves (described in \S~\ref{ssub:mock_binary_population} and \S~\ref{ssub:mock_binary_light_curves}) that have eclipses narrow in phase (i.e., eclipse widths extending up to $\approx 10\%$ of the phase) injected into our sample’s $K'$-band stellar light curves (described in \S~\ref{ssub:injection_of_binary_signals}) to evaluate the method's effectiveness at detection of young binaries in our dataset. In particular, we focused on mock binaries injected into the light curves of bright stars in our sample ($m_{K'} \lesssim 12$) that do not exhibit flux variability. These stellar light curves have low magnitude uncertainties and stable flux across the experiment's time baseline, making them the most ideal targets in which mock binaries may be detected via the BLS method. In our simulated light curves, binary systems with eclipses narrow in phase are typically systems where the component stars are further separated with orbital periods on the order of $\gtrsim 10$ days. In each injected light curve where we performed a test of the BLS method, orbital periods were $\approx 10$ days and $K'$-band eclipse depths were $\approx 0.5$ mag.

We were not able to recover any of the injected binary signals using the BLS method at the injected binary period for our test binaries with narrow eclipses. In each of these test cases, $\approx 5$ -- 10 $K'$-band observations were during an eclipse. However, the BLS periodicity search was not able to identify a period with these eclipses, and instead detected much stronger powers at a period of 1 day, originating from our daily observation cadence and associated aliases. On fainter stars in our experiment where the photometric uncertainty is larger, we expect that the BLS method will perform worse due to higher noise.
In order for the BLS periodicity search to be effective for detections of short eclipse binary systems at the GC, a much more frequent observation cadence and more observations overall will be required than our current experiment.

\begin{figure}[b]
  \centering
    \includegraphics[width=0.49\textwidth]{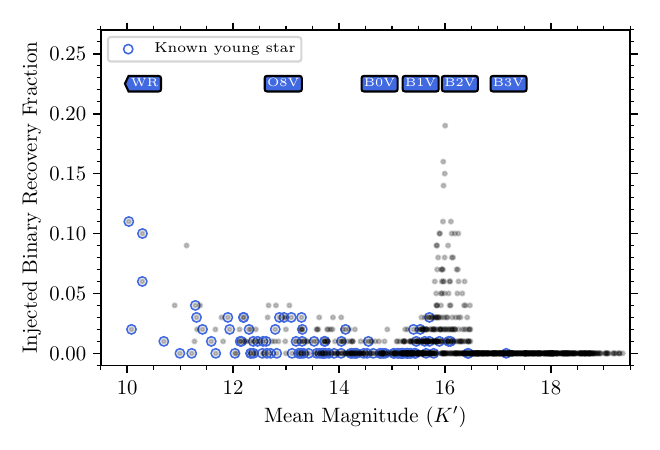}
  \caption{Recovery fraction for injected binaries into each of our sample star's light curves, with points corresponding to known young stars circled in blue. The recovery fraction of binary signals in the known young star light curves was used to determine the intrinsic binary fraction of the GC young star population.}
  \label{fig:recovery_frac}
\end{figure}

\begin{figure*}[htb!]
  \centering
    \includegraphics[width=\textwidth]{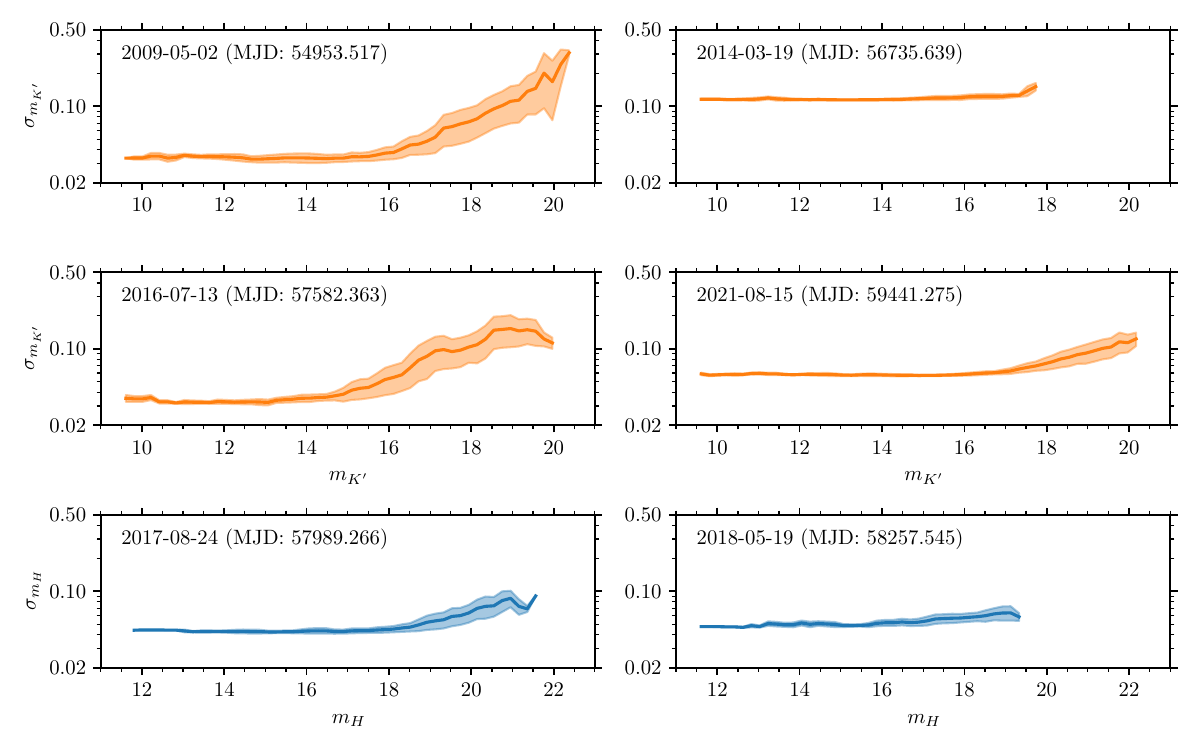}
  \caption{Examples of the flux vs. flux uncertainty relationship for six example observation nights is plotted as the solid line, indicated as magnitude ($m$) and median magnitude error ($\sigma_m$). The shaded area indicates the median absolute deviation around the median. Orange indicates $K'$-band observations, while blue indicates $H$-band observations. These trends were used to include flux uncertainty for each observation when injecting mock binary signals into observed stellar fluxes. For a given stellar flux to inject, the corresponding flux uncertainty on the observation date was sampled from this relationship, modulated by the median absolute deviation.}
  \label{fig:obs_unc_examples}
\end{figure*}

\begin{figure*}[p]
  \centering
    \includegraphics[width=\textwidth]{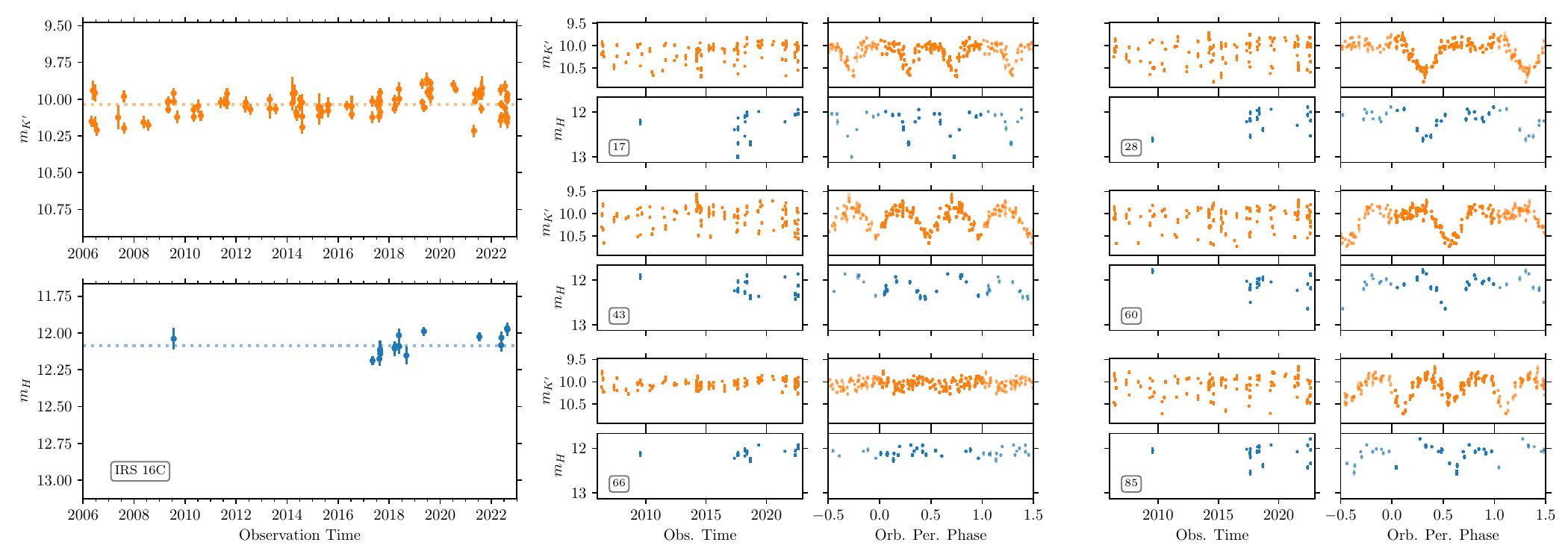}
    \\---\\
    \includegraphics[width=\textwidth]{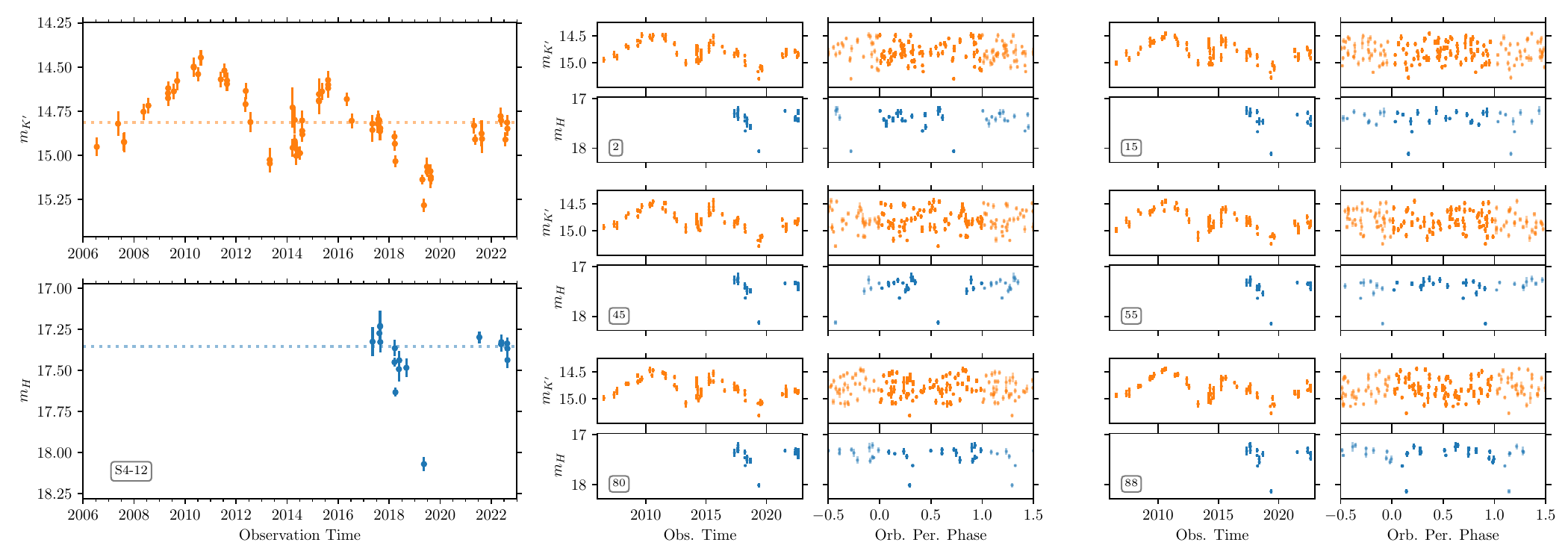}
    \\---\\
    \includegraphics[width=\textwidth]{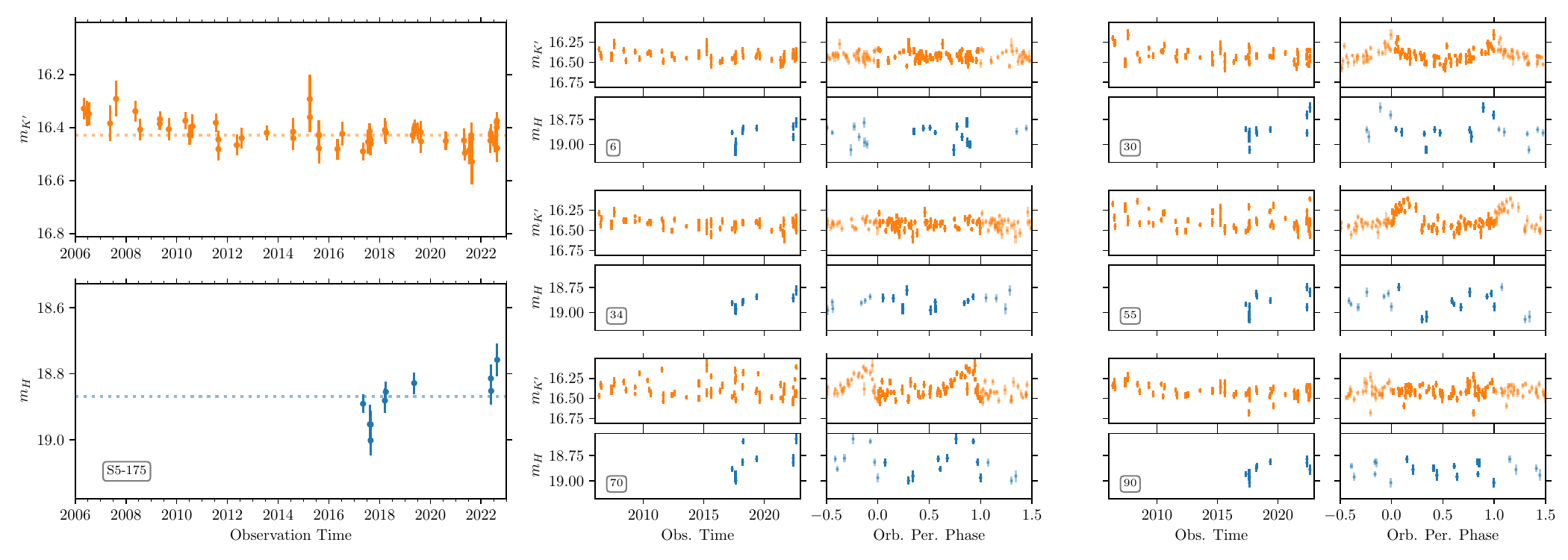}
  \caption{Examples of 3 of our sample's stellar light curves (IRS~16C, S4-12, and S5-175; panels in the left column), each with 6 of the 100 light curves with injected binary signals (right 4 columns). Orange and blue points indicate $K'$- and $H$-band observations, respectively. The left column panels in each example of an injected light curve show all observations arranged by observation time, while the right column panels in each example arrange all observations by the injected mock binary's orbital period.}
  \label{fig:sbv_examples}
\end{figure*}

\section{Additional details about the injection and recovery of mock binary signals} 
\label{sec:additional_deets_injection_recovery}
In order to modulate mock binary light curves before injecting them into sample star observations, we calculated the flux vs. flux uncertainty relationship for each observation in bins of half magnitude. In each bin, we calculated the median magnitude uncertainty (median $\sigma_{m}$) and the median absolute deviation in the magnitude uncertainty. The flux vs. flux uncertainty relationship for six example observations from our dataset is shown in Figure~\ref{fig:obs_unc_examples}.

Three examples from the mock binary variability injection procedure used in our experiment are shown in Figure~\ref{fig:sbv_examples}.
Of these examples, IRS~16C is a bright star with no long-term flux trends, so we expect any binary variability to be obvious in the phased light curves. Furthermore, we expect many large, contact systems to reside at IRS~16C's bright flux, and injected light curves in such bright stars often exhibit the quasi-sinusoidal variability expected from such contact binary systems.
On the other hand, the long-term flux variability in S4-12's light curve masks almost all binary variability if present. The binary variability in its phased light curves is masked by the high amplitude long-term variability.
Finally, S5-175's light curve is generally stable over the experiment time baseline, but we expect fewer contact binaries at its dimmer flux. In this regime, binary systems with flux variability will often have narrow features (like eclipses) in phase that our experiment's time sampling is likely to miss.

Figure~\ref{fig:recovery_frac} plots the recovery fraction of injected binaries for all stars and for the stars used for our binary fraction results. These recovery fractions are also listed for every star in our complete sample in Table~\ref{tab:stellar_sample} in Appendix~\ref{sec:stellar_sample}.


\section{Complete Stellar Sample} 
\label{sec:stellar_sample}

Table~\ref{tab:stellar_sample} is a list of all stars making up this experiment's sample. Notably, we list the injected binary recovery fraction for each star in our sample. This recovery fraction will allow this experiment's periodicity search results to place tighter constraints on the young star binary fraction once the age of additional stars in the experiment field of view is confirmed with deeper spectroscopic observations \citep[e.g., spectroscopic observations with the James Webb Space Telescope and ``extremely large telescope'' facilities, as described by][]{Do:2019}.


\begin{longrotatetable}
\begin{deluxetable}{l|DD|r|rr|DD|cc|D|DDD}
    \tablehead{
        \colhead{Star} &
        \multicolumn2c{$\overline{m}_{K'}$} & \multicolumn2c{$\overline{m}_{H}$} &
        \colhead{Star Age} &
        \colhead{$K'$} & \colhead{$H$} &
        \multicolumn2c{$\chi^2_{\text{red, }K'}$} & \multicolumn2c{$\chi^2_{\text{red, }H}$} &
        \colhead{$K'$} & \colhead{$H$} &
        \multicolumn2c{Injected Binary} &
        \multicolumn2c{$x_0$} & \multicolumn2c{$y_0$} & \multicolumn2c{$t_0$} \\
        \colhead{} &
        \multicolumn2c{} & \multicolumn2c{} &
        \colhead{} &
        \colhead{Nights} & \colhead{Nights} &
        \multicolumn2c{} & \multicolumn2c{} &
        \colhead{Var.?} & \colhead{Var.?} &
        \multicolumn2c{Recovery Fraction} &
        \multicolumn2c{($''$ E of}  & \multicolumn2c{($''$ N of}    & \multicolumn2c{}  \\
        \colhead{}  &
        \multicolumn2c{} & \multicolumn2c{} &
        \colhead{}  &
        \colhead{}  & \colhead{}    &
        \multicolumn2c{}    & \multicolumn2c{}  &
        \colhead{}  & \colhead{}    &
        \multicolumn2c{}    &
        \multicolumn2c{Sgr A*)} & \multicolumn2c{Sgr A*)}   & \multicolumn2c{}
    }
    \tablecaption{Stellar sample\label{tab:stellar_sample}}
\decimals
\startdata
IRS 16C & 10.03 & 12.07 & Known Young & 100 & 19 & 4.20 & 3.48 & Yes & No & 0.11 & 1.05 & 0.55 & 2009.989 \\
IRS 16SW & 10.08 & 12.16 & Known Young & 100 & 19 & 16.02 & 26.02 & Yes & Yes & 0.02 & 1.11 & -0.95 & 2009.820 \\
IRS 16NW & 10.28 & 12.34 & Known Young & 100 & 19 & 1.89 & 0.71 & Yes & No & 0.06 & 0.08 & 1.22 & 2010.047 \\
IRS 33E & 10.29 & 12.51 & Known Young & 100 & 19 & 1.22 & 0.61 & No & No & 0.10 & 0.71 & -3.14 & 2010.182 \\
S2-17 & 10.69 & 12.74 & Known Young & 100 & 19 & 1.10 & 2.48 & No & No & 0.01 & 1.34 & -1.88 & 2010.154 \\
S5-89 & 10.90 & 13.30 & Known Old & 99 & 19 & 1.23 & 1.27 & No & No & 0.04 & -0.79 & -5.25 & 2010.162 \\
IRS 16CC & 10.99 & 13.46 & Known Young & 100 & 19 & 80.36 & 67.19 & Yes & Yes & 0.00 & 1.98 & 0.60 & 2010.135 \\
S3-22 & 11.12 & 13.46 & Known Old & 100 & 19 & 0.38 & 0.53 & No & No & 0.09 & -0.34 & -3.21 & 2010.201 \\
IRS 16SW-E & 11.22 & 14.36 & Known Young & 92 & 19 & 5.61 & 12.51 & Yes & Yes & 0.00 & 1.90 & -1.12 & 2010.045 \\
S6-12 & 11.27 & 13.74 & Unknown Age & 93 & 19 & 3.49 & 3.39 & Yes & No & 0.01 & -0.21 & -6.08 & 2010.161
\enddata
\tablecomments{Table~\ref{tab:stellar_sample} is published in its entirety in the machine-readable format.
      A portion is shown here for guidance regarding its form and content.}
\end{deluxetable}
\end{longrotatetable}

\end{document}